\documentclass[aps,prd,superscriptaddress,amsmath,amsfont,amssymb,preprint,nofootinbib]{revtex4}
\pdfoutput=1
\allowdisplaybreaks
\usepackage{graphicx}
\usepackage{ bbold }
\usepackage{subfigure}
\usepackage{slashed}
\usepackage[english]{babel}

\newcommand{\C}{{\cal C}}
\newcommand{\Q}{{\cal Q}}
\newcommand{\todo}[1]{{\color{red} \ifmmode\else[todo]\fi #1}}
\usepackage[usenames,dvipsnames]{xcolor}
     \definecolor{hgreen}{rgb}{0,.3,0}
     \definecolor{hred}{rgb}{.3,0,0}
     \definecolor{hblue}{rgb}{0,0,.3}
     \definecolor{LightGray}{gray}{0.95}
\usepackage[backref=page,
            colorlinks=true,
            linkcolor=hblue,
            citecolor=hgreen,
            filecolor=hblue,
            urlcolor=hred]{hyperref}

\renewcommand*{\backref}[1]{}

\newcommand{\beq}{\begin{equation} }
\newcommand{\eeq}{\end{equation}} 
\newcommand{\bi}{\begin{itemize} }
\newcommand{\ei}{\end{itemize} }

\definecolor{Red}{rgb}{1.,0.,0.}
\definecolor{Grn}{rgb}{0.,0.75,0.}
\definecolor{Blu}{rgb}{0.,0.,1.}

\DeclareMathOperator{\diag}{diag}
\DeclareMathOperator{\Tr}{Tr}

\definecolor{pan624}{rgb}{0.482,0.635,0.588} 
\definecolor{pan576}{rgb}{0.412,0.569,0.231} 
\definecolor{pan129}{rgb}{0.961,0.812,0.278}
\definecolor{pan5405}{rgb}{0,0.129,0.278} 
\definecolor{shadecolor}{rgb}{0.482,0.635,0.588}
\definecolor{mygray}{HTML}{666666}
\definecolor{x11steelblue}{HTML}{4682B4}
\definecolor{x11firebrick}{HTML}{B22222}
\definecolor{x11forestgreen}{HTML}{228B22}

\usepackage{tikz}
\tikzstyle{every picture}+=[remember picture]
\usetikzlibrary{shapes.geometric}
\usetikzlibrary{calc}
\usetikzlibrary{decorations.markings}
\usetikzlibrary{decorations.text}
\usetikzlibrary{patterns}
\usetikzlibrary{backgrounds}
\usetikzlibrary{positioning}
\tikzstyle arrowstyle=[scale=2]
\tikzstyle directed=[postaction={decorate,decoration={markings,
		mark=at position 0.6 with {\arrow[arrowstyle]{>}}}}]
\tikzstyle rarrow=[postaction={decorate,decoration={markings,
		mark=at position 0.999 with {\arrow[arrowstyle]{>}}}}]
		
\newcommand{\lrpartial}{\negthickspace\stackrel{\leftrightarrow}{\partial}\negthickspace{}}

\newcommand{\lrnabla}{\negthickspace\stackrel{\leftrightarrow}{\nabla}\negthickspace{}}
\newcommand{\lnabla}{\negthickspace\stackrel{\leftarrow}{\nabla}\negthickspace{}}

\graphicspath{{./figs/}}

\newcommand{\1}{\mathbb{1}}
\newcommand{\ncdot}{\negthinspace \cdot \negthinspace}

\begin{document}

\title{Chiral Effective Theory of  Dark Matter Direct Detection}

\def\Cincy{Department of Physics, University of Cincinnati, Cincinnati, Ohio 45221,USA}
\def\UCSD{Department of Physics, University of California-San Diego, La Jolla, CA 92093, USA}
\def\Mainz{PRISMA Cluster of Excellence \& Mainz Institute for Theoretical
Physics, Johannes Gutenberg University, 55099 Mainz, Germany}
\def\TUD{Fakult\"at f\"ur Physik, TU Dortmund, D-44221 Dortmund, Germany} 
\def\CERN{CERN, Theory Division, CH-1211 Geneva 23, Switzerland}
\def\Oxford{Rudolf Peierls Centre for Theoretical Physics, University of Oxford OX1 3NP Oxford, United Kingdom}

\author{\textbf{Fady Bishara}}
\email{fady.bishara AT physics.ox.ac.uk}
\affiliation{\Oxford}

\author{\textbf{Joachim Brod}}
\email{joachim.brod AT tu-dortmund.de}
\affiliation{\TUD}

\author{\textbf{Benjamin Grinstein}}
\email{bgrinstein AT ucsd.edu}
\affiliation{\UCSD}

\author{\textbf{Jure Zupan}} 
\email{zupanje AT ucmail.uc.edu}
\affiliation{\Cincy}
\affiliation{\CERN}

\date{\today}

\begin{abstract}
We present the effective field theory for dark matter interactions
with the visible sector that is valid at scales of ${\mathcal
  O}(1\,{\rm GeV})$. Starting with an effective theory describing the
interactions of fermionic and scalar dark matter with quarks, gluons
and photons via higher dimension operators that would arise from
dimension-five and dimension-six operators above electroweak scale, we
perform a nonperturbative matching onto a heavy baryon chiral
perturbation theory that describes dark matter interactions with light
mesons and nucleons. This is then used to obtain the coefficients of
the nuclear response functions using a chiral effective theory
description of nuclear forces. Our results consistently keep the
leading contributions in chiral counting for each of the initial
Wilson coefficients.
\end{abstract}

\pacs{--pacs--}

\preprint{DO-TH 16/28}
\preprint{OUTP-16-24P}

\maketitle
\tableofcontents


\section{Introduction}
\label{sec:Intro}
Dark Matter (DM) scattering in direct detection lends itself well to
an Effective Field Theory (EFT) description \cite{Fan:2010gt,
  Fitzpatrick:2012ix, Cirigliano:2012pq, Fitzpatrick:2012ib,
  DelNobile:2013sia, Anand:2013yka, Barello:2014uda, Hill:2014yxa,
  Hoferichter:2015ipa, Catena:2014uqa, Kopp:2009qt, Hill:2013hoa,
  Hill:2011be, Hoferichter:2016nvd, Kurylov:2003ra, Pospelov:2000bq,
  Bagnasco:1993st}. DM scattering on nuclei can be taken to be
nonrelativistic, since, in order to be gravitationally bound in the DM
halo, the DM velocity needs to be below about $600$~km/s. The typical
DM velocity in the halo is thus $|\vec v_\chi| \sim 10^{-3}$. The
maximal recoil momentum transfer depends on the reduced mass of the
DM-nucleus system and on the range of recoil energies, $E_R$, that the
experiments are measuring. The recoil energy is typically kept in the
range of a few keV to few tens of keV, while the heaviest nuclei have
masses of $m_A\sim 100$ GeV. This gives a maximal momentum transfer of
\begin{equation}
q_{\rm max}\lesssim 200{\rm ~MeV}.
\label{eq:qmax}
\end{equation}
This is also a typical size of the momenta exchanged between the
nucleons bound inside the nucleus. The maximal recoil momentum is much
smaller than the proton and neutron masses, $q\ll m_N$, so that the
nucleons remain nonrelativistic also after scattering and the nucleus
does not break apart. One can then use the chiral EFT (ChEFT) approach
to nuclear forces to organize different terms using an expansion in
$q/\Lambda_{\rm ChEFT}\sim m_\pi/\Lambda_{\rm ChEFT}\sim 0.3$.

In this paper we perform such a systematic treatment of DM direct
detection. We start from an EFT that describes couplings of DM to
quarks, gluons and photons through higher dimension operators, keeping
only the terms that would arise from dimension-five and dimension-six
operators above the electroweak scale. We then match nonperturbatively
onto a theory that describes DM interactions with light mesons, i.e.,
Chiral Perturbation Theory (ChPT) with DM, and to a theory that also
includes DM interactions with protons and neutrons, i.e., Heavy Baryon
Chiral Perturbation Theory (HBChPT).  A single insertion of DM
interaction with either a light meson, or with a nucleon, then induces
the scattering of DM on the nucleus. We are able to compare the
parametric sizes of different contributions by using chiral counting
within ChEFT of nuclear forces. We keep the leading contributions in
chiral counting and calculate the resulting coefficients that multiply
the nuclear response functions of
Ref. \cite{Anand:2013yka,Fitzpatrick:2012ix}, treating $q^2$ as an
external parameter.

The EFT description of DM -- nucleus scattering is valid if the
mediators between the DM and the visible sector are heavier than
${\mathcal O}(1{\rm GeV})$, and therefore covers a wide range of
UV-complete theories of DM.  Our expressions extend previous results
on direct detection scattering rates. We cover both fermionic and
scalar DM, systematically keeping the leading terms in chiral
counting. Special care is needed, for instance, in the evaluation of
the product of axial-vector DM and vector quark currents, as well as
the product of vector DM and axial-vector quark currents. These
products vanish in the long wavelength limit where both the relative
velocity between DM and nucleus, $\Delta v$, and the momentum
exchange, $q$, are becoming arbitrarily small ($\Delta v, q\to
0$). The leading contributions thus follow from higher orders in a
derivative expansion of the interactions.

The chiral counting also allows for a systematic assignment of
uncertainties on the predictions. Since we restrict the analysis to
the leading order in chiral counting, the errors on the predictions
are expected to be of ${\mathcal O}(30\%)$. Furthermore, we use the
chiral counting to discuss higher-order corrections in the direct
detection rates. The short-distance scattering on two nucleons is, for
instance, suppressed by ${\mathcal O}(q^3)$ compared to the scattering
on a single nucleon. However, the long-distance corrections due to DM
scattering on a pion exchanged between two nucleons can already start
at ${\mathcal O}(q)$ \cite{Cirigliano:2012pq,Hoferichter:2015ipa}.

This paper is organized as follows. In
Sections~\ref{sec:EFTops}-\ref{sec:discussion:matching:ChEFT} we focus
on fermionic DM, while we give the results for scalar DM in
Section~\ref{sec:scalar:DM}. In Section~\ref{sec:EFTops} we first
introduce the EFT for DM coupling to quarks, gluons and photons
through higher dimension operators. We treat the DM mass as heavy,
$m_\chi\gg q$, leading to a Heavy Dark Matter Effective Theory
(HDMET). The DM interactions with mesons and nucleons are constructed
in Section~\ref{sec:DM:mesons:nucleons}, while
Section~\ref{sec:discussion:matching:ChEFT} contains the calculation
of the form factors for the nuclear response functions. The analysis
is repeated for scalar DM in Section~\ref{sec:scalar:DM}. We draw our
conclusions in Section~\ref{sec:conclusions}. In
Appendix~\ref{app:dictionary} we give the translation of our results
to the basis of Ref. \cite{Anand:2013yka,Fitzpatrick:2012ix}, while in
Appendix~\ref{app:low:eng:const} we provide the values of the required
low-energy constants. Appendix~\ref{app:furtherDMChPT} contains
further details on DM interactions with mesons and nucleons.

\section{Nonrelativistic dark matter interactions}
\label{sec:EFTops}
We first focus on fermionic DM and its interactions with quarks,
gluons and photons at the scale $\mu\sim 1~{\rm GeV}$. These
interactions are generated by mediators that couple to both the DM and
the visible sector. The DM interactions can be described by an EFT as
long as the mediators are much heavier than ${\mathcal O}(1{\rm
  GeV})$,
\begin{equation}\label{eq:lightDM:Lnf5}
{\cal L}_\chi=\sum_{a,d}
\hat \C_{a}^{(d)} {\cal Q}_a^{(d)}, 
\qquad {\rm where}\quad 
\hat \C_{a}^{(d)}=\frac{\C_{a}^{(d)}}{\Lambda^{d-4}}\,.
\end{equation}
Here, the $\C_{a}^{(d)}$ are dimensionless Wilson coefficients, while
$\Lambda$ can be identified with the mediator mass. For later
convenience of notation we also introduced dimensionful Wilson
coefficients, $\hat \C_{a}^{(d)}$. In our analysis we only keep those
operators that would arise from dimension-five and dimension-six
operators above the scale of electroweak symmetry
breaking~\cite{BBGZ:2016b}.

We first consider the case where DM is relativistic. There are two
dimension-five operators,
\begin{equation}
\label{eq:dim5:nf5:Q1Q2:light}
{\cal Q}_{1}^{(5)} = \frac{e}{8 \pi^2} (\bar \chi \sigma^{\mu\nu}\chi)
 F_{\mu\nu} \,, \qquad {\cal Q}_2^{(5)} = \frac{e }{8 \pi^2} (\bar
\chi \sigma^{\mu\nu} i\gamma_5 \chi) F_{\mu\nu} \,,
\end{equation}
where $F_{\mu\nu}$ is the electromagnetic field strength tensor. The
magnetic dipole operator $\Q_1^{(5)}$ is CP even, while the electric
dipole operator $\Q_2^{(5)}$ is CP odd. The dimension-six operators
are
\begin{align}
{\cal Q}_{1,q}^{(6)} & = (\bar \chi \gamma_\mu \chi) (\bar q \gamma^\mu q),
 &{\cal Q}_{2,q}^{(6)} &= (\bar \chi\gamma_\mu\gamma_5 \chi)(\bar q \gamma^\mu q), \label{eq:dim6EW:Q1Q2:light}
  \\ 
{\cal Q}_{3,q}^{(6)} & = (\bar \chi \gamma_\mu \chi)(\bar q \gamma^\mu \gamma_5 q)\,,
  & {\cal Q}_{4,q}^{(6)}& = (\bar
\chi\gamma_\mu\gamma_5 \chi)(\bar q \gamma^\mu \gamma_5 q)\,,\label{eq:dim6EW:Q3Q4:light}
\end{align}
and we also include a subset of the dimension-seven operators, namely
\begin{align}
{\cal Q}_1^{(7)} & = \frac{\alpha_s}{12\pi} (\bar \chi \chi)
 G^{a\mu\nu}G_{\mu\nu}^a\,, 
 & {\cal Q}_2^{(7)} &= \frac{\alpha_s}{12\pi} (\bar \chi i\gamma_5 \chi) G^{a\mu\nu}G_{\mu\nu}^a\,,\label{eq:dim7:Q1Q2:light}
 \\
{\cal Q}_3^{(7)} & = \frac{\alpha_s}{8\pi} (\bar \chi \chi) G^{a\mu\nu}\widetilde
 G_{\mu\nu}^a\,, 
& {\cal Q}_4^{(7)}& = \frac{\alpha_s}{8\pi}
(\bar \chi i \gamma_5 \chi) G^{a\mu\nu}\widetilde G_{\mu\nu}^a \,, \label{eq:dim7:Q3Q4:light}
\\
{\cal Q}_{5,q}^{(7)} & = m_q (\bar \chi \chi)( \bar q q)\,, 
&{\cal
  Q}_{6,q}^{(7)} &= m_q (\bar \chi i \gamma_5 \chi)( \bar q q)\,,\label{eq:dim7EW:Q5Q6:light}
  \\
{\cal Q}_{7,q}^{(7)} & = m_q (\bar \chi \chi) (\bar q i \gamma_5 q)\,, 
&{\cal Q}_{8,q}^{(7)} & = m_q (\bar \chi  \gamma_5 \chi)(\bar q \gamma_5
q)\,. \label{eq:dim7EW:Q7Q8:light}  
\end{align}
Here, $q=u,d,s$ denote the light quarks (we limit ourselves to flavor
conserving operators), $G_{\mu\nu}^a$ is the QCD field strength
tensor, while $\widetilde G_{\mu\nu} = \tfrac{1}{2}
\varepsilon_{\mu\nu\rho\sigma} G^{\rho\sigma}$
is its dual, and $a=1,\dots,8$ are the adjoint color indices. The
strong coupling constant $\alpha_s$ is taken at $\mu\sim 1$ GeV. We
also assumed that DM is a Dirac fermion in the expressions
above. However, our results will also apply for a Majorana fermion DM
with the exception that the operators ${\cal Q}_{1,2}^{(5)}$ and
$Q_{1,q;3,q}^{(6)}$ vanish in this case, and with straightforward
modifications in the matching onto the nonrelativistic theory, see
Appendix~\ref{app:Majorana}. Matching the UV theory to the EFT may
then require the inclusion of higher dimension operators which is
beyond the scope of the present paper. In
\eqref{eq:dim7EW:Q5Q6:light}, \eqref{eq:dim7EW:Q7Q8:light} we included
a factor of quark mass, $m_q$, in the definitions of the operators
because it arises from the flavor structure of many of the models of
DM.  In our analysis we keep the operators involving scalar currents
of the form $m_q(\bar \chi \chi) (\bar q q)$, but not those involving
tensor currents, such as $m_q(\bar \chi \sigma_{\mu\nu}\chi) (\bar q
\sigma^{\mu\nu}q)$, etc. The former can arise from the dimension-five
UV operator $(\bar \chi \chi) H^\dagger H$ by integrating out the
Higgs at the electroweak scale, see \cite{BBGZ:2016b}. The latter
requires a dimension-seven operator in the UV, such as $(\bar \chi
\sigma_{\mu\nu}\chi) (\bar Q_L \sigma^{\mu\nu}u_R) H$.

The DM in the galactic halo is nonrelativistic with a typical velocity
$|\vec v_\chi| \sim 10^{-3}$ so that the momenta exchanges are much
smaller than the DM mass, $q\ll m_\chi$. DM scattering in direct
detection experiments is thus described by a Heavy Dark Matter
Effective Theory (HDMET) in which the DM mass is integrated
out~\cite{Hill:2013hoa,BBGZ:2016b,Berlin:2015njh}, giving an expansion
in $1/m_\chi$. The leading term in the Lagrangian then describes the
motion of DM in the limit of infinite DM mass. To derive it we factor
out of the DM field $\chi$ the large momenta due to the propagation of
the heavy DM mass, defining (here $\chi$ is a Dirac fermion, for
Majorana fermions see Appendix~\ref{app:Majorana})
\begin{equation}\label{eq:chi-field-def}
  \chi (x) = e^{-i m_\chi v \cdot x} \big( \chi_v (x) + X_v (x) \big) \,,
\end{equation}
where 
\begin{equation}
  \chi_v (x) =  e^{i m_\chi v \cdot x} \frac{1 + {\slashed v}}{2} \chi
  (x) \,, \qquad X_v (x) = e^{i m_\chi v \cdot x} \frac{1 - \slashed
    v}{2} \chi (x) \,.
\end{equation}
This defines the heavy-particle field $\chi_v(x)$ in analogy to the
heavy quark field in Heavy Quark Effective Theory
\cite{Neubert:1993mb, Grinstein:1990mj, Eichten:1989zv,
  Georgi:1990um}. The remaining $x$ dependence is due to the soft
momenta. For instance, direct detection scattering changes the soft
momentum of the DM by $q$ but does not change the DM velocity label
$v$. The velocity label $v^\mu$ can be identified with either the
incoming or outgoing DM velocity four-vector, or any other velocity
four-vector that is nonrelativistically close to these two. In the
following section we will identify $v^\mu$ with the lab frame velocity
so that $v^\mu=(1,\vec 0\,)$; but, for now, we leave it in its
four-vector form.

The ``small-component'' field $X_v$ describes the antiparticle
modes. To excite an antiparticle mode requires the absorption of a
hard momentum of order ${\mathcal O}(2 m_\chi)$. In building HDMET the
antiparticle modes are integrated out, giving the tree-level
relation~\cite{Neubert:1993mb}
\begin{equation}
\chi=e^{-i m_\chi v \cdot x} \Big(1 +\frac{i \slashed
  \partial_\perp}{i v\cdot \partial+2 m_\chi-i \epsilon}\Big)
\chi_v\,,\label{eq:chi:rel}
\end{equation}
where $\gamma_\perp^\mu = \gamma^\mu - v^\mu \slashed{v}$. The HDMET
Lagrangian is thus given by 
\begin{equation} 
\label{eq:HDMET}
{\cal L}_{\rm HDMET}= \bar \chi_v (i v \cdot \partial) \chi_v
+\frac{1}{2m_\chi}\bar \chi_v (i \partial_\perp)^2 \chi_v+\cdots+{\cal
  L}_{\chi_v}.
\end{equation}
The first term is the leading-order (LO) HDMET Lagrangian and contains
no explicit dependence on $m_\chi$. The coefficient of the ${\mathcal
  O}(1/m_\chi)$ term is fixed by reparametrization invariance
\cite{Luke:1992cs}, and the ellipsis denotes higher-order terms. The
effective Lagrangian ${\cal L}_{\chi_v}$ gives the interactions of DM
with the SM. The expansion in powers of $1/m_\chi$ and $1/\Lambda$ can
be made explicit by defining
\begin{equation}\label{eq:ewDM:Lnf5}
{\cal L}_{\chi_v}=\sum_{d,m}
\hat \C_{a}^{(d,m)} {\cal Q}_a^{(d,m)}, 
\qquad{\rm where}\qquad
\hat \C_{a}^{(d,m)}=\frac{\C_{a}^{(d,m)}}{\Lambda^{d-m-4}
  m_\chi^{m}}\,.  
\end{equation}
Here, the operators ${\cal Q}_a^{(d,m)}$ arise as the terms of order
$1/m_\chi^m$ in the HDMET expansion of the UV operators ${\cal
  Q}_a^{(d)}$. For instance, we have (neglecting radiative corrections
to the matching conditions) 
\begin{align}
\label{eq:HDMETlimit:scalar}
\bar \chi \chi&\to \bar \chi_v \chi_v+\cdots, 
\\
\label{eq:HDMETlimit:pscalar}
 \bar \chi i \gamma_5
\chi &\to \frac{1}{ m_\chi}\partial_\mu \big(\bar \chi_v S_\chi^\mu \chi_v \big) + \ldots\,,
\\
\label{eq:vecDM:expand}
\bar \chi \gamma^\mu \chi &
\to  v^\mu \bar \chi_v \chi_v +\frac{1}{2 m_\chi}\bar \chi_v i \lrpartial_{\perp}^{\mu} \chi_v
+\frac{1}{2 m_\chi} \partial_\nu\big(\bar \chi_v \sigma_\perp^{\mu\nu} \chi_v\big)+\cdots, 
\\
\label{eq:axialDM:expand}
\bar \chi \gamma^\mu \gamma_5 \chi & 
\to  2 \bar \chi_v S_\chi^\mu \chi_v -\frac{i}{m_\chi} v^\mu \bar \chi_v S_\chi\cdot \lrpartial \chi_v
+\cdots,
\\
\label{eq:tensorDM:expand}
\bar \chi \sigma^{\mu\nu} \chi& \to  \bar \chi_v \sigma_\perp^{\mu\nu}
\chi_v + \frac{1}{2m_\chi} \Big( \bar\chi_v i v_{\phantom{\perp}}^{[\mu}
  \sigma_\perp^{\nu]\rho} \lrpartial_{\rho} \chi_v - v^{[\mu}
  \partial^{\nu]} \bar\chi_v \chi_v\Big) + \ldots\,,
  \\
  \label{eq:axialtensorDM:expand}
  \bar
\chi \sigma^{\mu\nu} i\gamma_5 \chi&\to 
 2\bar
\chi_v S_\chi^{[\mu}v^{\nu]} \chi_v+\cdots,
\end{align}
where $\sigma_\perp^{\mu\nu}=i [\gamma_\perp^\mu,
  \gamma_\perp^\nu]/2$, $\bar \chi_v \lrpartial^\mu \chi_v=\bar \chi_v
(\partial^\mu \chi_v)- (\partial^\mu\bar \chi_v) \chi_v$, and
$S^\mu=\gamma_\perp^\mu \gamma_5/2 $  is the
spin operator. The square brackets in the last line denote
antisymmetrization in the enclosed indices, while the ellipses denote
higher orders in $1/m_\chi$.

We group the operators in HDMET in terms of their $d-m$ values and
only display those $1/m_\chi$-suppressed operators that will be needed
to obtain all LO terms in chiral EFT description of DM scattering on
nuclei. The two dimension-five operators in
\eqref{eq:dim5:nf5:Q1Q2:light} get replaced by the HDMET operators
\begin{align}
\label{eq:dim5:HDM:nf5:Q1Q2}
{\cal Q}_{1}^{(5,0)} &= \frac{e}{4 \pi^2} \epsilon_{\mu\nu\alpha\beta} (\bar \chi_v S_\chi^{\alpha} v^\beta \chi_v)
 F^{\mu\nu}\,,\qquad
&{\cal Q}_2^{(5,0)}& = \frac{e}{2 \pi^2} (\bar
\chi_v S_\chi^{\mu}v^{\nu} \chi_v) F_{\mu\nu}\,,
\\
\label{eq:dim5:HDM:nf5:Q1NLO}
{\cal Q}_{1}^{(6,1)} &= \frac{ie}{8\pi^2} \big( \bar\chi_v v^{\mu} \sigma_\perp^{\nu\rho} \lrpartial_{\rho} \chi_v \big)
 F_{\mu\nu}\,,\qquad
&{\cal Q}_{2}^{(6,1)}& = -\frac{e}{8\pi^2} \big( v^{\mu} \partial^{\nu} \bar\chi_v \chi_v\big) F_{\mu\nu}\,.
\end{align}
We used the relation 
\begin{equation}\label{eq:sigma-to-epsilon-S}
\bar \chi_v \sigma_\perp^{\mu\nu}\chi_v=
-2 \epsilon^{\mu\nu\alpha\beta}v_{\alpha}\big(\bar \chi_v S_{\chi,\beta}\chi_v\big)\,,
\end{equation}
where $\epsilon^{\mu\nu\alpha\beta}$ is the totally antisymmetric
Levi-Civita tensor, with $\epsilon^{0123}=1$. If the matching from the
UV theory of DM interactions is done at tree level at $\mu\sim
m_\chi$, we have the following relations~\cite{BBGZ:2016b} 
\begin{equation}\label{eq:rel:C:dim5}
{\cal C}_{1}^{(5)}\overset{\rm tree}{=}{\cal
  C}_{1}^{(5,0)}\overset{\rm tree}{=}{\cal C}_{1}^{(6,1)}\overset{\rm
  tree}{=}{\cal C}_{2}^{(6,1)}\,, \qquad {\cal
  C}_{2}^{(5)}\overset{\rm tree}{=}{\cal C}_{2}^{(5,0)}\,,
\end{equation}
so that below $\mu\sim m_\chi $ the operators always appear in the
combination
\begin{equation}
\label{eq:Q5relations}
{\cal Q}_{1}^{(5,0)}+\frac{1}{m_\chi}\Big({\cal Q}_{1}^{(6,1)}+{\cal Q}_{2}^{(6,1)}\Big)+\cdots\,.
\end{equation}
The relations~\eqref{eq:rel:C:dim5} would receive corrections if the
matching is performed at loop level. Note that in our analysis we will
not need the $1/m_\chi$ corrections to the CP odd operator ${\cal
  Q}_2^{(5,0)}$.

The dimension-six operators to LO in $1/m_\chi$ are 
\begin{align}
{\cal Q}_{1,q}^{(6,0)} & = (\bar \chi_v  \chi_v) (\bar q \slashed v q), 
& {\cal Q}_{2,q}^{(6,0)} &= 2 (\bar \chi_vS_{\chi,\mu} \chi_v)(\bar q \gamma^\mu q),
\label{eq:dim6EW:Q1Q2:HDM}
  \\ 
{\cal Q}_{3,q}^{(6,0)} & = (\bar \chi_v \chi_v)(\bar q \slashed v \gamma_5 q)\,,
\qquad 
& {\cal Q}_{4,q}^{(6,0)}& = 2 (\bar
\chi_v S_{\chi,\mu} \chi_v)(\bar q \gamma^\mu \gamma_5 q).
 \label{eq:dim6EW:Q3Q4:HDM}
\end{align}
The $1/m_\chi$-suppressed operators that we need to
consider\footnote{In fact only the operators in
  \eqref{eq:dim6EW:Q3Q4:HDM1} and \eqref{eq:dim6EW:Q5Q6:HDM1} will
  enter the phenomenological analysis but we keep the other operators
  for completeness and transparency of notation.} are 
\begin{align}
{\cal Q}_{1,q}^{(7,1)} & = \frac{1}{2}(\bar \chi_v i \lrpartial_{\perp}^\mu \chi_v) (\bar q \gamma_\mu q), 
& {\cal Q}_{2,q}^{(7,1)} &= -i (\bar \chi_vS_\chi \cdot \lrpartial \chi_v)(\bar q \slashed v q),
\label{eq:dim6EW:Q1Q2:HDM1}
  \\ 
{\cal Q}_{3,q}^{(7,1)} & = \frac{1}{2}(\bar \chi_v i \lrpartial_{\perp}^\mu \chi_v) (\bar q \gamma_\mu \gamma_5 q)\,,
\qquad 
& {\cal Q}_{4,q}^{(7,1)}& = -i (\bar
\chi_v S_\chi \cdot \lrpartial \chi_v)(\bar q \slashed v \gamma_5 q),
 \label{eq:dim6EW:Q3Q4:HDM1}
 \\
 {\cal Q}_{5,q}^{(7,1)} & = \frac{1}{2}\partial_\nu (\bar \chi_v \sigma_\perp^{\mu\nu} \chi_v) (\bar q \gamma_\mu q), 
& {\cal Q}_{6,q}^{(7,1)} & = \frac{1}{2}\partial_\nu (\bar \chi_v \sigma_\perp^{\mu\nu} \chi_v) (\bar q \gamma_\mu \gamma_ 5q),
\label{eq:dim6EW:Q5Q6:HDM1}
\end{align}
where our convention is that the derivatives act only within the
brackets or on the nearest bracket. For matching from the UV theory at
scale $\mu\sim m_\chi$, we would have the following relations
\begin{equation}
\label{eq:rel:C:dim6}
{\cal C}_{i,q}^{(6)}\overset{\rm tree}{=}{\cal C}_{i,q}^{(6,0)}={\cal
  C}_{i,q}^{(7,1)}\overset{\rm tree}{=}{\cal C}_{(i+9)/2,q}^{(7,1)}\,,
\quad i=1,3\,; \qquad {\cal C}_{i,q}^{(6)}\overset{\rm tree}{=}{\cal
  C}_{i,q}^{(6,0)}={\cal C}_{i,q}^{(7,1)}\,, \quad i=2,4\,.
\end{equation}
Note that the equality denoted by ``tree'' is only valid for
tree-level matching, while the remaining relations are valid to all
orders due to reparametrization invariance, cf. Eqs. \eqref{eq:RPI}
and \eqref{eq:RPI:external}. Hence, in the EFT below $\mu\sim m_\chi$,
the following linear combinations of operators would appear with the
same coefficient,
\begin{equation}
\begin{split}
\label{eq:Q6relations}
{\cal Q}_{1,q}^{(6,0)}+\frac{1}{m_\chi}\Big({\cal
  Q}_{1,q}^{(7,1)}+{\cal Q}_{5,q}^{(7,1)}\Big)+\cdots\,, 
\qquad {\cal Q}_{2,q}^{(6,0)}+\frac{1}{m_\chi}{\cal
  Q}_{2,q}^{(7,1)}+\cdots\,, 
\\
{\cal Q}_{3,q}^{(6,0)}+\frac{1}{m_\chi}\Big({\cal
  Q}_{3,q}^{(7,1)}+{\cal Q}_{6,q}^{(7,1)}\Big)+\cdots\,, 
\qquad {\cal Q}_{4,q}^{(6,0)}+\frac{1}{m_\chi}{\cal
  Q}_{4,q}^{(7,1)}+\cdots\,,
\end{split}
\end{equation}
with the ellipses denoting higher-order terms. Note that the
coefficient in front of ${\cal Q}_{5,q}^{(7,1)}$ and ${\cal
  Q}_{6,q}^{(7,1)}$ in the two sums can differ from unity at loop
level in the matching.

The relevant dimension-seven
operators~\eqref{eq:dim7:Q1Q2:light}-\eqref{eq:dim7EW:Q7Q8:light}
involve scalar and pseudoscalar DM currents. The HDMET scalar current
operator starts at ${\mathcal O}(1/m_\chi^0)$, while the pseudoscalar
current starts at ${\mathcal O}(1/m_\chi)$.  We thus define the HDMET
operators 
\begin{align}
{\cal Q}_1^{(7,0)} & = \frac{\alpha_s}{12\pi} (\bar \chi_v \chi_v)
 G^{a\mu\nu}G_{\mu\nu}^a\,, 
 & {\cal Q}_2^{(8,1)} &= \frac{\alpha_s}{12\pi} \partial_\rho \big(\bar \chi_v S_\chi^\rho \chi_v\big) G^{a\mu\nu}G_{\mu\nu}^a\,,\label{eq:dim7:Q1Q2:HDM}
\\
{\cal Q}_3^{(7,0)} & = \frac{\alpha_s}{8\pi} (\bar \chi_v \chi_v) G^{a\mu\nu}\widetilde
 G_{\mu\nu}^a\,, 
& {\cal Q}_4^{(8,1)}& = \frac{\alpha_s}{8\pi}
\partial_\rho \big(\bar \chi_v S_\chi^\rho \chi_v\big) G^{a\mu\nu}\widetilde G_{\mu\nu}^a \,, \label{eq:dim7:Q3Q4:HDM}
\\
{\cal Q}_{5,q}^{(7,0)} & = m_q (\bar \chi_v \chi_v)( \bar q q)\,, 
&{\cal
  Q}_{6,q}^{(8,1)} &= m_q \partial_\mu \big(\bar \chi_v S_\chi^\mu \chi_v\big) ( \bar q q)\,,\label{eq:dim7EW:Q5Q6:HDM}
  \\
{\cal Q}_{7,q}^{(7,0)} & = m_q (\bar \chi_v \chi_v) (\bar q i \gamma_5 q)\,, 
&{\cal Q}_{8,q}^{(8,1)} & = - m_q \partial_\mu \big(\bar \chi_v S_\chi^\mu \chi_v\big)(\bar q i \gamma_5
q)\,, \label{eq:dim7EW:Q7Q8:HDM} 
\end{align}
so that we have the following tree-level matching conditions
\begin{equation}\label{eq:rel:C:dim7}
{\cal C}_i^{(7)}\overset{\rm tree}{=}{\cal C}_i^{(7,0)}\,, \quad i=1,3,5,7\,;\qquad
{\cal C}_i^{(7)}\overset{\rm tree}{=}{\cal C}_i^{(8,1)}\,, \quad i=2,4,6,8\,.
\end{equation}

\section{Dark matter interactions with mesons and nucleons}
\label{sec:DM:mesons:nucleons}
\subsection{QCD with external currents}
\label{sec:QCD}
As far as QCD interactions are concerned the DM currents can be viewed
as classical external fields. The quark level DM-SM interaction
Lagrangian can thus be written in a form familiar from the ChPT
literature~\cite{Gasser:1984gg} as 
\begin{equation}
\begin{split}\label{eq:appC:QCDLagr}
{\cal L}&={\cal L}_{\rm QCD}^{0}+s_G(x)\frac{\alpha_s}{12 \pi} G_{\mu\nu}^a G^{a\mu\nu}+\theta(x)\frac{\alpha_s}{8\pi}G_{\mu\nu}^a\tilde G^{a\mu\nu}\\
&+\bar q (x) \gamma^\mu \big[ \nu_\mu(x)+\gamma_5 a_\mu(x)\big] q(x)-\bar q (x)\big[ s(x)-i \gamma_5 p(x)\big] q(x),
\end{split}
\end{equation}
where $q=(u,d,s)$ is a vector of light quark fields.  Here ${\cal
  L}_{\rm QCD}^{0}$ is the QCD+QED Lagrangian in the limit of zero
quark masses and no interactions with DM. We treat the quark masses
and insertions of DM currents as perturbations. They are collected in
six spurions which, for relativistic
DM~\eqref{eq:dim6EW:Q1Q2:light}-\eqref{eq:dim7EW:Q7Q8:light}, are
given by 
\begin{align}
\nu_\mu(x)&=- e \bar Q_q A_\mu^{e}+\nu_{\chi, \mu}=- e \bar Q_q A_\mu^e + \bar \C_{1}^{(6)}\big(\bar \chi \gamma_\mu \chi\big)+\bar \C_{2}^{(6)}\big(\bar \chi \gamma_\mu \gamma_5 \chi\big), \label{eq:vmu}\\
a_\mu(x)&=\bar\C_{3}^{(6)}\big(\bar \chi \gamma_\mu \chi\big)+\bar \C_{4}^{(6)}\big(\bar \chi \gamma_\mu \gamma_5 \chi\big),\\
s(x)&={\cal M}_q +s_\chi={\cal M}_q-{\cal M}_q\, \bar \C_{5}^{(7)}\big(\bar \chi \chi\big) - {\cal M}_q \,\bar\C_{6}^{(7)}\big(\bar \chi  i\gamma_5 \chi\big),
\\
p(x)&={\cal M}_q\, \bar\C_{7}^{(7)}\big(\bar \chi \chi\big)- {\cal M}_q  \,\bar\C_{8}^{(7)}\big(\bar \chi  i \gamma_5 \chi\big),\\
s_G(x)&=
\hat \C_{1}^{(7)}\big(\bar \chi \chi\big)+\hat \C_{2}^{(7)}\big(\bar \chi i\gamma_5 \chi\big)
, \label{sG:matching}\\
\theta(x)&=
\hat\C_{3}^{(7)}\big(\bar \chi \chi\big)+\hat\C_{4}^{(7)} \big(\bar \chi i \gamma_5 \chi\big). \label{eq:theta(x):matching}
\end{align}
Here, we introduced $3\times 3$ diagonal matrices of Wilson
coefficients and electromagnetic charges 
\begin{equation}\label{eq:Wilson-diag-mat}
\begin{split}
\bar \C_{i}^{(d)}&=\diag\big(\hat \C_{i,u}^{(d)},\hat \C_{i,d}^{(d)},\hat \C_{i,s}^{(d)}\big), 
\\
 \bar Q_q&=\diag\big( Q_q\big)=\diag(2/3,-1/3,-1/3).
 \end{split}
\end{equation}
The general HDMET expressions for spurions are somewhat lengthier,
\begin{align}
\begin{split}
\nu^\mu(x)&=- e \bar Q_q A^{e,\mu}+\nu_{\chi}^{\mu}=- e \bar Q_q A^{e,\mu} + \bar \C_{1}^{(6,0)}v^\mu \bar \chi_v \chi_v +\frac{1}{2}\bar \C_{1}^{(7,1)}\bar \chi_v i \lrpartial_{\perp}^{\mu} \chi_v
\\
&\qquad+\frac{1}{2}\bar \C_{5}^{(7,1)}\partial_\nu\big(\bar \chi \sigma_\perp^{\mu\nu} \chi_v\big)+2 \bar \C_{2}^{(6,0)} \bar \chi_v S_\chi^\mu \chi_v-i \bar \C_{2}^{(7,1)} v^\mu \bar \chi_v S_\chi\cdot \lrpartial \chi_v +\cdots, 
\label{eq:vmu:HDMET}
\end{split}
\\
\begin{split}
a^\mu(x)&= \bar \C_{3}^{(6,0)}v^\mu \bar \chi_v \chi_v +\frac{1}{2}\bar \C_{3}^{(7,1)}\bar \chi_v i \lrpartial_{\perp}^{\mu} \chi_v+\frac{1}{2}\bar \C_{6}^{(7,1)}\partial_\nu\big(\bar \chi \sigma_\perp^{\mu\nu} \chi_v\big)+2 \bar \C_{4}^{(6,0)} \bar \chi_v S_\chi^\mu \chi_v
\\
&\qquad-i \bar \C_{4}^{(7,1)} v^\mu \bar \chi_v S_\chi\cdot \lrpartial \chi_v +\cdots\,, 
\label{eq:amu:HDMET}
\end{split}
\\
s(x)&={\cal M}_q +s_\chi={\cal M}_q-{\cal M}_q\, \bar \C_{5}^{(7,0)}\big(\bar \chi_v \chi_v\big) - {\cal M}_q \,\bar\C_{6}^{(8,1)}\partial_\mu \big(\bar \chi_v  S_\chi^\mu \chi_v\big)+\cdots\,,
\label{eq:s:HDMET}
\\
p(x)&={\cal M}_q\, \bar\C_{7}^{(7,0)}\big(\bar \chi_v \chi_v\big)- {\cal M}_q  \,\bar\C_{8}^{(8,1)}\partial_\mu \big(\bar \chi_v  S_\chi^\mu \chi_v\big)+\cdots\,,
\label{eq:p:HDMET}
\\
s_G(x)&=
\hat \C_{1}^{(7,0)}\big(\bar \chi_v \chi_v\big)+\hat \C_{2}^{(8,1)}\partial_\mu \big(\bar \chi_v  S_\chi^\mu \chi_v\big)+\cdots
\,, \label{sG:matching:HDMET}\\
\theta(x)&= 
\hat\C_{3}^{(7,0)}\big(\bar \chi_v \chi_v\big)+ \hat\C_{4}^{(8,1)} \partial_\mu \big(\bar \chi_v  S_\chi^\mu \chi_v\big)+\cdots
\,, \label{eq:theta(x):matching:HDMET}
\end{align}
where the $\bar \C_{i}^{(d,m)}$ are defined in analogy to
Eq.~\eqref{eq:Wilson-diag-mat} and the ellipses denote higher orders
in the $1/m_\chi$ expansion. The scalar spurion $s(x)$ contains the
diagonal quark matrix, ${\cal M}_q=\diag(m_q)$, as well as the DM
scalar current $s_\chi$. Similarly, the vector current contains a
contribution due to quarks interacting with the QED gauge field, $e
\bar Q_q A_\mu^e$, as well as the DM vector current $\nu_{\chi,
  \mu}$. All the remaining spurions vanish in the limit of vanishing
DM interactions. The chiral counting of spurions is $\nu_\mu, a_\mu,
s_G, \theta\sim {\mathcal O}(p^0)$, and $s,p\sim {\mathcal
  O}(p^2)$. However, in HDMET the contributions from the pseudoscalar
DM current only start at ${\mathcal O}(p)$ in $s_G, \theta$ and at
${\mathcal O}(p^3)$ in $s,p$.

The QCD Lagrangian ${\cal L}_{\rm QCD}^0$ exhibits a global chiral
$U(3)_L\times U(3)_R$ symmetry that is spontaneously broken to the
vectorial $U(3)_V$ at low energies (the anomalous $U(1)_A$ can be
included because of the shift symmetry in $\theta$, see below). The
combined Lagrangian~\eqref{eq:appC:QCDLagr}, composed of the spurion
terms and the QCD Lagrangian, is still formally invariant under the
local chiral transformations 
\begin{equation}\label{eq:appC:VLR}
q(x)\to V_R(x) \frac{1}{2}(1+\gamma_5) q(x)+V_L(x) \frac12 (1-\gamma_5) q(x),
\end{equation}
if the spurions transform simultaneously as 
\begin{align}
\nu_\mu+a_\mu &\to V_R(\nu_\mu +a_\mu)V_R^\dagger +i V_R\partial_\mu V_R^\dagger\,,\label{eq:transf:laws:R}\\
\nu_\mu-a_\mu &\to V_L(\nu_\mu - a_\mu)V_L^\dagger +i V_L\partial_\mu V_L^\dagger\,,\label{eq:transf:laws:L}\\
s+ip &\to V_R(s+ip)V_L^\dagger\,,\label{eq:transf:laws:S}\\
s_G & \to s_G\,.\label{eq:transf:laws:G}
\end{align}
The $\theta(x)$ undergoes a shift transformation such that it cancels
the contribution due to the anomalous $U(1)_A$ axial part of the
transformations \eqref{eq:appC:VLR}. For chiral transformations
$V_{L,R}(x)=\exp\big(i\alpha(x)\mp i\beta(x)\big)$ this
gives~\cite{Gasser:1984gg} 
\begin{equation}
\theta \to \theta - 2 \Tr(\beta).
\end{equation}

Since the DM currents can be viewed as classical external fields as
far as the QCD interactions are concerned, we can use the $U(1)_A$
transformation with 
\begin{equation}\label{eq:elim-theta}
\beta(x)=\frac{\theta(x)}{2}  \frac{{\cal M}_q^{-1}}{\Tr({\cal M}_q^{-1})},
\end{equation} 
to eliminate the $\theta$ term in Eq.~\eqref{eq:appC:QCDLagr} and move
it to the axial and pseudo-scalar currents \cite{Georgi:1986df}. After
the transformation, the Lagrangian is given by 
\begin{equation}
\begin{split}\label{eq:appC:QCDLagr:modified}
{\cal L}=&{\cal L}_{\rm QCD}^{0}+s_G(x)\frac{\alpha_s}{12 \pi} G_{\mu\nu}^a G^{a\mu\nu}
+\bar q (x) \gamma^\mu \big[ \nu_\mu(x)+\gamma_5 a'_\mu(x)\big] q(x)\\
&-\bar q (x)\big[ s(x)-i \gamma_5 p'(x)\big] q(x),
\end{split}
\end{equation}
where 
\begin{equation}
a'_\mu =a_\mu +\frac{\partial_\mu \theta }{2}\frac{{\cal
    M}_q^{-1}}{\Tr({\cal M}_q^{-1})} \,, \qquad
p'=p+\frac{\theta}{\Tr({\cal M}_q^{-1})}
\end{equation}
where we kept only terms linear in the spurions, so that in this
approximation $s'=s$. We have omitted the primes on the transformed
quark fields in \eqref{eq:appC:QCDLagr:modified}. The primed spurions
$a_\mu'$ and $p'$ obey the same transformation laws as the unprimed
equivalents in~\eqref{eq:transf:laws:R}-\eqref{eq:transf:laws:S}.

\subsection{Chiral perturbation theory for dark matter interactions}
The formal invariance of ${\cal L}$ in \eqref{eq:appC:QCDLagr} under
the local transformations \eqref{eq:appC:VLR} constrains the allowed
DM interactions with pions and nucleons. We start with the ChPT
Lagrangian for DM--pion interactions which needs to be formally
invariant under the transformations~\eqref{eq:appC:VLR}, thus limiting
the possible spurion insertions. As usual, the ChPT is organized in
terms of a derivative expansion. The pseudo-Nambu-Goldstone bosons
(PNGBs) are collected in the Hermitian matrix $\Pi\equiv\sum_a
\lambda_a \pi_a$, given by
\begin{equation}\label{eq:App:PiMat}
\Pi=
\begin{pmatrix}
\frac{\pi^0}{\sqrt2}+\frac{\eta_8}{\sqrt6} & \pi^+ &K^+ \\
\pi^- &  - \frac{\pi^0}{\sqrt2}+\frac{\eta_8}{\sqrt6} & K^0\\
K^- & \bar K^0 & -\frac{2\eta_8}{\sqrt6}
\end{pmatrix},
\end{equation} 
where $\lambda_a$ are the Gell-Mann matrices normalized as
$\Tr(\lambda_a \lambda_b)=\delta_{ab}$. We do not include $\eta'$ in
the ChPT Lagrangian due to its large mass, which therefore contributes
to the DM-nucleon contact terms. We thus also ignore $\eta-\eta'$
mixing, so that $\eta_8\simeq \eta$.

The PNGB degrees of freedom parametrize the coset space
$(SU(3)_L\times SU(3)_R)/SU(3)_V$. We use the exponential
parametrization of the coset space, given by the matrix $U(x)$. Under
chiral transformations
\begin{equation}
U\to V_R U V_L^\dagger.
\end{equation}
The $U$ matrix is unitary, $U U^\dagger=U^\dagger U=1$. Since the DM
$\theta(x)$ current has been moved to the axial and scalar currents,
it is consistent to impose the condition\footnote{Alternatively, we
  could have worked with untransformed interaction Lagrangian
  \eqref{eq:appC:QCDLagr} in which case the condition $\det U=e^{-i
    \theta(x)}$ would need to be imposed~\cite{Gasser:1984gg}.}  $\det
U=1$. Thus, in our convention the $U$ matrix is
\begin{equation}
U(x)=\exp\big(i \sqrt 2 \Pi/f\big),
\end{equation}
where $f\simeq 92$ MeV equals the pion decay constant at leading order
in ChPT (experimentally, we have
$f_\pi=92.21(14)$\,MeV~\cite{Agashe:2014kda}). Note that under a
parity transformation $\pi_a\to -\pi_a$, and thus $U\to U^\dagger$.

The ChPT Lagrangian at LO, i.e., at ${\mathcal O}(p^2)$, is given
by~\cite{Gasser:1984gg} 
\begin{equation}\label{eq:appC:ChPTLagr:noEta'}
\begin{split}
{\cal L}_{\rm ChPT}^{(2)}=&\frac{f^2}{4} \Tr\big(\nabla_\mu U^\dagger \nabla^\mu U\big)+\frac{B_0 f^2}{2} \Tr\big[(s-i p')U+(s+i p')U^\dagger\big],
\end{split}
\end{equation}
where $B_0$ is a low-energy constant. To ${\mathcal O}(m_q)$ it is
given by the quark condensate, and equals $\langle \bar q q\rangle
\simeq -f^2 B_0$. Using quark condensate from~\cite{McNeile:2012xh}
and the LO relation $f=f_\pi$ one has 
$B_0=2.666(57) {\rm ~GeV}$, evaluated at the scale $\mu=2\,$GeV. The
covariant derivative in \eqref{eq:appC:ChPTLagr:noEta'} is defined as
\begin{align}
\nabla_\mu U=& \partial_\mu U - i (\nu_\mu +a_\mu')U+i U(\nu_\mu -a_\mu'),
\end{align}
so that under chiral transformations
\begin{equation}
\nabla_\mu U\to  V_R \nabla_\mu UV_L^\dagger.
\end{equation}
Each of the terms in \eqref{eq:appC:ChPTLagr:noEta'} can be multiplied
by an arbitrary function of $s_G$.

To obtain the leading DM interactions with the pseudoscalar mesons we
expand \eqref{eq:appC:ChPTLagr:noEta'} up to linear order in the DM
currents. The zeroth order term gives the usual LO ChPT Lagrangian
\begin{equation}\label{eq:appC:ChPTLagrQCD}
\begin{split}
{\cal L}_{\rm ChPT}^{(2),{\rm QCD}}=&\frac{f^2}{4}
\Tr\big(\partial_\mu U^\dagger \partial^\mu U\big)+\frac{B_0 f^2}{2}
\Tr\big[{\cal M}_q(U+U^\dagger)\big], 
\end{split}
\end{equation}
while the QED interactions are 
\begin{equation}\label{eq:appC:ChPTLagrQED}
\begin{split}
{\cal L}_{\rm ChPT}^{(2),{\rm QED}}=&i \frac{e f^2}{2} A^{e,\mu}
\Tr\Big[\big(U\partial_\mu U^\dagger+U^\dagger \partial_\mu U\big)\bar
  Q_q \Big]. 
\end{split}
\end{equation}
The linear terms give the interactions of PNGBs with DM as
\begin{equation}\label{eq:appC:ChPTLagrDM:noEta'}
\begin{split}
{\cal L}_{\chi,{\rm ChPT}}=&- 
\frac{if^2}{2}\Tr\Big[\big(U\partial_\mu U^\dagger+U^\dagger
  \partial_\mu U\big)\nu_\chi^\mu +\big(U\partial_\mu
  U^\dagger-U^\dagger \partial_\mu U\big)a^\mu \Big]\\ 
&+\frac{B_0 f^2}{2} \Tr\Big[s_\chi(U+U^\dagger)-i
  p(U-U^\dagger) - \frac{i\theta}{\Tr({\cal M}_q^{-1})} (U-U^\dagger)
  \Big] + S_G (x) s_G\\ 
&-\frac{if^2}{4}\frac{\partial^\mu\theta}{\Tr({\cal
    M}_q^{-1})} \Tr\Big[\big(U\partial_\mu
  U^\dagger-U^\dagger \partial_\mu U\big) {\cal
    M}_q^{-1} \Big]\,.
\end{split}
\end{equation}
The scalar function $S_G(x)$ multiplying $s_G$ is chirally
invariant. To fix it to quadratic order in the derivative expansion we
require that the trace of the QCD energy-momentum tensor be reproduced
in the chiral effective theory. The general quadratic expansion has
the form
\begin{equation}
S_G(x)=a_1 \frac{f^2}{4} \Tr\big(\partial_\mu U^\dagger \partial^\mu
U\big) +a_2 \frac{B_0 f^2}{2}\Tr\big[{\cal M}_q(U+U^\dagger)\big]. 
\end{equation}
From the trace of the QCD energy momentum tensor, given at quark level
by 
$\theta_\mu^\mu=-\tfrac{9}{8 \pi} \alpha_s G^a_{\mu\nu}G^{a\mu\nu} +
\sum_q m_q \bar q q$, 
and at leading order in
the ChPT expansion by $\theta_\mu^{\mu{\rm \,
      eff}}=-\Tr(\partial_\mu\Pi \partial^\mu\Pi) + 4 B_0 \Tr({\cal
    M}_q \Pi^2)$,
  one obtains the LO expressions for
the low-energy coefficients $a_{1,2}$~\cite{Donoghue:1990xh},
\begin{equation}\label{eq:a12}
a_1=\frac{2}{3}a_2=\frac{4}{27}.
\end{equation}

Expanding \eqref{eq:appC:ChPTLagrDM:noEta'} to first nonzero order in
PNGB fields for each of the spurions gives 
\begin{equation}
\begin{split}\label{eq:appC:ChPTexpanded}
{\cal L}_{\rm ChPT}^{{\rm DM}}\supset& i \Tr\big(\big[\partial_\mu
  \Pi\, ,\Pi\big]\nu_\chi^\mu\big)-\sqrt 2 f\Tr\big(\partial^\mu \Pi
\,a_{\mu}\big)- B_0 \Tr\big(s_\chi \Pi^2\big)+\sqrt2 B_0 f \Tr\big(\Pi
\,p\big)\\ 
&+ \Big[ \frac{2}{27} \Tr\big( \partial_\mu \Pi \partial^\mu \Pi
  \big) - \frac{6}{27} B_0 \Tr \big( {\cal M}_q \Pi^2\big)\Big] s_G
+f\frac{\theta}{\sqrt{2}} \frac{\Tr(\partial^2 \Pi {\cal
      M}_q^{-1})}{\Tr({\cal M}_q^{-1})} + \cdots\,, 
\end{split}
\end{equation}
where the ellipses denote terms with more PNGBs. The $p$, $a_\mu$, and
$\theta$ spurions are flavor diagonal. The corresponding traces,
$\Tr\big(\partial^\mu \Pi\, a_\mu\big)$, $\Tr\big(\Pi\,p\big)$, and
$\Tr(\partial^2 \Pi {\cal M}_q^{-1})$ therefore lead to couplings of
DM axial and scalar currents to a single $\pi^0$ or $\eta$.
In contrast, the $\nu_{\chi}^\mu$, $s_\chi$, $s_G$, and $\theta$ DM
currents couple to at least two PNGBs. They thus enter the ChPT
description of the DM-nucleon scattering for the first time at
one-loop level.

Note that in~\eqref{eq:appC:ChPTexpanded} we do not display the terms
that contribute to the DM mass. Due to chiral symmetry breaking the DM
mass term is
\begin{equation}
\label{eq:Lchimass}
{\cal L}_{\chi}\supset -m_\chi \big(\bar \chi \chi\big) - \big(\bar
\chi \chi\big) \sum_q B_0 f^2 m_q \hat \C_{5,q}^{(7)} - \big(\bar \chi
i\gamma_5 \chi\big) \sum_q B_0 f^2 m_q \hat \C_{6,q}^{(7)}+\cdots\,,
\end{equation}
where we have displayed only the corrections to DM mass, $\delta
m_\chi$, due to scalar-spurion contribution in
Eq.~\eqref{eq:appC:ChPTLagrDM:noEta'}, with ellipsis denoting similar
terms due to the $s_G$ and $\theta$ spurions. Keeping only the leading
terms in $\delta m_\chi/m_\chi$, the last term in \eqref{eq:Lchimass}
can be eliminated by a small axial rotation of the DM field. The
second term, however, modifies the DM mass by a term of order
$\Lambda_{\rm QCD}^4/\Lambda^3$. This is a small correction for all
intents and purposes. For $\Lambda\gtrsim v_{\rm EW}$ one has $\delta
m_\chi \lesssim $ 1 eV. Similar comments apply to corrections due to
$s_G$ and $\theta$ spurions.

The various external DM currents in ${\cal L}_{\chi,{\rm ChPT}}$
\eqref{eq:appC:ChPTLagrDM:noEta'} have different chiral dimensions.
We thus organize the DM--meson interactions in terms of their overall
chiral suppression, including the derivative suppression of the DM
currents when expanded in $1/m_\chi$,
\begin{equation}
{\cal L}_{\chi,{\rm ChPT}}={\cal L}_{\chi,{\rm ChPT}}^{(1)}+{\cal L}_{\chi,{\rm ChPT}}^{(2)}+{\cal L}_{\chi,{\rm ChPT}}^{(3)}+\cdots.
\end{equation}
Keeping only the leading terms in chiral counting for each of the
Wilson coefficients in \eqref{eq:ewDM:Lnf5} gives 
\begin{align}
\begin{split}\label{eq:chiCHPT1:full}
{\cal L}_{\chi, {\rm ChPT}}^{(1)}&=-\frac{i f^2}{2} (\bar \chi_v \chi_v) v^\mu  \Tr\Big[(U\partial_\mu U^\dagger+U^\dagger\partial_\mu U)\,\bar \C_{1}^{(6,0)} +(U\partial_\mu U^\dagger-U^\dagger\partial_\mu U)\,\bar \C_{3}^{(6,0)}\Big]
\\
&-i f^2 (\bar \chi_v S_\chi^\mu\chi_v)  \Tr\Big[(U\partial_\mu U^\dagger+U^\dagger\partial_\mu U)\,\bar \C_{2}^{(6,0)} +(U\partial_\mu U^\dagger-U^\dagger\partial_\mu U)\,\bar \C_{4}^{(6,0)}\Big]\,,
\end{split}
\\
\begin{split}\label{eq:chiCHPT2:full}
{\cal L}_{\chi, {\rm ChPT}}^{(2)}&\supset-\frac{B_0f^2}{2}(\bar \chi_v \chi_v) \Tr \Big[ (U+U^\dagger){\cal M}_q  \bar \C_{5}^{(7,0)}+i (U-U^\dagger){\cal M}_q  \bar \C_{7}^{(7,0)}\Big]
\\
&+ \frac{f^2}{27} (\bar \chi_v \chi_v)  \Big[  \Tr\big(\partial_\mu U^\dagger \partial^\mu
U\big) + 3 B_0 \Tr\big[{\cal M}_q(U+U^\dagger)\big] \Big]
\hat \C_1^{(7,0)}
\\
&-\frac{i f^2}{4m_\chi}(\bar \chi_v i \lrpartial_{\perp}^{\mu} \chi_v)
\Tr\Big[(U\partial_\mu U^\dagger+U^\dagger\partial_\mu U)\,\bar
  \C_{1}^{(6,0)} +(U\partial_\mu U^\dagger-U^\dagger\partial_\mu
  U)\,\bar \C_{3}^{(6,0)}\Big]\\
&+\frac{if^2}{4}\frac{(\bar\chi_v \chi_v)}{\Tr({\cal
    M}_q^{-1})} \Tr\Big[\partial^\mu\big(U\partial_\mu
  U^\dagger-U^\dagger \partial_\mu U\big) {\cal
    M}_q^{-1} \Big]\hat \C_3^{(7,0)}\,,
\end{split}
\\
\begin{split}\label{eq:chiCHPT3:full}
{\cal L}_{\chi, {\rm ChPT}}^{(3)}&\supset-\frac{B_0f^2}{2} \partial_\mu(\bar \chi_v S_\chi^\mu \chi_v) \Tr \Big[ (U+U^\dagger){\cal M}_q  \bar \C_{6}^{(8,1)}-i (U-U^\dagger){\cal M}_q  \bar \C_{8}^{(8,1)}\Big]
\\
&+ \frac{f^2}{27} \partial_\nu(\bar \chi_v S_\chi^\nu \chi_v)
\Big[  \Tr\big(\partial_\mu U^\dagger \partial^\mu
U\big) + 3 B_0 \Tr\big[{\cal M}_q(U+U^\dagger)\big] \Big]
\hat \C_2^{(8,1)}\\
&+\frac{if^2}{4}\frac{\partial_\nu(\bar \chi_v S_\chi^\nu\chi_v)}{\Tr({\cal
    M}_q^{-1})} \Tr\Big[\partial^\mu\big(U\partial_\mu
  U^\dagger-U^\dagger \partial_\mu U\big) {\cal
    M}_q^{-1} \Big]\hat \C_4^{(8,1)}\,,
\end{split}
\end{align}
In \eqref{eq:chiCHPT2:full} we also kept part of the formally
subleading terms proportional to $\bar \C_{1}^{(6,0)}$ and
$\C_{3}^{(6,0)}$ because there is a cancellation with ${\cal L}_{\chi,
  {\rm ChPT}}^{(1)}$ that occurs after the expansion in meson fields.
The Lagrangians~\eqref{eq:chiCHPT1:full}-\eqref{eq:chiCHPT3:full}
expanded to first nonzero order in the meson fields are collected in
\eqref{eq:L1chiChPT:expand}-\eqref{eq:L3chiChPT:expand}.

\subsection{Heavy baryon chiral perturbation theory}
\label{App:HBChPT}
In order to describe the DM interactions including nucleons we use
Heavy Baryon Chiral Perturbation Theory (HBChPT)
\cite{Jenkins:1990jv}. This is the appropriate effective field theory
as long as $q\sim m_\pi \ll m_N$, where $m_N$ is the nucleon mass and
$q$ the typical momentum exchange. The baryon momentum can be split
into
\begin{equation}
p^\mu=m_N v^\mu +k^\mu,
\end{equation}
where $v^\mu$ is the four-velocity of the nucleon, while the soft
momentum $k^\mu\sim {\mathcal O}(q)$ gives the off-shellness of the
nucleon. The large momentum component due to the inertia of the heavy
baryon can be factored out from the dynamics. Generalizing to the
baryon octet, we introduce the HBChPT baryon field
\begin{equation}\label{eq:HBChPTfield}
B_v(x)=\exp(i m_N \slashed v v_{\mu} x^\mu) B(x), 
\end{equation}
where $m_N$ and $v$ are the baryon mass and velocity, respectively.
Some useful properties of the field $B_v$ are $\frac{1}{2}(1+\slashed
v)B_v=B_v$, $\bar B_v \gamma_5 B_v=0$, $\bar B_v \gamma_\mu B_v=
v_{\mu} \bar B_v B_v$, and $\bar B_v \gamma^\mu \gamma_5 B_v=2 \bar
B_v S_N^\mu B_v$, where $S_N^\mu$ is the spin
operator satisfying
\begin{equation}\label{eq:Svprop}
v\cdot S_N=0\,, \quad S_N^2 B_v = - \tfrac{3}{4} B_v \,, \quad
\{S_N^\mu, S_N^\nu\} = \tfrac{1}{2} \big( v^\mu v^\nu -
g^{\mu\nu}\big) \,, \quad [S_N^\mu, S_N^\nu] = -i
\epsilon^{\mu\nu\lambda\sigma} v_{\lambda} S_{N,\sigma}\,.
\end{equation}
As in HDMET, $v^\mu$ is just a label and is not changed by the QCD
interactions or by DM scattering that only lead to exchanges of soft
momenta of ${\mathcal O}(q)$. In the lab frame we have $v^\mu=(1,\vec
0\,)$.

The octet of baryons forms a $3\times 3$ matrix
\begin{equation}
B_v=
\begin{pmatrix}
\frac{1}{\sqrt2} \Sigma_v^0+\frac{1}{\sqrt  6}\Lambda_v & \Sigma_v^+ & p_v\\
\Sigma_v^- & -\frac{1}{\sqrt2}\Sigma_v^0+\frac{1}{\sqrt 6}\Lambda_v & n_v \\
\Xi_v^- & \Xi_v^0 & -\frac{2}{\sqrt6}\Lambda_v.
\end{pmatrix}.
\end{equation}
For tree-level contributions to DM-nucleon scattering, i.e. working at
LO, we need to keep only the $p_v$ and $n_v$ entries of the $B_v$
matrix, while the remaining entries can be set to zero. In order to
write down HBChPT it is useful to define the square root of the matrix
$U$,
\begin{equation}\label{eq:appC:defxi}
U(x)=\xi(x)^2.
\end{equation}
The $\xi(x)$ transforms under chiral rotations as\footnote{This
  differs from~\cite{Jenkins:1990jv} and
  follows~\cite{Gasser:1984gg}. The convention
  of~\cite{Jenkins:1990jv} is obtained by the replacement $\xi\to
  \xi^\dagger$.}
\begin{equation}
\xi(x)\to V_R(x) \xi(x) V^\dagger(x)=V(x)\xi(x) V_L^\dagger(x).
\end{equation}
This equation defines the vector transformation $V(x)$, an element of
the group $SU(3)_V$ that remains unbroken after the spontaneous
breaking of the chiral $SU(3)_L\times SU(3)_R$ symmetry. From the
scalar and pseudoscalar spurions appearing in
\eqref{eq:appC:QCDLagr:modified} we can construct a quantity that
transforms as an adjoint of $SU(3)_V$, 
\begin{equation}
\xi^\dagger (s+ip')\xi^\dagger\to V(x) \xi^\dagger (s+ip')\xi^\dagger V(x)^\dagger.
\end{equation}
A related parity-even spurion, 
\begin{equation}
s_+ \equiv \xi^\dagger (s+ip')\xi^\dagger +\xi (s-ip')\xi,
\end{equation}
is thus also in the adjoint of $SU(3)_V$, 
\begin{equation}
 s_+\to V s_+ V^\dagger.
\end{equation}
Note that $s_+$ contains, in addition to the DM scalar, pseudoscalar,
and $\theta$ currents, a contribution from the quark masses. For later
convenience we define a parity-even spurion that vanishes in the limit
of zero DM currents, 
\begin{equation}
s_+^\chi \equiv \xi^\dagger (s_\chi+ip)\xi^\dagger +\xi (s_\chi-ip)\xi+\frac{i\theta}{\Tr({\cal M}_q^{-1})}\big(U^\dagger-U\big)+\cdots\,.
\end{equation}
The ellipses denote terms that involve more than one insertion of the
DM currents.  One therefore has 
\begin{equation}
s_+ =s_+^\chi + \xi^\dagger {\cal M}_q \xi^\dagger +\xi {\cal M}_q \xi\,,
\end{equation}
where the last two terms arise purely from the QCD Lagrangian.

From the $a_\mu'$ and $\nu_\mu$ spurions
in~\eqref{eq:appC:QCDLagr:modified} we can form axial, $A_\mu$, and
vector, $V_\mu$, currents that transform under chiral rotations as
\begin{equation}
V_\mu \to V V_\mu V^\dagger +i V\partial_\mu V^\dagger, \qquad A_\mu \to V A_\mu V^\dagger.
\end{equation}
They are sums of DM and SM currents, 
\begin{equation}
V_\mu=V_\mu^\chi+i V_\mu^\xi, \qquad  A_\mu =A_\mu^\chi+A_\mu^\xi, 
\end{equation}
where\footnote{Our convention for the QED covariant derivative is
  $D_\mu = \partial_\mu + ieQA_\mu^e$, where $Q$ is the charge of the
  particle in terms of the positron charge, and $A_\mu^e$ is the
  photon field.}
\begin{align}
V_\mu^\chi&=\frac{1}{2}\big[\xi^\dagger (\nu_{\chi, \mu} +a_\mu') \xi+\xi (\nu_{\chi,\mu}-a_\mu')\xi^\dagger\big],
\\
V_\mu^\xi&=\frac{1}{2}\big(\xi^\dagger \partial_\mu \xi+\xi \partial_\mu \xi^\dagger)+i\frac{e A_\mu^e}{2} \big(\xi^\dagger \bar Q_q \xi+ \xi \bar Q_q  \xi^\dagger\big),
\end{align}
and
\begin{align}
A_\mu^\chi&=\frac{1}{2}\big[\xi^\dagger (\nu_{\chi, \mu} +a_\mu') \xi-\xi (\nu_{\chi, \mu}-a_\mu')\xi^\dagger\big],
\\
 A_\mu^\xi&=\frac{i}{2}\big(\xi^\dagger \partial_\mu \xi-\xi \partial_\mu \xi^\dagger)-\frac{e A_\mu^e}{2} \big(\xi^\dagger \bar Q_q \xi- \xi \bar Q_q  \xi^\dagger\big).
\end{align}
The $V_\mu^\xi$ and $A_\mu^\xi$ are pure QCD and QED currents, while
the dependence on the DM currents is included in $V_\mu^\chi$ and
$A_\mu^\chi$.

The pNGBs can be factored out of the baryon fields $B_v$ so that they
transform as
\begin{equation}
B_v\to V B_v V^\dagger.
\end{equation} 
We define the covariant derivative by 
\begin{equation}
\nabla_\mu B_v=\partial_\mu B_v- i [ V_\mu, B_v]=\partial_\mu B_v-i[V_\mu^\chi, B_v]+[V_\mu^\xi, B_v].
\end{equation}
Under chiral rotations it transforms as $\nabla_\mu B_v\to V
\nabla_\mu B_v V^\dagger$.

With the above notation in hand we can write down the HBChPT
Lagrangian. The ${\mathcal O}(p)$ terms are\footnote{Here we use the
  notation $p$ for the typical momenta exchange, $p\sim q$, since this
  is the usual notation in ChPT. It reduces the confusion with the
  quark indices.}
\begin{equation}
\begin{split}\label{eq:appC:HBChPT1}
{\cal L}_{\rm HBChPT}^{(1)}=&i \Tr\big(\bar B_v v\ncdot \nabla B_v\big) - \frac{2 m_G}{27} \Tr\big(\bar B_v B_v\big)s_G+ 2 D\Tr\big(\bar B_v S_N^\mu\{A_\mu, B_v\}\big)\\
&+2 F \Tr \big(\bar B_v S_N^\mu [A_\mu, B_v]\big)+2 G \Tr \big(\bar B_v S_N^\mu B_v\big) \Tr\big(A_\mu\big) +\Tr(V_\mu)\Tr\big(\bar B_v v^\mu B_v\big)
.\end{split}
\end{equation}
We included the dimension-seven $s_G$ contribution, formally of
${\mathcal O}(p^0)$, in the ${\mathcal O}(p)$ Lagrangian. Note that
the last two terms do not appear in \cite{Jenkins:1990jv} since the
QCD and QED parts of the currents vanish, $ \Tr V_\mu^\xi= \Tr
A_\mu^\xi=0$, while in our case $ \Tr V_\mu=\Tr(\nu_\mu)$ and $ \Tr
A_\mu=\Tr a_\mu +\partial_\mu \theta/2$ can be nonzero, depending on
the DM interactions.  The coefficient of the
last term is fixed by requiring that the vector current $\bar
q\gamma_\mu q$ counts the number of valence quarks in the
baryons. 

The scalar and pseudoscalar spurions first appear in the ${\mathcal
  O}(p^2)$ HBChPT Lagrangian.  The terms relevant for our analysis are
\begin{equation}
\begin{split}\label{eq:HBChPT2-vec}
{\cal L}_{\rm HBChPT}^{(2)}&\supset  
b_D \Tr\bar B_v\{ s_+,B_v\} +b_F\Tr \bar B_v [s_+, B_v]+b_0\Tr\big(\bar B_vB_v\big)\Tr\big(s_+^\chi\big)
-\frac{ \Tr\bar B_v \nabla^2 B_v}{2m_N}
\\
&  +\frac{1}{2m_N} \Tr\big(V_\mu\big)\Tr\big(\bar B_v  i\lrnabla^\mu B_v\big)-\frac{G}{m_N} \Tr\big(\bar B_v S_N\ncdot i\lrnabla B_v\big) \Tr\big(v \ncdot A\big)
\\
&-\frac{(D+F)}{2m_N} \Tr\big(\bar B_v \{S_N\ncdot i\lrnabla, v \ncdot A\} B_v\big)
- \frac{(D-F)}{m_N} \Tr\big[(\bar B_v S_N\ncdot i\lrnabla B_v) (v \ncdot A) \big]
\\
 &-i \epsilon^{\alpha\beta \lambda\sigma} v_{\alpha} \Big[ g_4 \Tr \big(\bar B_v S_{N\beta}   \nabla_\lambda \nabla_\sigma B_v \big) -i g_5  \Tr \big(\bar B_v S_{N\beta}  B_v \nabla_\lambda V_\sigma \big)
\\
 &\qquad\qquad \qquad\qquad\qquad\qquad\qquad\qquad +i g_4' \Tr \big(\bar B_v S_{N\beta}  B_v\big) \partial_\lambda \Tr (V_\sigma)\Big]
+\cdots,
\end{split}
\end{equation}
where we used reparametrization invariance to fix some of the
low-energy constants, see Eq.~\eqref{eq:RPI:c's}. The remaining
constants are given in Table \ref{tab:LEC}. They are related to the
proton and neutron magnetic moments, the nucleon sigma terms,
$\sigma_{u,d}^p$, and the axial-vector matrix elements, $\Delta q_p$,
as detailed in Appendix~\ref{app:low:eng:const}. The complete
expression for ${\cal L}_{\rm HBChPT}^{(2)}$ is given in Appendix
\ref{app:HBChPT2}.

The DM spurions in the above Lagrangian can be expanded in terms of
the PNGB fields. Keeping only the first nonzero terms, one has
\begin{align}
V_\mu^\chi&=\nu_{\chi,\mu}-\frac{i}{\sqrt2 f}[\Pi,a_\mu]-\frac{i
  \partial_\mu \theta\, [\Pi,{\cal M}_q^{-1}]}{2 \sqrt2 f \Tr({\cal
    M}_q^{-1})}+\cdots\,, \label{eq:appC:Vmuchi:expanded} 
\\
A_\mu^\chi&=a_\mu-\frac{i}{\sqrt 2
  f}[\Pi,\nu_{\chi,\mu}]+\frac{\partial_\mu \theta}{2} \frac{{\cal
    M}_q^{-1}}{\Tr({\cal
    M}_q^{-1})}+\cdots\,, \label{eq:appC:Amuchi:expanded} 
\\
s_+^\chi&=2 s_\chi+\frac{\sqrt2}{f}\{\Pi,p\}+\frac{2\sqrt
  2}{f}\frac{\theta\, \Pi}{\Tr({\cal M}_q^{-1})}+\cdots\,. 
\end{align}

\begin{table}
\begin{center}
\begin{tabular}{cccc}
\hline\hline
LE constant & value & LE constant & value \\
\hline
$D$&$0.812(30)$&$b_D$&$1.4\pm 0.8$\\
$F$&$0.462(14)$&$b_F$&$-1.8\pm 0.8$\\ 
$G$&$-0.376(28)$&$g_4$&$4.70/m_N$\\ 
$m_G$&$848(14)\,$MeV&$g_4'$&$1.03/m_N$\\
$b_0$&$-3.7\pm 1.4$&$g_5$&$5.95(6)/m_N$\\ 
\hline
\hline
\end{tabular}
\end{center}
\caption{Numerical values for the low-energy constants relevant for
  leading order DM scattering in ChEFT. Scale-dependent quantities are
  defined in the $\overline{\text{MS}}$ scheme at 2\,GeV. For more
  details and references, see the main text. 
}
\label{tab:LEC}
\end{table}

In our analysis we need the pure QCD interactions as well as the
interactions of nucleons with DM. Setting the DM currents to zero in
${\cal L}_{\rm HBChPT}$ gives the pure QCD part of HBChPT. This has
the following chiral expansion, ${\cal L}_{\rm HBChPT}^{\rm QCD}={\cal
  L}_{\rm HBChPT}^{(1),{\rm QCD}}+{\cal L}_{\rm HBChPT}^{(2),{\rm
    QCD}}+\cdots$, where 
\begin{align}\label{eq:appC:HBhPTQCD1}
{\cal L}_{\rm HBChPT}^{(1), {\rm QCD}}=&i \Tr\big(\bar B_v v\ncdot \nabla^\xi B_v\big) + 2 D\Tr\big(\bar B_v S_N^\mu\{A_\mu^\xi, B_v\}\big)+ 2 F \Tr \big(\bar B_v S_N^\mu [A_\mu^\xi, B_v]\big)\,,\\
\begin{split}\label{eq:appC:HBhPTQCD2}
{\cal L}_{\rm HBChPT}^{(2),{\rm QCD}}\supset &b_D \Tr\bar B_v\{\xi^\dagger {\cal M}_q \xi^\dagger +\xi {\cal M}_q \xi,B_v\} +b_F\Tr \bar B_v [\xi^\dagger {\cal M}_q \xi^\dagger +\xi {\cal M}_q \xi, B_v] +\cdots\,.
\end{split}
\end{align}
The QCD part of the HBChPT covariant derivative is
\begin{equation}
\nabla_\mu^\xi B_v=\partial_\mu B_v+[V_\mu^\xi, B_v]\,.
\end{equation}

The interactions between DM and nucleons have a chiral expansion that
starts at ${\mathcal O}(p^0)$, ${\cal L}_{\chi, {\rm HBChPT}}={\cal
  L}_{\chi, {\rm HBChPT}}^{(0)}+{\cal L}_{\chi, {\rm
    HBChPT}}^{(1)}+{\cal L}_{\chi, {\rm HBChPT}}^{(2)}+\cdots. $ In
our analysis we need terms up to ${\mathcal O}(p^3)$. Keeping only the
leading terms for each of the Wilson coefficients in
\eqref{eq:ewDM:Lnf5}, the HBChPT interaction Lagrangians are
\begin{align}
\begin{split}\label{eq:HBChPT0:full}
{\cal L}_{\chi, {\rm HBChPT}}^{(0)}&= (\bar \chi _v \chi_v ) \Big( 
\frac{1}{2} \Tr\bar B_v \big[ (\xi^\dagger\bar  \C_{1}^{(6,0)}\xi+\xi\bar  \C_{1}^{(6,0)} \xi^\dagger),B_v\big]
+\Tr\bar B_v B_v \Tr \bar  \C_{1}^{(6,0)} \Big)
\\
 &+2 (\bar \chi _v S_\chi^\mu \chi_v ) \Big(D  \Tr \bar B_v S_N^\mu \big\{\xi^\dagger\bar  \C_{4}^{(6,0)}\xi+\xi\bar  \C_{4}^{(6,0)} \xi^\dagger, B_v\big\}\\
& +F  \Tr \bar B_v S_N^\mu \big[\xi^\dagger\bar  \C_{4}^{(6,0)}\xi+\xi\bar  \C_{4}^{(6,0)} \xi^\dagger, B_v\big]+ 2 G \Tr \bar B_v S_N^\mu B_v \Tr \bar  \C_{4}^{(6,0)}\Big)
\\
&-\frac{2}{27}m_G\,
 (\bar \chi_v \chi_v)  \Tr ( \bar  B_v  B_v )\,\hat \C_{1}^{(7,0)}\,,
\end{split}
\\
\begin{split}
{\cal L}_{\chi, {\rm HBChPT}}^{(1)}&\supset 2(\bar \chi_v S_{\chi,\mu} \chi_v) \sum_q \tilde J_{q}^{V\mu, {\rm NLO}}\hat \C_{2,q}^{(6,0)}
+ \big(\bar \chi_v  \chi_v\big) \sum_q v \cdot \tilde J_{q,\mu}^{A,{\rm NLO}} \hat \C_{3,q}^{(6,0)} 
\\
&-i(\bar \chi_v S_\chi\cdot \lrpartial \chi_v) \Big( \frac{1}{2} \Tr\bar B_v \big[(\xi^\dagger \bar\C_{2}^{(7,1)}\xi+\xi\bar\C_{2}^{(7,1)} \xi^\dagger),B_v\big]+\Tr\bar B_v B_v \Tr\bar\C_{2}^{(7,1)} \Big)
\\
&+\frac{i}{2} \Big[\bar \chi_v  \lrpartial_{\perp}^{\mu} \chi_v
-i \partial_\nu\big(\bar \chi_v \sigma_\perp^{\mu\nu} \chi_v\big)\Big] \Big(\frac{1}{2} \Tr\bar B_v \big[v^\mu (\xi^\dagger\bar \C_{3}^{(7,1)}\xi-\xi\bar \C_{3}^{(7,1)} \xi^\dagger),B_v\big] \\
 & \quad +D  \Tr \bar B_v S_N^\mu \big\{\xi^\dagger\bar \C_{3}^{(7,1)}\xi+\xi\bar \C_{3}^{(7,1)} \xi^\dagger, B_v\big\}\\
 &\quad +F  \Tr \bar B_v S_N^\mu \big[\xi^\dagger\bar \C_{3}^{(7,1)}\xi+\xi\bar \C_{3}^{(7,1)} \xi^\dagger, B_v\big]+ 2 G \Tr \bar B_v S_N^\mu B_v \Tr\bar \C_{3}^{(7,1)}\Big)
\label{eq:HBChPT1:full}
\end{split}
 \\  \notag 
\begin{split}
&-\frac{2}{27}m_G\,
\partial_\mu (\bar \chi_v S_\chi^\mu \chi_v)  \Tr (\bar B_v  B_v )\, \hat \C_{2}^{(8,1)}
\\
&-
(\bar \chi _v \chi_v ) \frac{\hat \C_{3}^{(7,0)}}{2\Tr ({\cal M}_q^{-1}) }\biggr\{\frac{1}{2} v\ncdot \partial \Tr\bar B_v \big[ (\xi^\dagger  {\cal M}_q^{-1} \xi-\xi  {\cal M}_q^{-1}  \xi^\dagger),B_v\big]
\\
&+ \partial_\mu \Big(D  \Tr \bar B_v S_N^\mu
\big\{\xi^\dagger {\cal M}_q^{-1} \xi+\xi {\cal M}_q^{-1} \xi^\dagger, B_v\big\}\\
& \quad  +F  \Tr \bar B_v S_N^\mu \big[\xi^\dagger
  {\cal M}_q^{-1} \xi+\xi {\cal M}_q^{-1} \xi^\dagger, B_v\big]+ 2 G  \Tr ({\cal M}_q^{-1})
\Tr \bar B_v S_N^\mu B_v \Big) \biggr\} \,,  
\end{split}
\\
\begin{split}\label{eq:HBChPT2:full}
{\cal L}_{\chi, {\rm HBChPT}}^{(2)}&\supset- (\bar \chi_v\chi_v) 
\Big[ b_0 \Tr(\bar B_v B_v) 
\Tr {\cal M}_q \big( \bar \C_{5}^{(7,0)} (U^\dagger +U)
-i \bar \C_{7}^{(7,0)} (U^\dagger -U) \big)
\\
 &+b_D \Tr\bar B_v \big\{\xi^\dagger {\cal M}_q \big( \bar \C_{5}^{(7,0)}- i  \bar \C_{7}^{(7,0)}\big) \xi^\dagger+\xi {\cal M}_q \big( \bar \C_{5}^{(7,0)} +i  \bar \C_{7}^{(7,0)}\big) \xi, B_v\big\}
\\
&+b_F \Tr \bar B_v\big[\xi^\dagger {\cal M}_q \big(\bar \C_{5}^{(7,0)}
  -i \bar \C_{7}^{(7,0)}\big) \xi^\dagger+\xi {\cal M}_q\big(\bar
  \C_{5}^{(7,0)} +i \bar \C_{7}^{(7,0)}\big)\xi, B_v\big] \Big]\\
&-
 \partial_\nu (\bar \chi _v S_\chi^\nu \chi_v ) \frac{\hat
   \C_{4}^{(8,1)}}{2\Tr ({\cal M}_q^{-1}) } \biggr\{\frac{1}{2} v\ncdot \partial \Tr\bar B_v \big[ (\xi^\dagger  {\cal M}_q^{-1} \xi-\xi  {\cal M}_q^{-1}  \xi^\dagger),B_v\big]
\\
&+ \partial_\mu \Big(D  \Tr \bar B_v S_N^\mu
\big\{\xi^\dagger {\cal M}_q^{-1} \xi+\xi {\cal M}_q^{-1} \xi^\dagger, B_v\big\}\\
& \quad  +F  \Tr \bar B_v S_N^\mu \big[\xi^\dagger
  {\cal M}_q^{-1} \xi+\xi {\cal M}_q^{-1} \xi^\dagger, B_v\big]+ 2 G  \Tr ({\cal M}_q^{-1})
\Tr \bar B_v S_N^\mu B_v \Big) \biggr\} \,,
\end{split}
\\
\begin{split}\label{eq:HBChPT3:full}
{\cal L}_{\chi, {\rm HBChPT}}^{(3)}&\supset- \partial_\mu (\bar \chi_v S_\chi^\mu \chi_v )
\bigg[ b_0 \Tr(\bar B_v B_v) 
\Tr {\cal M}_q \big(\bar \C_{6}^{(8,1)} (U^\dagger +U) 
+i \bar \C_{8}^{(8,1)} (U^\dagger -U) \big)
\\
 &\quad+b_D \Tr\bar B_v \big\{\xi^\dagger {\cal M}_q \big( \bar \C_{6}^{(8,1)}+ i  \bar \C_{8}^{(8,1)}\big) \xi^\dagger+\xi {\cal M}_q \big( \bar \C_{6}^{(8,1)} -i  \bar \C_{8}^{(8,1)}\big) \xi, B_v\big\}
\\
&\quad+b_F \Tr \bar B_v\big[\xi^\dagger {\cal M}_q \big(\bar \C_{6}^{(8,1)}
  +i \bar \C_{8}^{(8,1)}\big) \xi^\dagger+\xi {\cal M}_q\big(\bar
  \C_{6}^{(8,1)} -i \bar \C_{8}^{(8,1)}\big)\xi, B_v\big] 
\bigg]\,.
\end{split}
\end{align}
The expressions for the NLO currents $\tilde J_{q,\mu}^{V,{\rm NLO}}$
and $\tilde J_{q,\mu}^{A,{\rm NLO}}$ appearing in
\eqref{eq:HBChPT1:full} can be found in~\eqref{eq:JVNLO}
and~\eqref{eq:JANLO:expanded}. The diagonal matrix of Wilson
coefficients $\bar \C_i$ was defined in
\eqref{eq:Wilson-diag-mat}. The DM HBChPT Lagrangians
\eqref{eq:HBChPT1:full}-\eqref{eq:HBChPT3:full}, expanded in the meson
fields, are collected in Section~\ref{subsec:DMinteraction}.

\section{Discussion and matching onto nuclear chiral EFT}
\label{sec:discussion:matching:ChEFT}
We are now in a position to calculate the scattering of DM on nuclei
using a chiral EFT description of nuclear forces.  We first briefly
review the results of the previous two sections, keeping only the
essential ingredients, and introduce a simplified notation. We rewrite
the HDMET interaction Lagrangian~\eqref{eq:HDMET} as
\begin{equation}
{\cal L}_\chi={\cal L}_\chi^{(5)}+{\cal L}_\chi^{(6)}+{\cal L}_\chi^{(7)}+\cdots,
\label{eq:Leff:nf=3}
\end{equation}
where we collect in each Lagrangian ${\cal L}_\chi^{(d)}$ the terms
that would come from relativistic DM operators with dimensionality $d$
in \eqref{eq:lightDM:Lnf5}-\eqref{eq:dim7EW:Q7Q8:light}, ${\cal
  L}_\chi^{(d)}=\sum_{a,m} \hat \C_{a}^{(d+m,m)} {\cal
  Q}_a^{(d+m,m)}$.  We work to tree-level order in the matching at the
scale $\mu\sim m_\chi$. The Wilson coefficients then satisfy the
relations~\eqref{eq:rel:C:dim5},~\eqref{eq:rel:C:dim6},~\eqref{eq:rel:C:dim7},
so that we have 
\begin{align}\label{eq:dim5:nf3}
\begin{split}
{\cal L}_\chi^{(5)}&=\hat {\cal C}_1^{(5,0)} \frac{e}{8 \pi^2} 
 F_{\mu\nu} J_\chi^{T,\mu\nu} + \hat {\cal C}_2^{(5,0)} \frac{e }{8 \pi^2}  F_{\mu\nu} J_\chi^{AT,\mu\nu}.
\end{split}
\\
\begin{split}
\label{eq:dim6Lagr:1GeV}
{\cal L}_\chi^{(6)}&= J_{\chi}^{V,\mu} \sum_{q=u,d,s}  \Big[
\hat \C_{1,q}^{(6,0)}\,  \big(\bar q\gamma_\mu q\big)  +  \hat \C_{3,q}^{(6,0)} \big(\bar q \gamma_\mu \gamma_5 q\big) \Big]  \\
&+J_{\chi}^{A,\mu} \sum_{q=u,d,s} 
 \Big[
\hat \C_{2,q}^{(6,0)}\, \big(\bar q\gamma_\mu q\big)  +  \hat \C_{4,q}^{(6,0)}\, \big(\bar q \gamma_\mu \gamma_5 q\big) \Big].
\end{split}
\\
\begin{split}
\label{eq:dim7Lagr:1GeV}
{\cal L}_\chi^{(7)}&=J_\chi^S \sum_{q=u,d,s} \Big[\hat \C_{5,q}^{(7,0)}\, m_q \big(\bar q q\big)+ \hat \C_{7,q}^{(7,0)} \, m_q\big(\bar q i\gamma_5 q\big)
\\
&\qquad\qquad\quad+ \hat \C_1^{(7,0)}\frac{\alpha_s}{12\pi}G_{\mu\nu}^aG^{a\mu\nu}+\hat \C_3^{(7,0)}\frac{\alpha_s}{8\pi}G_{\mu\nu}^a\tilde G^{a\mu\nu}\Big]
\\
&+ m_\chi J_\chi^P  \sum_{q=u,d,s} \Big[\hat \C_{6,q}^{(8,1)}\, m_q \big(\bar q q\big)- \hat \C_{8,q}^{(8,1)} \, m_q\big(\bar q i\gamma_5 q\big)
\\
&\qquad\qquad\quad+ \hat \C_2^{(8,1)}\frac{\alpha_s}{12\pi}G_{\mu\nu}^aG^{a\mu\nu}+\hat \C_4^{(8,1)}\frac{\alpha_s}{8\pi}G_{\mu\nu}^a\tilde G^{a\mu\nu}\Big].
\end{split}
\end{align}
Note that, for tree-level matching, the HDMET interactions are simply
a product of the DM and SM currents, with the DM currents taken
outside the sums over quark flavors.  The DM currents are given by
\begin{align}
\label{eq:tensorDM:define}
J_\chi^{T,\mu\nu}& = \bar \chi_v \sigma_\perp^{\mu\nu}
\chi_v + \frac{1}{2m_\chi} \Big( \bar\chi_v i v_{\phantom{\perp}}^{[\mu}
  \sigma_\perp^{\nu]\rho} \lrpartial_{\rho} \chi_v - v^{[\mu}
  \partial^{\nu]} \bar\chi_v \chi_v\Big) + \ldots \overset{\rm tree}{=}\bar \chi \sigma^{\mu\nu} \chi \,,
  \\
  \label{eq:axialtensorDM:define}
 J_\chi^{AT,\mu\nu}&= 
 2 \bar
\chi_v S_\chi^{[\mu}v^{\nu]} \chi_v+\cdots 
\overset{\rm tree}{=}
\bar
\chi \sigma^{\mu\nu} i\gamma_5 \chi,
\\
\label{eq:vecDM:define}
J_\chi^{V,\mu}&=v^\mu \bar \chi_v \chi_v +\frac{1}{2 m_\chi}\bar \chi_v i \lrpartial_{\perp}^{\mu} \chi_v
+\frac{1}{2 m_\chi} \partial_\nu\big(\bar \chi \sigma_\perp^{\mu\nu} \chi_v\big)+\cdots
\overset{\rm tree}{=}\bar \chi \gamma^\mu \chi\,,
\\
\label{eq:axialDM:define}
J_\chi^{A,\mu}&= 2 \bar \chi_v S_\chi^\mu \chi_v -\frac{i}{m_\chi} v^\mu \bar \chi_v S_\chi\cdot \lrpartial \chi_v
+\cdots \overset{\rm tree}{=}\bar \chi \gamma^\mu\gamma_5\chi\,,
\\
\label{eq:scalarDM:define}
J_\chi^S&=\bar \chi_v\chi_v+\cdots
\overset{\rm tree}{=}\bar \chi \chi\,,
\\
\label{eq:pscalarDM:define}
J_\chi^P&=\frac{1}{m_\chi}\partial_\mu \bar \chi_v S_\chi^\mu \chi_v+\cdots\,\,\overset{\rm tree}{=}\bar\chi i \gamma_5\chi\,.
\end{align}
The notation in \eqref{eq:dim5:nf3}-\eqref{eq:dim7Lagr:1GeV} will
prove useful we discuss the leading contributions in chiral counting
for each of the Wilson coefficients, as it makes it easy to see where
the $q/m_\chi$-suppressed terms come from. For matching at higher loop
orders at scale $\mu\sim m_\chi$ one could generalize the above
notation by making the DM currents quark-flavor dependent and move
them inside the quark flavor sums, although the notation would not be
simpler than in \eqref{eq:ewDM:Lnf5}.\footnote{For instance one could define 
 \begin{align}
\label{eq:Jchi:vector:expand}
J_{\chi,j,q}^{V,\mu}&=v^\mu \big(\bar \chi_v \chi_v\big)+\left[r_{j,q} (\bar \chi_v i \lrpartial_{\perp}^{\mu} \chi_v)+r_{(j+9)/2,q} \partial_\nu\big(\bar \chi \sigma_\perp^{\mu\nu} \chi_v\big)\right]/2,  &j=&1,3,
\\
\label{eq:Jchi:axial-vector:expand}
J_{\chi,j,q}^{A,\mu}&=2  \big(\bar \chi_v S_\chi^\mu \chi_v\big)- \,  r_{j,q} v^\mu  (\bar \chi_v i S\cdot \lrpartial \chi_v),  &j=&2,4,
\end{align}
with $r_{j,q}={\hat \C_{j,q}^{(7,1)}}/{\hat \C_{1,q}^{(6,0)}}$.
}

The Lagrangian ${\cal L}_\chi^{(5)}$, Eq.~\eqref{eq:dim5:nf3},
contains only QED interactions of DM with the SM. On the other hand,
we have seen in the previous two sections that the ${\cal
  L}_\chi^{(6,7)}$ interactions involving quarks and gluons,
Eq.~\eqref{eq:dim6Lagr:1GeV} and~\eqref{eq:dim7Lagr:1GeV}, match onto
an effective Lagrangian with mesons and nucleons, ${\cal L}_{\chi,
  {\rm ChPT}}+{\cal L}_{\chi, {\rm HBChPT}}$. Here ${\cal L}_{\chi,
  {\rm ChPT}}$ contains only the light pseudoscalar mesons $\pi$, $K$,
and $\eta$ as QCD asymptotic states, while ${\cal L}_{\chi, {\rm
    HBChPT}}$ contains, in addition, the protons and neutrons.  One
can organize the different terms using chiral counting since the
momentum transfer is small, $q\leq q_{\rm max}\ll 4\pi f_{\pi}$
(cf. Eq.~\eqref{eq:qmax}), where $f_\pi$ is the pion decay constant.
The chiral expansion corresponds to an expansion in momenta exchanges,
$p\sim q$, where the meson masses are counted as $m_\pi\sim {\mathcal
  O}(p)$. As a consequence the quark masses scale as $m_q\sim
{\mathcal O}(p^2)$ since $m_q\propto m_\pi^2$. The interactions of DM
with mesons start at ${\mathcal O}(p)$,
\begin{equation}\label{eq:chiChPTexpan}
{\cal L}_{\chi, {\rm ChPT}}={\cal L}_{\chi, {\rm ChPT}}^{(1)}+{\cal
  L}_{\chi, {\rm ChPT}}^{(2)}+\cdots, 
\end{equation}
while the interactions of DM with nucleons start at ${\mathcal
  O}(p^0)$
\begin{equation}\label{eq:hiHBChPTexpan}
{\cal L}_{\chi, {\rm HBChPT}}={\cal L}_{\chi, {\rm
    HBChPT}}^{(0)}+{\cal L}_{\chi, {\rm HBChPT}}^{(1)}+{\cal L}_{\chi,
  {\rm HBChPT}}^{(2)}+\cdots. 
\end{equation}
The QCD interactions among pions have an expansion in $p^2$, while the
interactions between pions and nucleons have an expansion in $p$,
\begin{equation}
{\cal L}_{\rm ChPT}^{\rm QCD}={\cal L}_{{\rm ChPT}}^{(2)}+{\cal
  L}_{{\rm ChPT}}^{(4)}+\cdots,\qquad {\cal L}_{\rm HBChPT}^{\rm
  QCD}={\cal L}_{\rm HBChPT}^{(1)}+{\cal L}_{\rm
  HBChPT}^{(2)}+\cdots\,. 
\end{equation}
The explicit forms of the above Lagrangians are given in
\eqref{eq:appC:ChPTLagrQCD}, \eqref{eq:appC:ChPTLagrQED},
\eqref{eq:chiCHPT1:full}-\eqref{eq:chiCHPT3:full},
\eqref{eq:appC:HBhPTQCD1}, \eqref{eq:appC:HBhPTQCD2}, and
\eqref{eq:HBChPT0:full}-\eqref{eq:HBChPT3:full}. 
The LO QCD interactions are schematically
\begin{equation}\label{eq:LChPT:expand}
\begin{split}
{\cal L}_{{\rm ChPT}}^{(2)}&\sim (\partial_\mu \pi)^2+(\pi \partial_\mu\pi)^2+\cdots, \qquad
\\
 {\cal L}_{\rm HBChPT}^{(1)}&\sim \bar N v\ncdot \partial N + \bar N N v\ncdot \partial \pi+ \bar N S_N^\mu N \pi \partial_\mu \pi +\cdots,
 \end{split}
\end{equation}
where we expanded in the meson fields, $\pi$, with $N$ denoting a
nucleon field.

The leading few terms in chiral counting for the DM--meson
interactions are 
\begin{align}
\begin{split}\label{eq:chiCHPT1}
{\cal L}_{\chi, {\rm ChPT}}^{(1)}&=J_\chi^{V,\mu}\sum_q \Big(J_{q,\mu}^V \hat \C_{1,q}^{(6,0)}+J_{q,\mu}^A \hat \C_{3,q}^{(6,0)}\Big)\\
&+J_\chi^{A,\mu}\sum_q \Big(J_{q,\mu}^V \hat \C_{2,q}^{(6,0)}+J_{q,\mu}^A \hat \C_{4,q}^{(6,0)}\Big)\,,
\end{split}
\\
\begin{split}\label{eq:chiCHPT2}
{\cal L}_{\chi, {\rm ChPT}}^{(2)}&=J_\chi^S \Big(\sum_q J_q^S \hat
\C_{5,q}^{(7,0)}+\sum_q J_q^P \hat \C_{7,q}^{(7,0)}+J^G \hat \C_1^{(7,0)} +
J^\theta \hat \C_3^{(7,0)} \Big)\,,
\end{split}
\\
\label{eq:chiCHPT3}
{\cal L}_{\chi, {\rm ChPT}}^{(3)}&= m_\chi J_\chi^P \Big(\sum_q J_q^S
\hat \C_{6,q}^{(8,1)}-\sum_q J_q^P \hat \C_{8,q}^{(8,1)}+J^G \hat
\C_2^{(8,1)} + J^\theta \hat \C_4^{(8,1)}\Big)\,.
\end{align}
Here, we used tree-level matching expressions, and have thus factored
out the DM currents
\eqref{eq:vecDM:define}-\eqref{eq:pscalarDM:define}. We took into
account the scaling $J_\chi^P\sim {\mathcal O}(p)$,
c.f. Eq. \eqref{eq:pscalarDM:define}, while all the other DM currents
are ${\mathcal O}(p^0)$. The quark level currents were hadronized into
the corresponding mesonic currents, 
\begin{equation}\label{eq:Jmeson:hadron}
\begin{split}
\bar q\gamma_\mu q&\to\, J_{q,\mu}^V\sim \pi \partial_\mu \pi +\cdots\,, \qquad \quad \bar q\gamma_\mu \gamma_5 q\to J_{q,\mu}^A\sim \partial_\mu \pi +\cdots\,,\\
m_q \, \bar q q&\to J_{q}^S \sim m_q \pi^2 +\cdots\,,\qquad \qquad  \bar qi\gamma_5 q\,\to \,J_{q}^P\sim  m_q \pi +\cdots\,,
\\
\frac{\alpha_s}{12\pi}G_{\mu\nu} G^{\mu\nu}&\to J^G \sim \partial^2
\pi^2+\cdots\,, \qquad \qquad \frac{\alpha_s}{8\pi}G_{\mu\nu}
\tilde G^{\mu\nu} \to J^\theta \sim \partial^2 \pi+\cdots\,.
\end{split}
\end{equation}
Again, we showed their schematic structure when expanded in the meson
fields, keeping only the first nonzero terms. The full form of the
currents are given in Appendix~\ref{app:furtherDMChPT}.

The DM--nucleon interactions, keeping only the leading terms in chiral
counting for each effective operator, are given by 
\begin{align}
\begin{split}\label{eq:HBChPT0}
{\cal L}_{\chi, {\rm HBChPT}}^{(0)}&= \sum_q \Big( J^{V}_{\chi} \ncdot \tilde J_{q}^V\,\hat \C_{1,q}^{(6,0)} +J^{A}_\chi \ncdot \tilde J_{q}^A\, \hat \C_{4,q}^{(6,0)} \Big)
+J_\chi^S\tilde J^G \hat \C_{1}^{(7,0)},
\end{split}
\\
\begin{split}\label{eq:HBChPT1}
{\cal L}_{\chi, {\rm HBChPT}}^{(1)}&\supset\sum_q \Big( J^{A}_{\chi} \ncdot \tilde J_{q}^V\,\hat \C_{2,q}^{(6,0)} +J^{V}_\chi \ncdot \tilde J_{q}^A\, \hat \C_{3,q}^{(6,0)} \Big)
+m_\chi J_\chi^P\tilde J^G\, \hat \C_{2}^{(8,1)}
\\
&\qquad + J_\chi^S \tilde J^\theta\, \hat \C_{3}^{(7,0)}, 
\end{split}
\\
\begin{split}\label{eq:HBChPT2}
{\cal L}_{\chi, {\rm HBChPT}}^{(2)}&\supset J_\chi^S \sum_q
\Big( \tilde J_q^S \,\hat \C_{5,q}^{(7,0)} +\tilde J_q^P \hat \C_{7,q}^{(7,0)}\Big)+m_\chi J_\chi^P \tilde J^\theta\, \hat \C_{4}^{(8,1)},
\end{split}
\\
\begin{split}
{\cal L}_{\chi, {\rm HBChPT}}^{(3)}&\supset m_\chi J^P_\chi \sum_q \Big( \tilde J_q^S\, \hat \C_{6,q}^{(8,1)} -\tilde J_q^P \, \hat \C_{8,q}^{(8,1)}\Big),\label{eq:HBChPT3}
\end{split}
\end{align}
where in each term one should keep only the leading nonzero terms in
the HDMET expansion of the DM currents. The explicit form of the
Lagrangians are given in
Eqs.~\eqref{eq:HBChPT0:full}-\eqref{eq:HBChPT3:full}, while the
expressions expanded in meson fields are given in
Eqs.~\eqref{eq:HBChPT0:expand}-\eqref{eq:HBChPT3:expand}.  The
quark-level currents get hadronized to nucleon currents. They are
schematically 
\begin{equation}
\begin{split}
\bar q \gamma^\mu q &\to \tilde J_{q}^{V,\mu}\sim v^\mu \bar N N+\cdots, 
\qquad \qquad\,\,\,\,\bar q\gamma^\mu \gamma_5 q \to \tilde J_{q}^{A,\mu}\sim \bar N S_{N}^{\mu} N+\cdots,
\\
m_q\, \bar q q  &\to \tilde J_q^S\sim m_q \bar N N+\cdots, 
\qquad\qquad
 m_q\, \bar qi\gamma_5 q  \to \tilde J_q^P\sim m_q \bar N N \pi+\cdots, 
 \\
\frac{\alpha_s}{12\pi}G G &\to \tilde J^G\sim \bar N N+\cdots, 
\qquad \qquad \qquad
\frac{\alpha_s}{8\pi}G\tilde G\to \tilde J^\theta\sim q^\mu \bar N S_{N,\mu} N +\cdots,
\end{split}
\end{equation}
with their explicit forms given in
Eqs. \eqref{eq:tildeJV}-\eqref{eq:tildeJP},
Eqs. \eqref{eq:JVNLO}-\eqref{eq:JANLO:expanded}, and
Eqs.~\eqref{eq:tildeJqV}-\eqref{eq:Jtildetheta}. Using the expressions
\eqref{eq:vecDM:define}-\eqref{eq:pscalarDM:define} for the DM
currents expanded in $1/m_\chi$, and the fact that $v\cdot S_N=0$,
$v\cdot S_\chi=0$, we see that the ${\mathcal O}(p^0)$ terms cancel in
the products $J_\chi^V \cdot \tilde J_q^A$ and $J_\chi^A \cdot \tilde
J_q^V$. These are then part of ${\cal L}_{\chi, {\rm HBChPT}}^{(1)}$,
see Eq. \eqref{eq:HBChPT1}. Schematically, we have for the products of
currents in \eqref{eq:HBChPT0}-\eqref{eq:HBChPT3} 
\begin{equation}
\begin{split}\label{eq:Jtilde:products}
 J^{V}_{\chi} \ncdot \tilde J_{q}^V&\sim  J_\chi^S\tilde J^G \sim (\bar \chi_v \chi_v)\, (\bar N N),
\qquad\qquad\qquad\,\,\,\,
J^{A}_\chi \ncdot \tilde J_{q}^A \sim (\bar \chi_v S_\chi \chi_v)\,\ncdot (\bar N S_{N} N),
 \\
 J^{A}_{\chi} \ncdot \tilde J_{q}^V&\sim J_\chi^P\tilde J^G\sim   (\bar \chi_v \partial\ncdot S_\chi \chi_v)\, (\bar N N), 
 \qquad \quad
J^{V}_\chi \ncdot \tilde J_{q}^A \sim  J_\chi^S \tilde J^\theta \sim  (\bar \chi_v  \chi_v)\, (\bar N \partial\ncdot S_N N), 
\\
J_\chi^S  \tilde J_q^S  \sim m_q & (\bar \chi_v \chi_v)\, (\bar N N), \quad
J_\chi^S \tilde J_q^P \sim m_q (\bar \chi_v \chi_v)\, (\bar N N) \pi, \quad
  J_\chi^P \tilde J^\theta\, \sim (\bar \chi_v \partial \ncdot S_\chi \chi_v)\, (\bar N \partial \ncdot S_N N),
\\
 J_\chi^P \tilde J_q^S &\sim  (\bar \chi_v \partial\ncdot S_\chi  \chi_v) m_q(\bar N N),\qquad \qquad \qquad
 J_\chi^P \tilde J_q^P\sim (\bar \chi_v \partial\ncdot S_\chi  \chi_v) m_q(\bar N N) \pi,
\end{split}
\end{equation}
where in addition the $ J^{A}_{\chi} \ncdot \tilde J_{q}^V$ and $
J^{V}_{\chi} \ncdot \tilde J_{q}^A$ contain the operator
$\epsilon^{\alpha\beta\mu\nu} v_\alpha q_\beta (\bar \chi_v S_\mu
\chi_v)\, (\bar N S_\nu N)$.  In accordance with the chiral counting,
the products of currents \eqref{eq:Jtilde:products} entering ${\cal
  L}_{\chi, {\rm HBChPT}}^{(d)}$ have chiral dimension $d$, that is,
they either have $d$ derivatives, or have $d-2$ derivatives and one
factor of $m_q\sim {\mathcal O}( p^2)$. Note that the hadronization of
the pseudoscalar current, $\bar q i\gamma_5 q$, requires the emission
of at least one meson.

\subsection{DM--nucleus scattering in chiral EFT}

The above DM--nucleon interactions are the building blocks for
predicting the DM--nucleus scattering rates using the ChEFT-based
description of nuclear forces. DM scattering is described by a single
insertion of the interaction Lagrangian ${\cal L}_\chi$,
Eq. \eqref{eq:lightDM:Lnf5}, in the scattering amplitude. Our goal is
to obtain the leading contribution to the DM-nucleus scattering rate
for all the interactions in
Eq.~\eqref{eq:dim5:nf5:Q1Q2:light}-\eqref{eq:dim7EW:Q7Q8:light}. Each
of the operators in
\eqref{eq:dim5:nf5:Q1Q2:light}-\eqref{eq:dim7EW:Q7Q8:light} induces
both a coupling of DM to the light mesons only and a coupling of DM to
nucleons and mesons. In order to gauge the importance of each of these
two types of contributions we use the chiral counting for nuclear
forces within ChEFT.

The ChEFT description of nuclear forces is based on Weinberg's insight
that the $N$-body nucleon potentials can be obtained from $N$-nucleon
irreducible amplitudes \cite{Weinberg:1990rz,Weinberg:1991um}. The
$N$-nucleon irreducible amplitudes consist of those diagrams that
cannot be disconnected by cutting $N$ nucleon lines, i.e., there must
be at least one pion exchange. The internal pion and nucleon
propagators are off-shell by $E\sim {\mathcal O}(p)\sim {\mathcal
  O}(m_\pi)$. As such they allow for a consistent chiral counting. The
properties of the nucleus can then be obtained by solving the
Schr\"odinger equation involving the $2$, $3$, $\dots, N$-nucleon
potentials. This is equivalent to resumming the reducible diagrams
where some of the internal nucleon lines are close to being on-shell,
with $E\sim {\mathcal O}(p^2/m_N)$.

We are interested in DM scattering on a nucleus with atomic number
$A$. The scattering operator follows from a sum of $A$-nucleon
irreducible amplitudes, $M_{A,\chi}$, with one insertion of the DM
interaction. A given $A$-nucleon irreducible amplitude scales as
$M_{A,\chi}\sim (p/\Lambda_{\rm ChEFT})^\nu$,
with~\cite{Bedaque:2002mn, Weinberg:1991um, Cirigliano:2012pq}
\begin{equation}\label{eq:counting:rule}
\nu=4-A-2C+2L+\sum_i V_i \epsilon_i + \epsilon_{\chi},
\end{equation}
for a diagram with $C$ connected parts, $L$ loops, $V_i$
strong-interaction vertices of type $i$, and one DM interaction
vertex. The effective chiral dimension $\epsilon_i$ of the vertex of
type $i$ is given by $\epsilon_i=d_i+n_i/2-2$, where $d_i$ is the
chiral dimension of the vertex and $n_i$ the number of nucleon legs
attached to the vertex. We explicitly isolated the contribution
$\epsilon_{\chi}$ due to the external DM current since each amplitude
will only have one such insertion~\cite{Cirigliano:2012pq}. For
instance, the effective chiral dimension of a vertex from ${\cal
  L}_{\chi, {\rm ChPT}}^{(d)}$ is $\epsilon_{\chi}=d-2$, while the
DM-nucleon interactions in ${\cal L}_{\chi, {\rm HBChPT}}^{(d)}$ have
effective chiral dimension $\epsilon_{\chi}=d-1$. The leading QCD
interactions from ${\cal L}_{{\rm HBChPT}}^{(1)}$ and ${\cal L}_{{\rm
    ChPT}}^{(2)}$ have $\epsilon_i=0$. This means that one can insert
an arbitrary number of these strong vertices without affecting the
$p^\nu$ power scaling.

The chiral loop counting in irreducible amplitudes suggests that the
cut-off of the ChEFT is the same as in ChPT, $\Lambda_{\rm ChEFT}\sim
\Lambda_{\chi}\sim 4\pi f_\pi\sim 1 $ GeV. 
The resummation of the bubble diagrams in the reducible amplitudes, on
the other hand, leads to the appearance of the observed shallow bound
states for $p\, m_N / \Lambda_{\rm ChEFT}^2\sim 1 $, and thus for
$\Lambda_{\rm ChEFT}\sim 0.5$ GeV.\footnote{This scaling would imply
  that the nucleon mass is parametrically larger than $\Lambda_{\rm
    ChEFT}$, so that $p/m_N\sim{\mathcal O}(p^2)$, where $p\sim m_\pi$
  \cite{Weinberg:1991um, Epelbaum:2010nr}.  In the derivation of the
  nuclear potentials using ChEFT one counts $p/m_N\sim {\mathcal
    O}(p)$, the same as in HBChPT \cite{Epelbaum:2008ga}. The
  Weinberg's counting is fully consistent when deriving the nuclear
  potentials. Renormalization of the potentials when solving the
  Schr\"odinger equation, however, may require counterterms of
  formally higher chiral order \cite{Bedaque:2002mn, Epelbaum:2008ga,
    Epelbaum:2010nr, Machleidt:2016rvv}.  For instance, divergences
  due to iterations of leading-order interactions may not be absorbed
  by the leading-order operators themselves \cite{Bedaque:2002mn}.
  While conceptually discomforting, this problem is numerically small
  when using momentum cut-off regularization for modes above $\sim
  1$GeV. The alternative KSW counting \cite{Kaplan:1996xu,
    Kaplan:1998tg, Kaplan:1998we}, treating the NLO corrections
  perturbatively, is fully consistent, but leads to poorly convergent
  results \cite{Fleming:1999ee}.  } Conservatively, we will use in the
numerical estimates $p/\Lambda_{\rm ChEFT}\sim m_\pi/\Lambda_{\rm
  ChEFT}\sim 0.3$.

\begin{figure}
\includegraphics[scale=1]{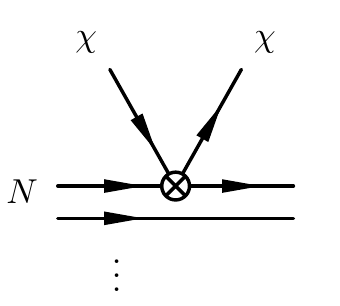}\hspace*{2cm}
\includegraphics[scale=1]{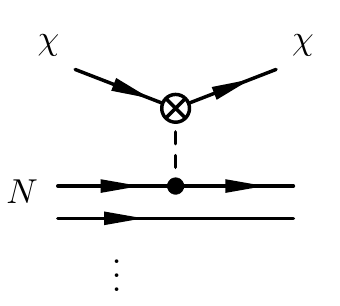}
\caption{The leading order diagrams for the DM-nucleus scattering. The
  effective DM--nucleon and DM--meson interactions is denoted by a
  circle, the dashed lines denote mesons, and the dots represent the
  remaining $A-2$ nucleon lines.  }
         \label{fig:LOChPT}
\end{figure}

The LO diagrams for DM-nucleon scattering are shown in
Fig.~\ref{fig:LOChPT}.  The left diagram gives the leading
contribution for the hadronization of the $\bar q \gamma_\mu q$, $\bar
q q$, and $GG $ currents.  The
right diagram is the leading contribution for the hadronization of the
$\bar q i\gamma_5 q$ current (the insertion is the mesonic $J^P$
current), in which case the left diagram is absent. Finally, for $\bar
q \gamma_\mu \gamma_5 q$ and $G\tilde G$ both the left and the right
diagrams are leading and contribute at the same order (for $J_\chi^V
\ncdot J_q^A$ the left diagram dominates).  In terms of the $p^\nu$
scaling we have for the leading contributions proportional to the
$\hat \C_a^{(n,m)}$ Wilson coefficients
\begin{equation}
\begin{split}
\label{eq:chiral:scaling}
\nu_{\rm min}&=\nu_{\rm LO}[J_\chi^V\ncdot \tilde J_q^V]=\nu_{\rm LO}[J_\chi^A\ncdot \tilde J_q^A]=\nu_{\rm LO}[J_\chi^S\tilde J^G],
\\
\nu_{\rm min}+1&=\nu_{\rm LO}[J_\chi^V \ncdot \tilde J_q^A]=\nu_{\rm LO}[J_\chi^A\ncdot \tilde J_q^V]= \nu_{\rm LO}[ J_\chi^P \tilde J^G]= \nu_{\rm LO}[J_\chi^S \tilde J^\theta]=\nu_{\rm LO}[J_\chi^S J_q^P],
 \\
\nu_{\rm min}+2&= \nu_{\rm LO}[J_\chi^S \tilde J_q^S]\,\,\,\,=\,\, \nu_{\rm LO}[ J_\chi^P \tilde J^\theta]\,=\nu_{\rm LO}[ J_\chi^P J_q^P],
 \\
\nu_{\rm min}+3&= \nu_{\rm LO}[ J_\chi^P \tilde J_q^S].
\end{split}
\end{equation}
Here $\nu_{\rm min}=3-3A$ simply reflects our normalization of the
$A$-nucleon state, where $A$ is the atomic number of the nucleus. In
the brackets we displayed the leading products of currents that
multiply the $\hat \C_a^{(n,m)}$ in
Eqs. \eqref{eq:HBChPT0}-\eqref{eq:HBChPT3}. As already mentioned
above, for most products of currents the left diagram in
Fig. \ref{fig:LOChPT} gives the dominant contribution. The resulting
$p^{\nu_{\rm LO}}$ suppression then follows directly from the chiral
suppression of the corresponding interaction Lagrangian, ${\cal
  L}_{\chi, {\rm HBChPT}}^{(0,1,2,3)}$. The exceptions are $J_\chi^S
J_q^P$ and $J_\chi^P J_q^P$, for which the right diagram in
Fig.~\ref{fig:LOChPT} dominates. These products have a chiral
suppression that is smaller than naively expected from the
dimensionality of the corresponding term in ${\cal L}_{\chi, {\rm
    HBChPT}}^{(0,1,2,3)}$ since the single pion exchange reduces
$\nu_{\rm LO}$ by one. From the general counting
rule~\eqref{eq:counting:rule} a similar conclusion would be reached
also for the product $J_\chi^V \ncdot \tilde J_q^A$. However, in this
case the single pion coupling to the DM current vanishes due to vector
current conservation, so that the formally leading contribution from
the right diagram in Fig.~\ref{fig:LOChPT} is zero. Special cases are $J_\chi^A\ncdot \tilde J_q^A$,
  $J_\chi^P\ncdot \tilde J_q^\theta$, and $J_\chi^S\ncdot \tilde
  J_q^\theta$, for which both diagrams in Fig. \ref{fig:LOChPT}
  contribute at the same order.

\begin{figure}
\includegraphics[scale=1]{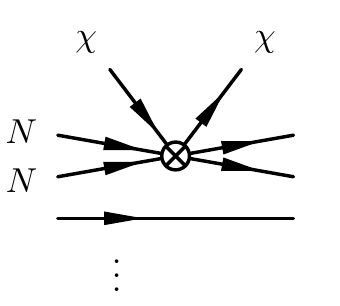}\hspace*{5mm}
\includegraphics[scale=1]{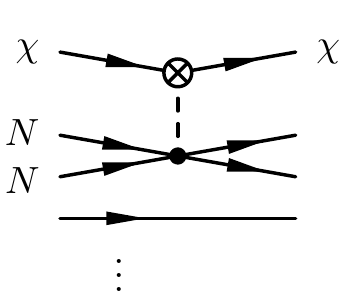}\hspace*{5mm}
\includegraphics[scale=1]{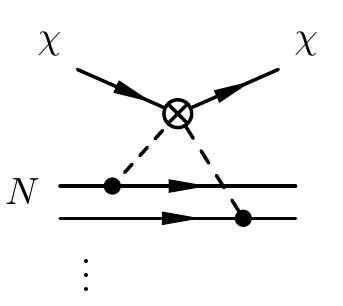}\hspace*{5mm}
\includegraphics[scale=1]{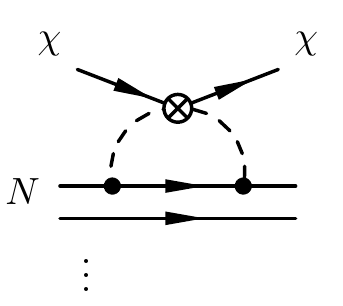}
\caption{Sample NLO diagrams for the DM-nucleon scattering inside
  nuclei. The effective DM--nucleon or DM--meson interaction is
  denoted by a box, the dashed lines denote mesons.  }
         \label{fig:NLOChPT}
\end{figure}

Note that at LO in chiral counting DM interacts with a single nucleon,
either directly through the short distance operator, or through a
single pion exchange. An interesting question is at which order in $p$
the two-body interactions do become important. Examples of the
relevant subleading contributions are shown in
Fig. \ref{fig:NLOChPT}. The first two diagrams are due to DM coupling
to a short distance two-nucleon current. These contributions always
scale as $p^{\nu_{\rm LO}+3}$. There are also contributions, shown in
the third diagram of Fig.~\ref{fig:NLOChPT}, where the DM attaches to
the meson exchanged between two nucleons, leading to long-distance
two-nucleon currents. For DM interactions originating from ${\cal
  O}_{2,q}^{(6)} \sim J_\chi^A \ncdot J_q^V$, ${\cal O}_{5,q}^{(7)}
\sim J_\chi^S \ncdot J_q^S$, and ${\cal O}_{6,q}^{(7)} \sim J_\chi^P
\ncdot J_q^S$, this contribution scales as $p^{\nu_{\rm LO}+1}$, while
for DM interactions originating from ${\cal O}_{1,q}^{(6)} \sim
J_\chi^V\ncdot J_q^V$ it scales as $p^{\nu_{\rm LO}+2}$. In these
cases the long distance two-nucleon contributions are parametrically
larger than the short distance ones. For the remaining operators the
long-distance contributions are of the same order or power suppressed
compared to the short-distance ones.

In addition, there are higher-order corrections that involve
single-nucleon interactions with DM. The last diagram in
Fig.~\ref{fig:NLOChPT} shows an example of such an one-loop
contribution. In addition there are also power suppressed
single-nucleon current insertions. These include the counterterms that
cancel the 1-loop divergences. For the DM interactions $J_\chi^A
\ncdot J_q^V$, $J_\chi^S \ncdot J_q^S$, and $J_\chi^P \ncdot J_q^S$,
the one-loop contributions scale as $p^{\nu_{\rm LO}+1}$, and are,
together with the long-distance pion exchange from the third diagram
in Fig.~\ref{fig:NLOChPT}, the leading chiral corrections.

In this work we are satisfied with LO matching and neglect relative
${\mathcal O}(p)$-suppressed terms. Our results thus have a relative
${\mathcal O}(q/\Lambda_{\rm ChEFT})\sim 30\%$ accuracy. At this order
the effective DM interactions involve only single nucleon currents. At
NLO, i.e., at relative ${\mathcal O}((q/\Lambda_{\rm ChEFT})^2)\sim
10\%$ accuracy, the DM is still interacting with a single nucleon
current for almost all DM--nucleon effective operators. The exceptions
are the DM--nucleon interactions $J_\chi^A \ncdot J_q^V$, $J_\chi^S
\ncdot J_q^S$, and $J_\chi^P \ncdot J_q^S$. For these the two-nucleon
contributions are a long-distance effect so that the corrections are
still calculable in ChPT. The results for scalar quark currents in the
case of Xe are available in
\cite{Cirigliano:2012pq,Hoferichter:2015ipa}, and are of the expected
size.  The genuine short-distance two-nucleon currents, for which one
would require lattice QCD calculations, appear only at NNNLO in chiral
counting, i.e., below few-percent accuracy.

\subsection{Form factors for dark matter--nucleon interactions}
\label{eq:pionlessEFT}
We can use the formalism in the previous section to calculate the form
factors for the DM--nucleon interactions. We perform the leading order
matching, shown in Fig. \ref{fig:LOChPT}. The hadronized $\bar q
\gamma_\mu q$, $\bar q q$, $G\tilde G$, and $GG $ currents receive
contributions from the left diagram, the $\bar q i\gamma_5 q$ current
from the right diagram, while $\bar q \gamma_\mu \gamma_5 q$ receives
contributions from both diagrams. The expanded hadronic currents are
collected in Appendix \ref{app:expanded:meson:fields}. We include the
contributions from single $\pi^0$ and $\eta$ exchanges in the
$q^2$-dependent coefficients of the nonrelativistic operators defined
below. The momenta exchanges are small enough that the DM-nucleon
interactions cannot lead to the dissociation of nuclei and the
production of on-shell pions.

The resulting effective Lagrangian is
\begin{equation}\label{eq:LNN}
{\cal L}_{\rm eff}= \sum_{i,d} \Big( c_{i,p}^{(d)}(q^2)\, Q_{i,p}^{(d)}+c_{i,n}^{(d)}(q^2)\, Q_{i,n}^{(d)}\Big),
\end{equation}
with $d$ counting the number of derivatives in the operators.  It is
understood that ${\cal L}_{\rm eff}$ is to be used only at tree
level. The operator basis is, for $d=0$,
\begin{align}\label{eq:Q1p0}
Q_{1,p}^{(0)}&= \big(\bar \chi_v \chi_v\big)\big(\bar p_v p_v\big), 
&Q_{2,p}^{(0)}&= \big(\bar \chi_v S_\chi^\mu \chi_v\big)\big(\bar p_v S_{N,\mu} p_v\big),
\end{align}  
with a similar set of operators for neutrons, with $p\to n$.  The
$Q_{1,p}^{(0)}$ operator will induce spin-independent DM scattering on
the nucleus, while $Q_{2,p}^{(0)}$ will induce the spin-dependent
scattering. In our analysis we also include all the $d=1$ operators,
\begin{align}\label{eq:Q12p1}
Q_{1,p}^{(1)}&= \big(\bar \chi_v  \chi_v\big)\big(\bar p_v i q\ncdot  S_N p_v\big), 
&Q_{2,p}^{(1)}&=\big(\bar \chi_v i q\ncdot  S_\chi \chi_v\big)\big(\bar p_v  p_v\big) ,
\\
\label{eq:Q34p1}
Q_{3,p}^{(1)}&= m_N \big(\bar \chi_v \chi_v\big)\big(\bar p_v \, v_\perp \ncdot S_N\, p_v\big) ,
 &Q_{4,p}^{(1)}&=m_N \big(\bar \chi_v\, v_\perp\ncdot S_\chi\, \chi_v\big)\big(\bar p_v p_v\big),
\\
\label{eq:Q56p1}
Q_{5,p}^{(1)}&= i \epsilon^{\alpha\beta\mu\nu}v_\alpha q_\beta \big(\bar \chi_v S_{\chi,\mu}\chi_v\big)\big(\bar p_v S_{N,\nu} p_v\big), 
&Q_{6,p}^{(1)}&= m_N \epsilon^{\alpha\beta\mu\nu}v_\alpha v_{\perp,\beta} \big(\bar \chi_v S_{\chi,\mu}\chi_v\big)\big(\bar p_v S_{N,\nu} p_v\big).
\end{align}  
The related operators for neutrons are obtained with a $p\to n$
replacement. From the $d=2$ set of operators we need only 
\begin{align}\label{eq:Q1p2}
Q_{1,p}^{(2)}&= \big(\bar \chi_v i q\ncdot S_\chi \chi_v\big)\big(\bar p_v \, i q \ncdot S_N\, p_v\big) ,
&Q_{2,p}^{(2)}&= i m_N \, \epsilon^{\alpha\beta\mu\nu}  v_\alpha q_\beta v_{\perp,\mu} \big(\bar \chi_v S_{\chi,\nu} \chi_v\big) \big(\bar p_v p_v\big) .
\end{align}

\begin{figure}
\includegraphics[scale=0.5]{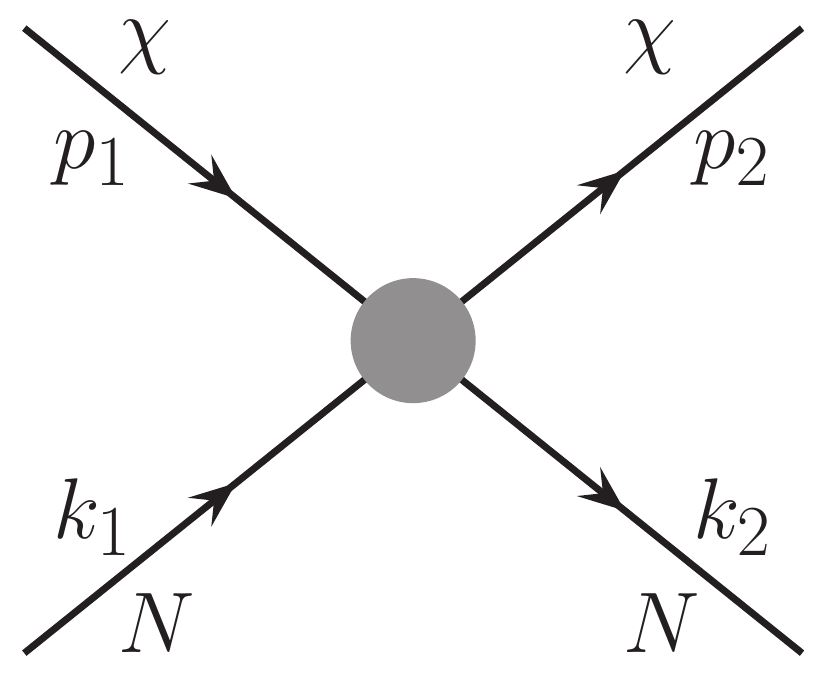}~~~~~~~~~~
\caption{The kinematics of DM scattering on nucleons,
  $\chi(p_1)N(k_1)\to \chi(p_2) N(k_2)$.  }
         \label{fig:scattering_kin}
\end{figure}

Above, we have defined several kinematic quantities for DM-nucleon
scattering, $\chi(p_1)N(k_1)\to \chi(p_2) N(k_2)$, see
Fig. \ref{fig:scattering_kin}. The momentum exchange is
\begin{equation}\label{eq:momentum:exchange}
q^\mu \equiv k_2^\mu-k_1^\mu=p_1^\mu-p_2^\mu,  \qquad q^\mu=\big(q^0,\vec q\big).
\end{equation}
The definition of momentum exchange three vector is thus\footnote{This
  differs by a sign from \cite{Anand:2013yka}, a difference that we
  will keep track of in our definitions of the NR operators.}
\begin{equation}
\vec q = \vec k_2-\vec k_1=\vec p_1 -\vec p_2. 
\end{equation}
It is also useful to define the four-component perpendicular relative
velocity (see also \cite{Anand:2013yka})
\begin{equation} \label{eq:vperp}
v_\perp^\mu=\frac{1}{2}\Big(\frac{p_1^\mu}{m_\chi}+\frac{p_2^\mu}{m_\chi}-\frac{k_1^\mu}{m_N}-\frac{k_2^\mu}{m_N}\Big)=\Delta v^\mu -\frac{q^\mu}{2\mu_N}\,,
\end{equation}
where $\Delta v^\mu$ is the initial relative velocity between DM and
nucleon,
\begin{equation}
\Delta v^\mu=\frac{p_1^\mu}{m_\chi}-\frac{k_1^\mu}{m_N},
\end{equation}
and $\mu_N=m_\chi m_N/(m_\chi +m_N)$ the reduced mass of the
DM--nucleon system (we work in the isospin limit, $m_N=m_p=m_n$). Note
the difference in our notation between $v^\mu$, the HBChPT velocity
label, and $\Delta v^\mu$, the initial relative velocity. In the lab
frame we have $v^\mu=(1,\vec 0\,)$, while $\Delta v^\mu\sim {\mathcal
  O}(\Lambda_{\rm QCD}/m_N)$ arises primarily due to the movement of
nucleons inside the nucleus.  The perpendicular relative velocity
obeys $v_\perp\cdot q=0$. Furthermore, in the lab frame one has
$v_\perp^\mu=(0,\vec v_\perp)$ so that also $\vec v_\perp \cdot \vec
q=0$.

The $d=0$ Wilson coefficients for interactions of DM with protons are
given by 
\begin{align}
\begin{split}\label{eq:c1p0}
c_{1,p}^{(0)}&=2\,\hat \C_{1,u}^{(6,0)}+\hat \C_{1,d}^{(6,0)}-\frac{2m_G}{27}
 \, \hat \C_{1}^{(7,0)}
    +\sigma_u^p\, \hat \C_{5,u}^{(7,0)}   +\sigma_d^p\, \hat \C_{5,d}^{(7,0)}
      +\sigma_s \hat \C_{5,s}^{(7,0)} 
   - \frac{\alpha Q_p}{2\pi m_\chi}  \hat \C_{1}^{(5,0)}\,,
\end{split}
\\
\begin{split}\label{eq:c2p0}
c_{2,p}^{(0)}&=4\big(\,\Delta u_p\, \hat \C_{4,u}^{(6,0)}+\,\Delta d_p\, \hat \C_{4,d}^{(6,0)}+\Delta s\, \hat \C_{4,s}^{(6,0)} \big)
 + \frac{2\alpha}{\pi} \frac{\mu_p}{m_N} \hat \C_{1}^{(5,0)} \,,
\end{split}
\end{align}
while the contributions for the neutrons are obtained through the
replacement $p\to n$, $u\leftrightarrow d$. For convenience of
notation we assumed that HDMET matching was done at tree level (see
the end of this Section for the general case).  The above results can
then be used directly also for the relativistic form of the DM EFT
\eqref{eq:dim5:nf5:Q1Q2:light}-\eqref{eq:dim7EW:Q7Q8:light} by simply
replacing $\hat C_i^{(d,0)}\to \hat C_i^{(d)}$. The terms proportional
to $\hat \C_{1}^{(5,0)}$ come from a single photon exchange. For the
photon propagator we used that $v\cdot q={\mathcal O}(q^2)$, so that
$(q^2-(v\cdot q)^2)/q^2=1+{\mathcal O}(q^2)$. The low-energy constant
$m_G$ is the gluon contribution to the nucleon mass. The remaining
HBChPT constants have been converted to nucleon sigma terms,
$\sigma_{u,d}^p$, axial vector matrix elements, $\Delta q_p$, and
nuclear magnetic moments, $\mu_N$, using the leading-order expressions
in Appendix~\ref{app:low:eng:const}. Their values are given in
Appendix~\ref{app:low:eng:const} and are collected in
Tab.~\ref{tab:numinput}. $Q_{p(n)}=1(0)$ is the proton (neutron)
electric charge.

\begin{table}
\begin{center}
\begin{tabular}{cccccc}
\hline\hline
LE constant & value & LE constant & value & LE constant & value \\
\hline
$\Delta u_p = \Delta d_n$&$0.897(27)$&$\Delta d_p = \Delta u_n$&$-0.376(27)$&$\Delta s$&$-0.026(4)$\\
$\mu_p$&$2.79$&$\mu_n$&$-1.91$&$\mu_s$&$-0.073(19)$\\
$B_0 m_u$&$(6200\pm400)\,\text{MeV}^2$&$B_0 m_d$&$(13300\pm400)\,\text{MeV}^2$&$B_0 m_s$&$(0.27\pm0.01)\,\text{GeV}^2$\\
$\sigma_u^p$&$(17\pm 5)\,\text{MeV}$&$\sigma_d^p$&$(32\pm 10)\,\text{MeV}$&$\sigma_s$&$(41.3\pm 7.7)\,\text{MeV}$\\
$\sigma_u^n$&$(15\pm 5)\,\text{MeV}$&$\sigma_d^n$&$(36\pm 10)\,\text{MeV}$&&\\
$g_A$&$1.2723(23)$&$m_G$&$848(14)\,$MeV&&\\
\hline
\hline
\end{tabular}
\end{center}
\caption{Numerical input values for the non-perturbative
  constants. For more details and references, see
  Appendix~\ref{app:low:eng:const}.}
\label{tab:numinput}
\end{table}

The $d=1$ Wilson coefficients are 
\begin{align}
\begin{split}\label{eq:c1p1}
c_{1,p}^{(1)}&=
 \frac{B_0 \, g_A}{m_\pi^2-q^2}\big( m_u\hat \C_{7,u}^{(7,0)}-m_d \hat \C_{7,d}^{(7,0)}\big)
 \\
&\quad+\frac{B_0}{3}\frac{(\Delta u_p+\Delta d_p -2 \Delta s)}{ m_\eta^2-q^2} \big(m_u \hat \C_{7,u}^{(7,0)}+ m_d\hat \C_{7,d}^{(7,0)}-2 m_s\hat \C_{7,s}^{(7,0)}\big)
\\
&\quad 
- \tilde m\bigg[ \frac{\Delta u_p}{m_u} +\frac{\Delta d_p}{m_d} + \frac{\Delta s}{m_s} 
 + \frac{g_A}{2} \bigg( \frac{1}{m_u} - \frac{1}{m_d} \bigg) \frac{q^2}{m_\pi^2-q^2} \\
  & \qquad \quad + \frac{1}{6}\big(\Delta u_p + \Delta d_p - 2\Delta s\big) \bigg( \frac{1}{m_u} + \frac{1}{m_d} - \frac{2}{m_s} \bigg) \frac{q^2}{m_\eta^2-q^2} 
  \bigg] \hat \C_3^{(7,0)}\,, 
\end{split}
\\
\begin{split}\label{eq:c2p1}
c_{2,p}^{(1)}&= -\sigma_u^p \hat \C_{6,u}^{(8,1)} - \sigma_d^p \hat \C_{6,d}^{(8,1)}- \sigma_s \hat \C_{6,s}^{(8,1)} 
+ \frac{2 m_G}{27}
\hat \C_2^{(8,1)} - \frac{2\alpha Q_p}{\pi q^2} \hat \C_{2}^{(5,0)}\,,
\end{split}
\\
\label{eq:c3p1}
c_{3,p}^{(1)}&=\frac{2}{m_N}\Big[\Delta u_p\, \hat \C_{3,u}^{(6,0)}+ \Delta d_p\, \hat \C_{3,d}^{(6,0)}+ \Delta s\, \hat \C_{3,s}^{(6,0)}\Big]\,,
\\
\label{eq:c4p1}
c_{4,p}^{(1)}&=-\frac{2}{m_N}\Big[2\,\hat \C_{2,u}^{(6,0)}+\,\hat \C_{2,d}^{(6,0)}\Big],
\\
\begin{split}
\label{eq:c5p1}
 c_{5,p}^{(1)}&= 
  \frac{4\hat \mu_u^p}{m_N}\hat \C_{2,u}^{(6,0)} + \frac{2\hat \mu_d^p}{m_N} \hat \C_{2,d}^{(6,0)} 
 -\frac{6}{m_N} \mu_s\, \hat \C_{2,s}^{(6,0)}\\
&\quad +\frac{2}{m_\chi} \Big(\Delta u_p \, \hat \C_{3,u}^{(6,0)} +
\Delta d_p\, \hat \C_{3,d}^{(6,0)}+\Delta s\, \hat \C_{3,s}^{(6,0)}\Big)\,,
\end{split}
\\
\label{eq:c6p1}
c_{6,p}^{(1)}&=0\,,
\end{align}
while the Wilson coefficients for the interactions of DM with
neutrons, $c_i^n$, are obtained by the replacements $p\to n,
u\leftrightarrow d$. We have defined
\begin{equation}
\tilde m = \bigg(\frac{1}{m_u}+\frac{1}{m_d}+\frac{1}{m_s}\bigg)^{-1}\,.
\end{equation}
The above results apply to the relativistic form of the DM EFT
\eqref{eq:dim5:nf5:Q1Q2:light}-\eqref{eq:dim7EW:Q7Q8:light} by
replacing $\hat C_i^{(d,0)}\to \hat C_i^{(d)}$ and $\hat
C_i^{(8,1)}\to \hat C_i^{(7)}$, as long as the matching to HDMET was
performed at tree level (see the end of this Section for the general
case). In Eq.~\eqref{eq:c1p1} and in Eq.~\eqref{eq:c1p2} below we use
$\Delta u_p - \Delta d_p = \Delta d_n - \Delta u_n = g_A$ as this
combination is determined more precisely, see
Eq.~\eqref{eq:App:u-d}. The $B_0$ coefficient is related to the quark
condensate so that $B_0 m_q\sim m_\pi^2$, see
Eq.~\eqref{eq:B0mq:num}. We have also defined the contributions to the
proton and neutron magnetic moments from the $u$- and $d$-quark
currents, $\hat \mu_u^p=\hat \mu_d^n=1.84, \hat \mu_d^p=\hat
\mu_u^n=-1.03$ (see Eq.~\eqref{eq:nuc:mm:quarks}), while $\mu_s$ is
the $s$ quark contribution to the proton and neutron magnetic moments,
see Eq.~\eqref{eq:mus}.

The $d=2$ Wilson coefficients are 
\begin{align}
\begin{split}\label{eq:c1p2}
c_{1,p}^{(2)}&= \frac{2\alpha}{\pi q^2} \frac{\mu_p}{m_N} \, \hat \C_{1}^{(5,0)}
-\frac{2}{3}\frac{(\Delta u_p+\Delta d_p- 2 \Delta s)}{m_\eta^2-q^2}\Big( \hat \C_{4,u}^{(6,0)}+ \hat \C_{4,d}^{(6,0)}- 2 \hat \C_{4,s}^{(6,0)}\Big)
\\
&\quad-\frac{2g_A}{m_\pi^2-q^2}\Big(\hat \C_{4,u}^{(6,0)}- \hat \C_{4,d}^{(6,0)}\Big) +\frac{B_0}{m_\chi} \frac{g_A}{m_\pi^2-q^2}  \big( m_u\,\hat \C_{8,u}^{(7,0)}-m_d \,\hat \C_{8,d}^{(7,0)}\big)
 \\
&\quad+\frac{B_0}{3 m_\chi}\frac{(\Delta u_p +\Delta d_p- 2\Delta s)}{ m_\eta^2-q^2}  \big(m_u \,\hat \C_{8,u}^{(7,0)}+ m_d\,\hat \C_{8,d}^{(7,0)}-2 m_s\hat \C_{8,s}^{(7,0)}\big) 
\\ 
&\quad+ \tilde m\bigg[ \frac{\Delta u_p}{m_u} +\frac{\Delta d_p}{m_d} + \frac{\Delta s}{m_s} 
 + \frac{g_A}{2} \bigg( \frac{1}{m_u} - \frac{1}{m_d} \bigg) \frac{q^2}{m_\pi^2-q^2} 
 \\
  & \qquad \quad + \frac{1}{6}\big(\Delta u_p + \Delta d_p - 2\Delta s\big) \bigg( \frac{1}{m_u} + \frac{1}{m_d} - \frac{2}{m_s} \bigg) \frac{q^2}{m_\eta^2-q^2} 
  \bigg] \hat \C_4^{(8,1)} ,
\end{split}
\\
c_{2,p}^{(2)}&= - \frac{2\alpha Q_p}{\pi m_N q^2} \, \hat \C_{1}^{(5,0)},\label{eq:c2p2}
\end{align}
while the expressions for the neutron follow from the replacements
$p\to n, u\leftrightarrow d$. As before the above results also apply
to the relativistic form of the DM EFT
\eqref{eq:dim5:nf5:Q1Q2:light}-\eqref{eq:dim7EW:Q7Q8:light} by
replacing $\hat C_i^{(d,0)}\to \hat C_i^{(d)}$, $\hat
\C_{4}^{(8,1)}\to \hat \C_{4}^{(7)}/m_\chi $ 
(for tree level matching to HDMET). Note that, due to the photon
$1/q^2$ pole, the $\hat \C_{1}^{(5,0)}$ contributions are of the same
order as in \eqref{eq:c1p0}, \eqref{eq:c1p1}, even though they
multiply operators that are ${\mathcal O}(q^2)$ suppressed.
Similarly, due to the meson poles, the contributions proportional to
the Wilson coefficients $\hat \C_{4,q}^{(6,0)}$ in $c_{1,p}^{(2)}$,
coming from the right diagram in Fig.~\ref{fig:LOChPT} are of the same
chiral order as the $\hat \C_{4,q}^{(6,0)}$ terms in $c_{2,p}^{(0)}$,
coming from the left diagram in Fig.~\ref{fig:LOChPT}.

The coefficients $c_{1,N}^{(1)}$, $c_{2,p}^{(1)}$, $c_{1,N}^{(2)}$
have a $q^2$-dependence from pion, $\eta$, and photon exchanges, i.e.,
they are non-local at the scale $q\sim m_\pi$. This signals that the
above effective description of DM--nucleon interactions is not an
effective field theory in the usual sense, and ${\cal L}_{\rm eff}$
from~\eqref{eq:LNN} may only be used at tree level. The effective
description does make sense, though, since the pion and the $\eta$
cannot be kinematically produced and never appear as asymptotic
states. In the scattering process $q^2$ is spacelike, so that one
never reaches the pion or $\eta$ pole in the above expressions. The
single photon exchange similarly leads to a classical potential for
the DM-proton interactions.

The above results apply with trivial changes also if the matching to
HDMET is performed beyond tree level. In that case one needs to
replace $ \hat \C_{1}^{(5,0)}\to m_\chi \hat \C_{2}^{(6,1)}$ in
\eqref{eq:c1p0}, $ \hat \C_{1}^{(5,0)}\to m_\chi \hat \C_{1}^{(6,1)}$
in \eqref{eq:c2p2}, $\hat \C_{3,q}^{(6,0)}\to m_\chi \hat
\C_{3,q}^{(7,1)}$ in \eqref{eq:c3p1}, $\hat \C_{3,q}^{(6,0)}\to m_\chi
\hat \C_{6,q}^{(7,1)}$ in \eqref{eq:c5p1}, $\hat \C_{8,q}^{(7,0)}\to
m_\chi \hat \C_{8,q}^{(8,1)}$ in \eqref{eq:c1p2}. 

Using the results of Ref.~\cite{Anand:2013yka} for the nuclear
response in DM direct detection, the cross section for DM scattering
on the nucleus is given by\footnote{For the reader's convenience we
  translate our notation to the basis of~\cite{Anand:2013yka} in
  Appendix~\ref{app:dictionary}.} 
\begin{equation}
\frac{d\sigma}{dE_R}=\frac{m_A}{2\pi |\vec v_\chi|^2} \frac{1}{(2 J_\chi+1)}\frac{1}{(2 J_A+1)}\sum_{\rm spins}|{\cal M}|_{\rm  NR}^2,
\end{equation}
where $E_R$ is the recoil energy of the nucleus, $m_A$ its mass, and
$\vec v_\chi$ the initial DM velocity in the lab frame.  The
non-vanishing contributions to the matrix element squared
are~\cite{Anand:2013yka} 
\begin{equation}\label{eq:Nucl:response:matrixel}
\begin{split}
\frac{1}{2 J_\chi+1}\,\frac{1}{2 J_A+1}\sum_{\rm spins}|{\cal M}|_{\rm  NR}^2=
&\frac{4\pi}{2J_A+1} \sum_{\tau=0,1}\sum_{\tau'=0,1}
\Big\{R_M^{\tau\tau'} W_M^{\tau\tau'}(q)+R_{\Sigma''}^{\tau\tau'} W_{\Sigma''}^{\tau\tau'}(q)
\\
+&R_{\Sigma'}^{\tau\tau'} W_{\Sigma'}^{\tau\tau'}(q)
+\frac{\vec q^{\,\,2}}{m_N^2}\Big[
R_{\Delta}^{\tau\tau'} W_{\Delta}^{\tau\tau'}(q)+R_{\Delta\Sigma'}^{\tau\tau'} W_{\Delta\Sigma'}^{\tau\tau'}(q)\Big]\Big\}, 
\end{split}
\end{equation}
where $J_\chi=1/2$ is the spin of DM, and $J_A$ is the spin of the
target nucleus. The nonrelativistic matrix element ${\cal M}_{\rm NR}$
has the same normalization as the one in \cite{Anand:2013yka}.  The
coefficients $R_i^{\tau\tau'}$ depend on $\vec v_T^{\perp 2}$, $\vec
q^{\,\,2}/m_N^2$, as well as on the coefficients $c_{i,N}^{(d)}$
in~\eqref{eq:c1p0}-\eqref{eq:c2p2}, and are given
by~\cite{Anand:2013yka} 
\begin{align}
\begin{split}\label{eq:RM}
R_M^{\tau \tau'}&=  c_{1,\tau}^{(0)} c_{1,\tau'}^{(0)}+
\frac{m_N^2}{4}\Big[\frac{\vec q^{\,\,2}}{m_N^2}c_{2,\tau}^{(1)} c_{2,\tau'}^{(1)}+\vec v_T^{\perp2} \Big(c_{4,\tau}^{(1)} c_{4,\tau'}^{(1)} +  \vec q^{\,\,2}   c_{2,\tau}^{(2)} c_{2,\tau'}^{(2)}\Big) \Big],
\end{split}
\\
\begin{split}
R_{\Sigma''}^{\tau\tau'}&=  \frac{1}{16}\Big[c_{2,\tau}^{(0)} c_{2,\tau'}^{(0)}+ \vec q^{\,\,2} \Big(c_{2,\tau}^{(0)} c_{1,\tau'}^{(2)}+c_{1,\tau}^{(2)} c_{2,\tau'}^{(0)}+ 4    c_{1,\tau}^{(1)} c_{1,\tau'}^{(1)}\Big) +\vec q^{\,\,4}c_{1,\tau}^{(2)} c_{1,\tau'}^{(2)}\Big],\label{eq:RSigma''}
\end{split}
\\
R_{\Sigma'}^{\tau\tau'}&= \frac{m_N^2}{8} \vec v_T^{\perp 2}\,c_{3,\tau}^{(1)} c_{3,\tau'}^{(1)}
+\frac{1}{16}\Big(c_{2,\tau}^{(0)} c_{2,\tau'}^{(0)} +\vec q^{\,\,2} c_{5,\tau}^{(1)} c_{5,\tau'}^{(1)}\Big),\\
R_{\Delta}^{\tau\tau'}&= \frac{m_N^2}{4}\Big(c_{4,\tau}^{(1)} c_{4,\tau'}^{(1)}+\vec q^{\,\,2} c_{2,\tau}^{(2)} c_{2,\tau'}^{(2)}\Big),\label{eq:RDelta}\\
R_{\Delta\Sigma'}^{\tau\tau'}&= \frac{m_N^2}{4}\Big(c_{4,\tau}^{(1)} c_{5,\tau'}^{(1)}- c_{2,\tau}^{(2)} c_{2,\tau'}^{(0)}\Big).
\label{eq:RDeltaSigma'}
\end{align}
Note that \eqref{eq:RM}-\eqref{eq:RDeltaSigma'} are already specific
to the case of fermionic DM, $J_\chi=1/2$ (see~\cite{Anand:2013yka}
for the general expression).  Above,
\begin{equation}\label{eq:vTperp}
\vec v_T^\perp=\vec v_\chi -\vec q/(2 \mu_{\chi A}),
\end{equation}
is the component of initial DM velocity in the lab frame, $\vec v_\chi
$, that is perpendicular to $\vec q$, in complete analogy with the
single nucleon case \eqref{eq:vperp}. The typical value is $|\vec
v_T^\perp|\sim 10^{-3}$. Here $\mu_{\chi A}=1/(1/m_\chi+1/m_A)$ is the
reduced mass of the DM and the nucleus. The sum in
\eqref{eq:Nucl:response:matrixel} is over isospin values
$\tau=0,1$. The Wilson coefficients $c_{i,\tau}^{(d)}$ are related to
the proton and neutron Wilson coefficients through
\begin{equation}\label{eq:cirelations:isospin}
 c_{i,0}^{(d)}=\frac{1}{2}\big(c_{i,p}^{(d)}+c_{i,n}^{(d)}\big), \qquad  c_{i,1}^{(d)}=\frac{1}{2}\big(c_{i,p}^{(d)}-c_{i,n}^{(d)}\big)\,.
\end{equation}

\section{Scalar dark matter}
\label{sec:scalar:DM}
The above results are easily extended to the case of scalar
DM.\footnote{For operators, spurions, and Wilson coefficients we adopt
  the same notation for scalar DM as for fermionic DM. No confusion
  should arise as this abuse of notation is restricted to this
  section.} For relativistic scalar DM, denoted by $\varphi$, the
effective interactions with the SM start at dimension six,
\begin{equation}\label{eq:lightDM:Ln:scalar}
{\cal L}_\varphi=
\hat \C_{a}^{(6)} {\cal Q}_a^{(6)}+\cdots, 
\qquad {\rm where}\quad 
\hat \C_{a}^{(6)}=\frac{\C_{a}^{(6)}}{\Lambda^{2}}\,,  
\end{equation}
where ellipses denote higher dimension operators. The dimension-six
operators are 
\begin{align}
\label{eq:dim6:Q1Q2:light:scalar}
{\cal Q}_{1,q}^{(6)} & = \big(\varphi^* i\overset{\leftrightarrow}{\partial_\mu} \varphi\big) (\bar q \gamma^\mu q),
 &{\cal Q}_{2,q}^{(6)} &= \big(\varphi^* i\overset{\leftrightarrow}{\partial_\mu} \varphi\big)(\bar q \gamma^\mu \gamma_5 q), 
 \\
 \label{eq:dim6:Q3Q4:light:scalar}
 {\cal Q}_{3,q}^{(6)} & = m_q (\varphi^* \varphi)( \bar q q)\,, 
&{\cal
  Q}_{4,q}^{(6)} &= m_q (\varphi^* \varphi)( \bar q i\gamma_5 q)\,,
\\
\label{eq:dim6:Q5Q6:light:scalar}
  {\cal Q}_5^{(6)} & = \frac{\alpha_s}{12\pi} (\varphi^* \varphi)
 G^{a\mu\nu}G_{\mu\nu}^a\,, 
 & {\cal Q}_6^{(6)} &= \frac{\alpha_s}{8\pi} (\varphi^* \varphi) G^{a\mu\nu}\widetilde G_{\mu\nu}^a\,.
  \\
  \label{eq:dim6:Q7Q8:light:scalar} 
  {\cal Q}_{7}^{(6)} &= i \frac{e}{8 \pi^2} \big(\partial_\mu \varphi^* \partial_\nu \varphi\big)
 F^{\mu\nu} \,, &~ &
\end{align}
Here $\overset{\leftrightarrow}{\partial_\mu}$ is defined through
$\phi_1\overset{\leftrightarrow}{\partial_\mu} \phi_2=\phi_1
\partial_\mu \phi_2- (\partial_\mu \phi_1) \phi_2$, and $q=u,d,s$
again denote the light quarks. The strong coupling constant $\alpha_s$
is taken at $\mu\sim 1$ GeV. The operator ${\cal Q}_6^{(6)}$ is
CP-odd, while the other operators are CP-even. Note that because there
are also leptonic equivalents to the operators ${\cal Q}_{1,q}^{(6)}$
which we do not include in the analysis, the inclusion of ${\cal
  Q}_{7}^{(6)}$ is not redundant (the equations of motion relate
$\partial^\mu F_{\mu\nu}=\sum_f e Q_f \bar f \gamma_\nu f $ where $f$
are both quarks and leptons).

In \eqref{eq:dim6:Q1Q2:light:scalar}-\eqref{eq:dim6:Q7Q8:light:scalar}
we kept the leading operators that one would get from a UV theory of
complex scalar DM for each of the chiral and flavor structures. At
dimension six there are also the Rayleigh operators $(\varphi^*
\varphi) F^{\mu\nu}F_{\mu\nu}$ and $(\varphi^* \varphi)
F^{\mu\nu}\widetilde F_{\mu\nu}$ which, however, lead to scattering
rates suppressed by a factor of $\alpha$ compared to ${\cal
  Q}_{7}^{(6)} $~\cite{Ovanesyan:2014fha}. For real scalar DM the
operators ${\cal Q}_{1,q}^{(6)}$, ${\cal Q}_{2,q}^{(6)}$, and ${\cal
  Q}_{7}^{(6)}$ vanish, and one would need to consider subleading
operators. In this paper we limit ourselves to the case of complex
scalar DM.\footnote{We have also neglected the contributions of
  operators of dimension seven and higher that are promoted to
  dimension five or six in going to HDMET, e.g.,
  $\partial^\mu\varphi^*\partial_\mu\varphi\,\bar q q\to
  m_\varphi^2\varphi_v^*\varphi_v\,\bar q q $. These operators are
  suppressed by additional powers of $m_\varphi/\Lambda$ compared to
  the operators that we consider.}

The next step is to consider scalar DM interactions with the visible
sector in HDMET. To derive it we factor out the large momenta,
\begin{equation}
\varphi(x)=\frac{1}{\sqrt{2 m_\varphi}}e^{-i m_\varphi v\cdot x}\varphi_v\,.
\end{equation}
The HDMET for scalar DM is thus
\begin{equation}
{\cal L}_{\rm HDMET}=\varphi_v^* i v\ncdot \partial
\varphi_v^{\phantom{*}}+\frac{1}{2 m_\varphi}\varphi_v^*
(i\partial_\perp)^2\varphi_v^{\phantom{*}}+\cdots+{\cal
  L}_{\varphi_v}\,. 
\end{equation}
The first term is the LO HDMET for scalar fields. The $1/m_{\varphi}$
term is fixed by reparametrization invariance \cite{Luke:1992cs},
while the ellipses denote the higher-order terms. The interaction
Lagrangian ${\cal L}_{\varphi_v}$ is also expanded in $1/m_\varphi$,
\begin{equation}\label{eq:DM:Lnf5:mphi}
{\cal L}_{\varphi_v}=\sum_{d,m}
\hat \C_{a}^{(d,m)} {\cal Q}_a^{(d,m)}, 
\qquad{\rm where}\qquad
\hat \C_{a}^{(d,m)}=\frac{\C_{a}^{(d,m)}|_{n_f=5}}{\Lambda^{d-m-4}
  m_\varphi^{m}}\,.  
\end{equation}
As for fermionic DM, the operators ${\cal Q}_a^{(d,m)}$ arise as the
terms of order $1/m_\chi^m$ in the HDMET expansion of the UV operators
${\cal Q}_a^{(d)}$
in~\eqref{eq:dim6:Q1Q2:light:scalar}-\eqref{eq:dim6:Q7Q8:light:scalar}.
Because of the derivatives acting on scalar fields the index $m$ can
also be negative, since 
\begin{align}
i\big(\varphi^* \overset{\leftrightarrow}{\partial_\mu} \varphi\big) & \to 2 m_\varphi v_\mu 
\big(\varphi_v^*  \varphi_v\big) +i\big(\varphi_v^* \overset{\leftrightarrow}{\partial_\mu} \varphi_v\big)+\cdots, 
\\
\big(\partial_{[\mu}\varphi^*\partial_{\nu]} \varphi\big) & \to i m_\varphi v_{[\mu}\partial_{\nu]}
\big(\varphi_v^*  \varphi_v\big) +\big(\partial_{[\mu}\varphi_v^* \partial_{\nu]} \varphi_v\big)+\cdots. \label{eq:dphidphi-NR}
\end{align}
We thus have three HDMET operators that start at dimension five
\begin{align}
\label{eq:dim5:Q1Q2:heavy:scalar}
{\cal Q}_{1,q}^{(5,-1)} & = 2 (\varphi_v^*  \varphi_v) (\bar q \slashed v q),
 &{\cal Q}_{2,q}^{(5,-1)} &= 2  (\varphi_v^*  \varphi_v) (\bar q\slashed v \gamma_5 q), 
  \\
    \label{eq:dim5:Q3:heavy:scalar}
  {\cal Q}_{3}^{(5,-1)} &= -\frac{e}{8 \pi^2} v_\mu \partial_\nu \big( \varphi_v^*  \varphi_v\big)  F^{\mu\nu}.
 &~&
 \end{align}
The relevant dimension-six operators are 
\begin{align}
\label{eq:dim6:Q1Q2:heavy:scalar}
{\cal Q}_{1,q}^{(6,0)} & = \big(\varphi_v^* i\overset{\leftrightarrow}{\partial_\mu} \varphi_v\big) (\bar q \gamma^\mu q),
 &{\cal Q}_{2,q}^{(6,0)} &= \big(\varphi_v^* i\overset{\leftrightarrow}{\partial_\mu} \varphi_v\big)(\bar q \gamma^\mu \gamma_5 q), 
\\
 \label{eq:dim6:Q3Q4:heavy:scalar}
 {\cal Q}_{3,q}^{(6,0)} & = m_q (\varphi_v^* \varphi_v)( \bar q q)\,, 
&{\cal
  Q}_{4,q}^{(6,0)} &= m_q (\varphi_v^* \varphi_v)( \bar q i\gamma_5 q)\,,
\\
\label{eq:dim6:Q5Q6:heavy:scalar}
  {\cal Q}_5^{(6,0)} & = \frac{\alpha_s}{12\pi} (\varphi_v^* \varphi_v)
 G^{a\mu\nu}G_{\mu\nu}^a\,, 
 & {\cal Q}_6^{(6,0)} &= \frac{\alpha_s}{8\pi} (\varphi_v^* \varphi_v) G^{a\mu\nu}\widetilde G_{\mu\nu}^a\,,
  \\
  \label{eq:dim6:Q7Q8:heavy:scalar}
  {\cal Q}_{7}^{(6,0)} &= i\frac{e}{8 \pi^2} \big(\partial_\mu \varphi_v^* \partial_\nu \varphi_v\big)
 F^{\mu\nu}.
\end{align}
These are simple extensions of the relativistic operators in
\eqref{eq:dim6:Q1Q2:light:scalar}-\eqref{eq:dim6:Q7Q8:light:scalar},
but the derivatives are now all $ {\mathcal O}(q)$ since they act on
HDMET fields. Reparametrization invariance fixes
 \begin{equation}\label{eq:RPI:scalar}
\C_{1,q}^{(5,-1)}=\C_{1,q}^{(6,0)}\,, \qquad \C_{2,q}^{(5,-1)}=\C_{2,q}^{(6,0)}\,, \qquad \C_{3}^{(5,-1)}=\C_{7}^{(6,0)}
\end{equation}
to all loop orders in the matching at the scale $m_\chi$. In the
Lagrangian~\eqref{eq:DM:Lnf5:mphi} the operators thus always appear in
the linear combinations
\begin{equation}
\begin{split}
 m_\varphi {\cal Q}_{1,q}^{(5,-1)}+{\cal Q}_{1,q}^{(6,0)}, \qquad m_\varphi {\cal Q}_{2,q}^{(5,-1)}+{\cal Q}_{2,q}^{(6,0)}, \qquad  m_\varphi {\cal Q}_{3}^{(5,-1)}+{\cal Q}_{7}^{(6,0)}.
\end{split}
\end{equation}

It is now easy to obtain the ChPT and HBChPT Lagrangians. The external
spurions in the QCD Lagrangian \eqref{eq:appC:QCDLagr} are, for
relativistic DM, 
\begin{align}
\nu_\mu(x)&=- e \bar Q_q A_\mu^{e}+\nu_{\chi, \mu}=- e \bar Q_q A_\mu^e + \bar \C_{1}^{(6)} \big(\varphi^* i \overset{\leftrightarrow}{\partial_\mu} \varphi\big), \label{eq:vmu:scalar}\\
a_\mu(x)&=\bar\C_{2}^{(6)}\big(\varphi^* i \overset{\leftrightarrow}{\partial_\mu} \varphi\big),\\
s(x)&={\cal M}_q +s_\chi={\cal M}_q-{\cal M}_q\, \bar \C_{3}^{(6)}\big(\varphi^* \varphi\big),
\\
p(x)&={\cal M}_q\,  \bar \C_{4}^{(6)}\big(\varphi^* \varphi\big),\\
s_G(x)&=
\hat \C_{5}^{(6)}\big(\varphi^* \varphi\big), \label{sG:matching:scalar}\\
\theta(x)&=
\hat \C_{6}^{(6)}\big(\varphi^* \varphi\big). \label{eq:theta(x):matching:scalar}
\end{align}
For HDMET the external spurions are thus
\begin{align}
\nu_\mu(x)&=- e \bar Q_q A_\mu^{e}+\nu_{\chi, \mu}=- e \bar Q_q A_\mu^e + 2 v^\mu \bar \C_{1}^{(5,-1)}
  (\varphi_v^* \varphi_v)+ \bar \C_{1}^{(6,0)} \big(\varphi_v^* i \overset{\leftrightarrow}{\partial_\mu} \varphi_v\big)+\cdots, \label{eq:vmu:scalar:HDMET}
\\
a_\mu(x)&=2 v^\mu \bar \C_{2}^{(5,-1)}
  (\varphi_v^* \varphi_v)+ \bar \C_{2}^{(6,0)} \big(\varphi_v^* i \overset{\leftrightarrow}{\partial_\mu} \varphi_v\big)+\cdots,
\\
s(x)&={\cal M}_q +s_\chi={\cal M}_q-{\cal M}_q\, \bar \C_{3}^{(6,0)}\big(\varphi_v^* \varphi_v\big)+\cdots,
\\
p(x)&={\cal M}_q\,  \bar \C_{4}^{(6,0)}\big(\varphi_v^* \varphi_v\big)+\cdots ,\\
s_G(x)&=
\hat \C_{5}^{(6,0)}\big(\varphi_v^* \varphi_v\big)+\cdots, \label{sG:matching:scalar:HDMET}\\
\theta(x)&=
\hat \C_{6}^{(6,0)}\big(\varphi_v^* \varphi_v\big)+\cdots,\label{eq:theta(x):matching:scalar:HMDET}
\end{align}
with ellipses denoting higher order terms. We use the same notation
for Wilson coefficients as in \eqref{eq:Wilson-diag-mat}. Note that
the derivatives act on HDMET fields and are thus ${\mathcal O}(q)$.

From this we can immediately obtain the ChPT interactions for scalar
DM, 
\begin{align}
\begin{split}\label{eq:chiCHPT1:full:scalar}
{\cal L}_{\varphi, {\rm ChPT}}^{(1)}&=-i f^2 m_\varphi   (\varphi_v^* \varphi_v) v^\mu \Tr\Big[(U\partial_\mu U^\dagger+U^\dagger\partial_\mu U)\,\bar \C_{1}^{(6,0)} +(U\partial_\mu U^\dagger-U^\dagger\partial_\mu U)\,\bar \C_{2}^{(6,0)}\Big],
\end{split}
\\
\begin{split}\label{eq:chiCHPT2:full:scalar}
{\cal L}_{\varphi, {\rm ChPT}}^{(2)}&\supset-\frac{B_0f^2}{2}(\varphi_v^* \varphi_v) \Tr \Big[ (U+U^\dagger){\cal M}_q  \bar \C_{3}^{(6,0)}+i (U-U^\dagger){\cal M}_q  \bar \C_{4}^{(6,0)}\Big]
\\
&+ (\varphi_v^* \varphi_v)\Big[ \frac{2}{27} \Tr\big(\partial_\mu \Pi \partial^\mu\Pi\big)-\frac{6}{27} B_0 \Tr \big({\cal M}_q\Pi^2\big)\Big] \hat \C_5^{(6,0)}
\\
&-\frac{i f^2}{2}\big(\varphi_v^* i \overset{\leftrightarrow}{\partial_\mu} \varphi_v\big)  \Tr\Big[(U\partial_\mu U^\dagger+U^\dagger\partial_\mu U)\,\bar \C_{1}^{(6,0)} +(U\partial_\mu U^\dagger-U^\dagger\partial_\mu U)\,\bar \C_{2}^{(6,0)}\Big].
\end{split}
\end{align}
In \eqref{eq:chiCHPT1:full:scalar} we used the relations $\bar
\C_{1}^{(5,-1)}=m_\varphi\bar \C_{1}^{(6,0)}$ and $\bar
\C_{2}^{(5,-1)}=m_\varphi\bar \C_{2}^{(6,0)}$, valid to all orders in
perturbation theory (cf. Eq.~\eqref{eq:RPI:scalar}), in order to
explicitly show the $m_\varphi$ dependence. Compared to the fermionic
DM ChPT Lagrangians in
\eqref{eq:chiCHPT1:full}-\eqref{eq:chiCHPT3:full}, there are fewer
terms
in~\eqref{eq:chiCHPT1:full:scalar}-\eqref{eq:chiCHPT2:full:scalar}, as
there is no equivalent of the pseudoscalar and axial-vector DM
currents for scalar DM. Working at LO we thus do not need to consider
${\cal L}_{\varphi, {\rm ChPT}}^{(3)}$ at all. Note that in ${\cal
  L}_{\chi, {\rm ChPT}}^{(2)}$ we need to keep the ${\mathcal O}(p^2)$
terms proportional to $\bar \C_{1,2}^{(6,0)}$, even though these
Wilson coefficients appear already in ${\cal L}_{\varphi, {\rm
    ChPT}}^{(1)}$. Both of these terms give contributions of the same
order in ChEFT since \eqref{eq:chiCHPT1:full:scalar} gives $v\cdot
q\sim {\mathcal O}(p^2)$ suppressed contributions for typical external
momenta.

The HBChPT interaction Lagrangians for scalar DM are
\begin{align}
\begin{split}\label{eq:HBChPT0:full:scalar}
{\cal L}_{\varphi, {\rm HBChPT}}^{(0)}&= m_\varphi (\varphi _v^* \varphi_v ) \Big( 
 \Tr\bar B_v \big[ (\xi^\dagger\bar  \C_{1}^{(6,0)}\xi+\xi\bar  \C_{1}^{(6,0)} \xi^\dagger),B_v\big]
+ 2 \Tr\bar B_v B_v \Tr \bar  \C_{1}^{(6,0)} \Big)
\\
&-\frac{2}{27}m_G\,
(\varphi _v^* \varphi_v ) \Tr ( \bar  B_v  B_v )\,\hat \C_{5}^{(6,0)}\,,
\end{split}
\\
\label{eq:HBChPT1:full:scalar}
\begin{split}
{\cal L}_{\varphi, {\rm HBChPT}}^{(1)}&\supset 2 m_\varphi (\varphi _v^* \varphi_v ) \sum_q v \cdot \tilde J_{q,\mu}^{A,{\rm NLO}} \hat \C_{2,q}^{(6,0)} 
\\
&+i(\varphi_v^*  \lrpartial_\mu \varphi_v) \Big(\frac{1}{2} \Tr\bar B_v \big[v^\mu (\xi^\dagger\bar \C_{2}^{(6,0)}\xi-\xi\bar \C_{2}^{(6,0)} \xi^\dagger),B_v\big] \\
 & \quad +D  \Tr \bar B_v S_N^\mu \big\{\xi^\dagger\bar \C_{2}^{(6,0)}\xi+\xi\bar \C_{2}^{(6,0)} \xi^\dagger, B_v\big\}
 \\
 &\quad +F  \Tr \bar B_v S_N^\mu \big[\xi^\dagger\bar \C_{2}^{(6,0)}\xi+\xi\bar \C_{2}^{(6,0)} \xi^\dagger, B_v\big]+ 2 G \Tr \bar B_v S_N^\mu B_v \Tr\bar \C_{2}^{(6,0)}\Big)
\\
 &- (\varphi_v^* \varphi_v) 
 \frac{\hat \C_{6}^{(6,0)}}{2\Tr ({\cal M}_q^{-1}) }\biggr\{\frac{1}{2} v\ncdot \partial \Tr\bar B_v \big[ (\xi^\dagger  {\cal M}_q^{-1} \xi-\xi  {\cal M}_q^{-1}  \xi^\dagger),B_v\big]
\\
&\quad+ \partial_\mu \Big(D  \Tr \bar B_v S_N^\mu
\big\{\xi^\dagger {\cal M}_q^{-1} \xi+\xi {\cal M}_q^{-1} \xi^\dagger, B_v\big\}\\
& \quad  +F  \Tr \bar B_v S_N^\mu \big[\xi^\dagger
  {\cal M}_q^{-1} \xi+\xi {\cal M}_q^{-1} \xi^\dagger, B_v\big]+ 2 G  \Tr ({\cal M}_q^{-1})
\Tr \bar B_v S_N^\mu B_v \Big) \biggr\} \,,
\end{split}
\\
\begin{split}\label{eq:HBChPT2:full:scalar}
{\cal L}_{\varphi, {\rm HBChPT}}^{(2)}&\supset- (\varphi_v^*\varphi_v) 
\Big[ b_0 \Tr(\bar B_v B_v) 
\Tr {\cal M}_q \big( \bar \C_{3}^{(6,0)} (U^\dagger +U)
-i \bar \C_{4}^{(6,0)} (U^\dagger -U) \big)
\\
 &+b_D \Tr\bar B_v \big\{\xi^\dagger {\cal M}_q \big( \bar \C_{3}^{(6,0)}- i  \bar \C_{4}^{(6,0)}\big) \xi^\dagger+\xi {\cal M}_q \big( \bar \C_{3}^{(6,0)} +i  \bar \C_{4}^{(6,0)}\big) \xi, B_v\big\}
\\
&+b_F \Tr \bar B_v\big[\xi^\dagger {\cal M}_q \big(\bar \C_{3}^{(6,0)} -i \bar \C_{4}^{(6,0)}\big) \xi^\dagger+\xi {\cal M}_q\big(\bar \C_{3}^{(6,0)} +i \bar \C_{4}^{(6,0)}\big)\xi, B_v\big] \Big]\,.
\end{split}
\end{align}
The contribution of ${\cal O}(p^0)$ to the Wilson coefficient $\bar
\C_{2}^{(5,-1)}$ cancels because $v\ncdot \tilde J_{q,\mu}^{A,{\rm
    LO}}=0$, so that the leading contributions are given by ${\cal
  L}_{\varphi, {\rm HBChPT}}^{(1)}$ in \eqref{eq:HBChPT1:full:scalar},
where we used the all-order relation $\bar
\C_{2}^{(5,-1)}=m_\varphi\bar \C_{2}^{(6,0)}$,
Eq. \eqref{eq:RPI:scalar}, to make the dependence on $m_\varphi$
explicit. The expression for the NLO axial-vector current $\tilde
J_{q,\mu}^{A,{\rm NLO}}$ is given in~\eqref{eq:JANLO:expanded}. The
diagonal matrix of Wilson coefficients $\bar \C_i$ was defined in
\eqref{eq:Wilson-diag-mat}.  The A-nucleon irreducible amplitudes
follow the same scaling within ChEFT as for fermionic DM,
Eq. \eqref{eq:counting:rule}, with the trivial replacement
$\epsilon_\chi\to \epsilon_\varphi$. Here, the effective chiral
dimension $\epsilon_\varphi$ is the same as for the fermionic DM, as
we did not include the dimension of the external DM fields in its
definition,and we have $\epsilon_\varphi=d-2$ for ${\cal L}_{\varphi,
  {\rm ChPT}}^{(d)}$ and $\epsilon_\varphi=d-1$ for ${\cal
  L}_{\varphi, {\rm HBChPT}}^{(d)}$.

Accounting for the effect of $\pi^0$ and $\eta$ exchange through
$q$-dependent Wilson coefficients results in an effective Lagrangian
\begin{equation}\label{eq:LNN:scal}
{\cal L}_{\rm eff}= \sum_{i,d} \Big( c_{i,p}^{(d)}(q^2)\, Q_{i,p}^{(d)}+c_{i,n}^{(d)}(q^2)\, Q_{i,n}^{(d)}\Big),
\end{equation}
where $d$ denotes the number of derivatives.  For scalar DM we have,
for $d=0,1$,
\begin{align}
\label{eq:Q1p0:scalar}
Q_{1,p}^{(0)}&= \big(\varphi_v^* \varphi_v\big)\big(\bar p_v p_v\big), 
&~&
\\
\label{eq:Q12p1:scalar}
Q_{1,p}^{(1)}&=\big(\varphi_v^* \varphi_v\big)\big(\bar p_v i q\ncdot  S_N p_v\big),&Q_{2,p}^{(1)}&= m_N \big(\varphi_v^* \varphi_v\big)\big(\bar p_v \, v_\perp \ncdot S_N\, p_v\big) ,
\end{align}  
with a similar set of operators for neutrons, with $p\to n$. Unlike
fermionic DM we do not need the $d=2$ operators when working to
leading order. For fermionic DM, photon exchange and couplings of the
DM spin to axial-vector and pseudoscalar quark currents lead to
momentum-suppressed operators in the nonrelativistic limit. Because of
the enhancement by the photon and pion poles, respectively, these
contributions were of leading order. No such terms are possible for
scalar DM as it does not carry spin. The leading contributions from
the operators 
\eqref{eq:dim6:Q1Q2:light:scalar}-\eqref{eq:dim6:Q7Q8:light:scalar}
are thus already captured by the nonrelativistic
operators~\eqref{eq:Q1p0:scalar} and~\eqref{eq:Q12p1:scalar} with up
to one derivative. The matching calculation gives, for scalar DM,
\begin{align}
\begin{split}\label{eq:c1p0:scalar}
c_{1,p}^{(0)}&=2m_\varphi\Big(2\,\hat \C_{1,u}^{(6,0)}+\hat \C_{1,d}^{(6,0)}\Big)-\frac{2m_G}{27}
\hat \C_{5}^{(6,0)} \\
  & \quad  + \sigma_u^p\, \hat \C_{3,u}^{(6,0)} 
           + \sigma_d^p\, \hat \C_{3,d}^{(6,0)}
           + \sigma_s \, \hat \C_{3,s}^{(6,0)} 
   - \frac{\alpha Q_p}{2\pi} m_\varphi \hat \C_{7}^{(6,0)}\,,
\end{split}
 \\
 \begin{split}\label{eq:c1p1:scalar}
c_{1,p}^{(1)}&=
 \frac{B_0 \, g_A}{m_\pi^2+\vec q^{\,\,2}}\big( m_u\hat \C_{4,u}^{(6,0)}-m_d \hat \C_{4,d}^{(6,0)}\big)
\\
&\quad+\frac{B_0}{3}\frac{(\Delta u_p+\Delta d_p -2 \Delta s)}{ m_\eta^2+\vec q^{\,\,2}} \big(m_u \hat \C_{4,u}^{(6,0)}+ m_d\hat \C_{4,d}^{(6,0)}-2 m_s\hat \C_{4,s}^{(6,0)}\big)
\\
&\quad 
- \tilde m\bigg[ \frac{\Delta u_p}{m_u} +\frac{\Delta d_p}{m_d} + \frac{\Delta s}{m_s} 
 - \frac{g_A}{2} \bigg( \frac{1}{m_u} - \frac{1}{m_d} \bigg) \frac{\vec q^{\,\,2}}{m_\pi^2+\vec q^{\,\,2}} \\
  & \qquad \quad - \frac{1}{6}\big(\Delta u_p + \Delta d_p - 2\Delta s\big) \bigg( \frac{1}{m_u} + \frac{1}{m_d} - \frac{2}{m_s} \bigg) \frac{\vec q^{\,\,2}}{m_\eta^2+\vec q^{\,\,2}} 
  \bigg] \hat \C_6^{(6,0)}\,, 
\end{split}
\\
\label{eq:c2p1:scalar}
c_{2,p}^{(1)}&=\frac{4 m_\varphi}{m_N}\Big[\Delta u_p\, \hat \C_{2,u}^{(6,0)}+ \Delta d_p\, \hat \C_{2,d}^{(6,0)}+ \Delta s\, \hat \C_{2,s}^{(6,0)}\Big]\,.
\end{align}
Starting from the EFT for relativistic DM
\eqref{eq:dim6:Q1Q2:light:scalar}-\eqref{eq:dim6:Q7Q8:light:scalar},
the above results can be used by simply replacing $\hat
\C_{i}^{(6,0)}\to \hat \C_{i}^{(6)}=\C_{i}^{(6)}/\Lambda^2$. Because
of the reparametrization invariance relation \eqref{eq:RPI:scalar},
they are also valid if the masses of DM and mediators are comparable,
in which case the matching to HDMET is done at the same time as the
mediators are being integrated out, i.e., at $\mu\sim \Lambda\sim
m_\varphi$. Note that $B_0 m_q\sim {\mathcal O}(m_\pi^2)$, with the
explicit relations given in \eqref{eq:B0mq:num}. In terms of the
$p^\nu$ scaling we have for the leading contributions proportional to
the $\hat \C_a^{(n,m)}$ Wilson coefficients 
\begin{equation}
\begin{split}
\label{eq:chiral:scaling:scalar}
\nu_{\rm min}&=\nu_{\rm LO}[J_\varphi^V\ncdot \tilde J_q^V]=\nu_{\rm LO}[J_\varphi^S\tilde J^G]\,,
\\
\nu_{\rm min}+1&=\nu_{\rm LO}[J_\varphi^V \ncdot \tilde J_q^A]= \nu_{\rm LO}[J_\varphi^S \tilde J^\theta]=\nu_{\rm LO}[J_\varphi^S J_q^P]\,,
 \\
\nu_{\rm min}+2&= \nu_{\rm LO}[J_\varphi^S \tilde J_q^S]\,,
\end{split}
\end{equation}
where we follow the same notation as for the case of fermionic DM for
ease of comparison. The vector and scalar DM currents are
$J_\varphi^V= i \varphi^* \overset{\leftrightarrow}{\partial_\mu}
\varphi$ and $J_\varphi^S=\varphi^* \varphi$, respectively, with their
HDMET decomposition given in \eqref{eq:dim5:Q1Q2:heavy:scalar},
\eqref{eq:dim5:Q3:heavy:scalar}. The leading contributions for the
$J_\varphi^S J_q^P$ interaction comes from the right diagram in
Fig.~\ref{fig:LOChPT}. The pion exchange reduces the chiral scaling of
the resulting amplitude by one, compared to the contact interaction.
For all the other operators the leading contribution comes from the
left diagram in Fig.~\ref{fig:LOChPT}, so that the chiral scaling is
given by the chiral dimension of the corresponding HBChPT Lagrangian,
${\cal L}_{\varphi,{\rm HBChPT}}^{(d)}$. As before, $\nu_{\rm
  min}=3-3A$ simply reflects our normalization of the $A$-nucleon
state. Note that the
results~\eqref{eq:c1p0:scalar}-\eqref{eq:c1p1:scalar} are valid for
matching at $\mu\sim m_\varphi$ to all loop orders, but only to
leading order in the chiral expansion.

The cross section for scalar DM scattering on the nucleus
is~\cite{Anand:2013yka} 
\begin{equation}
\frac{d\sigma}{dE_R}=\frac{m_A}{2\pi |\vec v_\chi|^2}\frac{4\pi}{2J_A+1}\biggr[ \sum_{\tau,\tau'=0,1}
R_M^{\tau\tau'} W_M^{\tau\tau'}(q)+R_{\Sigma''}^{\tau\tau'} W_{\Sigma''}^{\tau\tau'}(q)+
R_{\Sigma'}^{\tau\tau'} W_{\Sigma'}^{\tau\tau'}(q)\biggr],
\end{equation}
where $E_R$ is the recoil energy of the nucleus, $m_A$ its mass, $\vec
v_\chi$ the initial DM velocity in the lab frame, and
$W_i^{\tau\tau'}(q)$ the nuclear response functions. The coefficients
multiplying them are given by 
\begin{equation}
\label{eq:RM:scalar}
R_M^{\tau \tau'}=  c_{1,\tau}^{(0)} c_{1,\tau'}^{(0)}, \qquad
R_{\Sigma''}^{\tau\tau'}=  \frac{1}{4}  \vec q^{\,\,4}  c_{1,\tau}^{(1)} c_{1,\tau'}^{(1)},
\qquad
R_{\Sigma'}^{\tau\tau'}= \frac{m_N^2}{8} \vec v_T^{\perp 2}\,c_{2,\tau}^{(1)} c_{2,\tau'}^{(1)}.
\end{equation}
The perpendicular velocity $\vec v_T^\perp$ is defined in
\eqref{eq:vTperp}, while the relations between the coefficients in
\eqref{eq:c1p0:scalar}-\eqref{eq:c2p1:scalar} and the coefficients in
the isospin basis $c_{i,\tau}^{(d)}$ are given in
\eqref{eq:cirelations:isospin}.

\section{Conclusions}
\label{sec:conclusions}
Dark Matter scattering in direct detection is naturally described by
an EFT if the mediators are heavier than about $\sim1$GeV. We
performed the leading order matching between the EFT with quark,
gluons and photons as the external states and the EFT that describes
DM interactions with light mesons and nucleons. We covered both
fermionic and scalar DM and analyzed the operators that correspond to
interactions between the visible and DM sector up to and including
dimension-six operators above the electroweak scale. The resulting EFT
was then used to obtain the coefficients that multiply the nuclear
response functions, see, e.g., Ref.~\cite{Anand:2013yka}. Our main
results for fermionic DM are given in
\eqref{eq:c1p0}-\eqref{eq:c2p2}. With these one can go directly from
the EFT with quarks, gluons and photons,
Eqs.~\eqref{eq:dim5:nf5:Q1Q2:light}-\eqref{eq:dim7EW:Q7Q8:light} to
the nuclear response functions and DM scattering rates.  The results
for scalar DM are given in
\eqref{eq:c1p0:scalar}-\eqref{eq:c1p1:scalar}. The translation to the
notation of Ref.~\cite{Anand:2013yka} for fermionic DM is given in
Appendix~\ref{app:dictionary}, in
Eqs.~\eqref{eq:CNR1:app}-\eqref{eq:CNR11:app}. Note that only a subset
of 9 out of 14 possible nonrelativistic operators with up to two
derivatives is generated in our set-up.

For each of the initial operators coupling DM to quarks and gluons we
derived the leading contributions when they hadronize. In order to
compare the size of different potential contributions we used chiral
power counting in the ChEFT of nuclear forces, where we counted the
momentum exchange between DM and the nucleus as $|\vec q| \sim
m_\pi\sim {\mathcal O}(200{\rm ~MeV})$.

Using this counting one can see, for instance, that for fermionic DM
the axial-axial operator induces two different leading contributions
to the spin-dependent scattering rate. The first contribution is due
to the scattering of DM on a single nucleon, while the second
contribution arises from a pion exchange between DM and the
nucleon. The pion exchange contribution involves a nonrelativistic
operator with two derivatives, $Q_{1,N}^{(2)}$ in Eq.~\eqref{eq:Q1p2}
that would naively give a ${\mathcal O}(q^2)$ suppressed contribution
to the scattering amplitude.  Its contribution is, however, enhanced
by the pion pole $1/(m_\pi^2+|\vec q\,|^2)$, giving a contribution of
${\mathcal O}(|\vec q\,|^2/m_\pi^2)\sim {\mathcal O}(1)$ for $|\vec
q\,|\sim m_\pi$. For this reason we needed to keep the nonrelativistic
operators with up to two derivatives.

Similar arguments apply to all the operators in
Eqs.~\eqref{eq:dim5:nf5:Q1Q2:light}-\eqref{eq:dim7EW:Q7Q8:light}. Pion
exchange is the leading contribution to the scattering amplitude for
operators with pseudoscalar quark currents, while it is of the same
order as the contact interactions with nucleons for the axial-axial
operator as well as those operators coupling the DM current to
$G\tilde G$. For the remaining operators, the DM-nucleon contact
interactions give the leading contributions. Moreover, obtaining
contributions of leading order in chiral counting requires some care
for the case of vector and axial-vector quark currents, since these
need to be expanded to NLO in chiral counting when they are multiplied
by axial-vector and vector DM currents, respectively.

The EFT we constructed in this paper is valid at $\mu\sim 1$GeV. A
different EFT analysis, valid all the way up to the scale of the
mediator much above the electroweak scale, can be useful when relating
direct detection to processes at much higher energies, the DM searches
at the LHC \cite{Haisch:2015ioa, Cotta:2012nj, Busoni:2013lha,
  Fox:2011fx, Rajaraman:2011wf, Fox:2011pm, Racco:2015dxa,
  Jacques:2015zha} or signals from DM annihilation
\cite{Kumar:2013iva, Cao:2009uw, Goodman:2010qn, Ciafaloni:2011sa,
  Cheung:2011nt, Cheung:2012gi}. When relating these with direct
detection it is important to use simplified models
\cite{Bauer:2016gys, Abdallah:2015ter, Bruggisser:2016nzw,
  DeSimone:2016fbz, Kahlhoefer:2015bea} and even to include loop
corrections \cite{Crivellin:2014qxa, D'Eramo:2014aba, D'Eramo:2016atc,
  Crivellin:2015wva, Haisch:2013ata, Haisch:2013uaa, Haisch:2012kf,
  Frandsen:2012db, Freytsis:2010ne}. In the present work we completed
the final step of this program, explicitly connecting the EFT
describing DM interactions with quarks and gluons with nuclear
physics.

Our results assume that that there are no large cancellations between
different Wilson coefficients in the UV. In the presence of
cancellations one would need to include terms of higher order in the
chiral expansion.  For instance, the pseudoscalar-pseudoscalar UV
operator ${\cal Q}_{8,q}^{(7)}$ in Eq.~\eqref{eq:dim7EW:Q7Q8:light}
contributes to the $(q\cdot S_\chi)(q\cdot S_N)$ nonrelativistic
operator in Eq.~\eqref{eq:Q1p2}. This contribution vanishes, however,
if $m_u \hat \C_{8,u}^{(7,0)} = m_d \hat \C_{8,d}^{(7,0)} = m_s \hat
\C_{8,s}^{(7,0)}$, cf., Eq.~\eqref{eq:c1p2}. The leading contribution
to the DM-nucleon scattering would then come from a contact term of
higher order in chiral counting which could be viewed as due to
$\eta'$ exchange. For contributions of this type one could easily
extend our analysis and include the $\eta'$ exchange contributions by
multiplying each term in ${\cal L}_{\rm ChPT}+{\cal L}_{\rm HBChPT}$
with an arbitrary function of the $\eta'$ field. However, since the
mass of the $\eta'$ is comparable with the cut-off of the theory, it
is consistent to integrate it out, as we did. The same is true for the
scalar-pseudoscalar UV operator ${\cal Q}_{7,q}^{(7)}$ in
Eq.~\eqref{eq:dim7EW:Q7Q8:light} whose contributions to the $(q\ncdot
S_N)$ nonrelativistic operator in Eq.~\eqref{eq:Q12p1} also vanish in
the limit $m_u \hat \C_{7,u}^{(7,0)} = m_d \hat \C_{7,d}^{(7,0)} = m_s
\hat \C_{7,s}^{(7,0)}$. A somewhat different situation is encountered
for the axialvector-axialvector operator ${\cal Q}_{4,q}^{(6)}$ in
Eq.~\eqref{eq:dim6EW:Q3Q4:light}. Its contributions to $Q_{1,N}^{(2)}$
vanish if $\hat \C_{4,u}^{(6,0)} = \hat \C_{4,d}^{(6,0)}= \hat
\C_{4,s}^{(6,0)}$. However, in this case the contact contributions of
${\cal Q}_{4,q}^{(6)}$ to $Q_{2,N}^{(0)}$ would still be nonzero, see
Eq.~\eqref{eq:c2p0}, and would be leading over the $\eta'$ exchange
contributions.

Similar situations can arise for all the other operators in
\eqref{eq:dim5:nf5:Q1Q2:light}-\eqref{eq:dim7EW:Q7Q8:light}, where,
through fine-tuning in the UV theory, one can cancel the leading
contributions in chiral counting. In such situations it would be
important to extend our analysis to higher orders in chiral counting,
as well as to analyze whether or not such fine-tunings are stable
under quantum corrections in the UV.  We postpone such an analysis to
future work~\cite{BBGZ:2016b}. 

Note that our analysis, while valid within the assumed power counting,
does not capture the leading contributions for all theories of DM,
even without considering fine tuning. For instance, dimension-seven
Rayleigh operators can be leading for Majorana DM~\cite{Weiner:2012cb,
  Weiner:2012gm}. For this particular case the EFT analysis is already
available \cite{Ovanesyan:2014fha}, while a more complete analysis of
dimension-seven and higher dimension operators is still called for.

{\bf Acknowledgements:} We would like to thank V. Cirigliano, E. Del
Nobile, M. Hoferichter, D. Phillips, S. Scherer, and M. Solon for
useful discussions, and F. Ertas and A. Gootjes-Dreesbach for comments
on the manuscript. J.Z. is supported in part by the U.S. National
Science Foundation under CAREER Grant PHY-1151392. This work was
performed in part at the Aspen Center for Physics, which is supported
by National Science Foundation grant PHY-1066293. F.B. is supported by
the STFC and and acknowledges the hospitality and support of the CERN
theory division.  BG is supported in part by the U.S. Department of
Energy under grant DE-SC0009919.\appendix

\section{Relation to the basis of Anand et al.}
\label{app:dictionary}
In this appendix we relate our nonrelativistic basis to the operator
basis from Ref.~\cite{Anand:2013yka}. The operators in
\eqref{eq:Q1p0}-\eqref{eq:Q1p2} are products of nonrelativistic DM and
nucleon currents, although still given in a Lorentz covariant
notation. We now pass to a manifestly nonrelativistic
notation\footnote{Our metric convention for the Lorentz vectors is
  $p^\mu=(p^0,\vec p\,)$, $p_\mu=(p^0,-\vec p\,)$.}, for which we use
the operator basis from Ref.~\cite{Anand:2013yka}. The operators with
up to two derivatives are 
\begin{align}
\label{eq:O1pO2p}
{\mathcal O}_1^N&= \mathbb{1}_\chi \mathbb{1}_N\,,
&{\mathcal O}_2^N&= \big(v_\perp\big)^2 \, \mathbb{1}_\chi \mathbb{1}_N\,,
\\
\label{eq:O3pO4p}
{\mathcal O}_3^N&= \mathbb{1}_\chi \, \vec S_N\cdot \Big(\vec v_\perp\negthickspace \times \frac{i\vec q}{m_N}\Big) \,,
&{\mathcal O}_4^N&= \vec S_\chi \cdot \vec S_N \,,
\\
\label{eq:O5pO6p}
{\mathcal O}_5^N&= \vec S_\chi \cdot \Big(\vec v_\perp \times \frac{i\vec q}{m_N} \Big) \, \mathbb{1}_N \,,
&{\mathcal O}_6^N&= \Big(\vec S_\chi \cdot \frac{\vec q}{m_N}\Big) \, \Big(\vec S_N \cdot \frac{\vec q}{m_N}\Big),
\\
\label{eq:O7pO8p}
{\mathcal O}_7^N&= \mathbb{1}_\chi \, \big( \vec S_N \cdot \vec v_\perp \big)\,,
&{\mathcal O}_8^N&= \big( \vec S_\chi \cdot \vec v_\perp \big) \, \mathbb{1}_N\,,
\\
\label{eq:O9pO10p}
{\mathcal O}_9^N&= \vec S_\chi \cdot \Big(\frac{i\vec q}{m_N} \times \vec S_N \Big)\,,
&{\mathcal O}_{10}^N&= - \mathbb{1}_\chi \, \Big(\vec S_N \cdot \frac{i\vec q}{m_N} \Big)\,,
\\
\label{eq:O11pO12p}
{\mathcal O}_{11}^N&= - \Big(\vec S_\chi \cdot \frac{i\vec q}{m_N} \Big) \, \mathbb{1}_N \,,
&{\mathcal O}_{12}^N&= \vec S_\chi \cdot \Big( \vec S_N \times \vec v_\perp \Big) \,,
\\
\label{eq:O13pO14p}
{\mathcal O}_{13}^N&= -\Big(\vec S_\chi \cdot \vec v_\perp \Big) \, \Big(\vec S_N\cdot \frac{i\vec q}{m_N} \Big) \,,
&{\mathcal O}_{14}^N&= -\Big(\vec S_\chi \cdot \frac{i\vec q}{m_N}  \Big) \, \Big(\vec S_N\cdot \vec v_\perp  \Big) \,,
\end{align}
with $N=p,n$. Note that each insertion of $\vec q$ is accompanied with
a factor of $1/m_N$, so that all of the above operators have the same
dimensionality. The minus signs and order changes in the cross
products for the definitions of some of the operators compensate the
relative sign difference between our convention for the momentum
exchange \eqref{eq:momentum:exchange} and the one
in~\cite{Anand:2013yka}.

The Wilson coefficients are in this basis given by 
\begin{align}
c_{{\rm NR},1}^N&=c_{1,N}^{(0)}\,, 
&c_{{\rm NR},2}^N&=0\,, 
& c_{{\rm NR},3}^N&=0\,,\label{eq:cNR1-3}
\\
 c_{{\rm NR},4}^N&=-c_{2,N}^{(0)}\,, 
&c_{{\rm NR},5}^N&=m_N^2 c_{2,N}^{(2)}\,,
 &c_{{\rm NR},6}^N&=-m_N^2c_{1,N}^{(2)}\,,\label{eq:cNR4-6}
 \\
   c_{{\rm NR},7}^N&=-m_Nc_{3,N}^{(1)}\,,
    &c_{{\rm NR},8}^N&=-m_Nc_{4,N}^{(1)}\,,
& c_{{\rm NR},9}^N&=m_Nc_{5,N}^{(1)}\,,\label{eq:cNR7-9}
 \\ 
c_{{\rm NR},10}^N&=m_Nc_{1,N}^{(1)}\,, 
&
  c_{{\rm NR},11}^N&=m_Nc_{2,N}^{(1)}\,,&c_{{\rm NR},12}^N&=-m_Nc_{6,N}^{(1)}\,.\label{eq:cNR10-12}
   \\ 
c_{{\rm NR},13}^N&=0\,, 
&
  c_{{\rm NR},14}^N&=0\,.&\label{eq:cNR13-14} 
\end{align}
With this dictionary one can go directly from the EFT with quark,
gluons and photons as external states,
\eqref{eq:dim5:nf5:Q1Q2:light}-\eqref{eq:dim7EW:Q7Q8:light}, to the
nuclear response functions, using the coefficients in
\eqref{eq:c1p0}-\eqref{eq:c2p2}.  In the expressions
\eqref{eq:c1p1}-\eqref{eq:c2p2} one also needs to replace $q^2 \to -
\vec q^{\,\,2}$, so that, e.g., the propagators due to pion exchange
are proportional to $1/(m_\pi^2+\vec q^{\,\,2})$.  For tree-level
matching onto HDMET we thus have in terms of the operators
\eqref{eq:dim5:nf5:Q1Q2:light}-\eqref{eq:dim7EW:Q7Q8:light} 
\begin{align}
\label{eq:CNR1:app}
c_{{\rm NR},1}^p&=2\,\hat \C_{1,u}^{(6)}+\hat \C_{1,d}^{(6)}-\frac{2m_G}{27}
 \hat \C_{1}^{(7)}
    +\sigma_u^p\, \hat \C_{5,u}^{(7)}   +\sigma_d^p\, \hat \C_{5,d}^{(7)}
      +\sigma_s \hat \C_{5,s}^{(7)} 
   - \frac{\alpha Q_p}{2\pi m_\chi}  \hat \C_{1}^{(5)}\,, 
\\
 c_{{\rm NR},4}^p&=-4\,\Big(\Delta u_p\, \hat \C_{4,u}^{(6)}+\,\Delta d_p\, \hat \C_{4,d}^{(6)}+\Delta s\, \hat \C_{4,s}^{(6)} \Big)
 - \frac{2\alpha}{\pi} \frac{\mu_p}{m_N} \hat \C_{1}^{(5)}\,, \label{eq:CNR4}
 \\
c_{{\rm NR},5}^p&= \frac{2\alpha Q_p m_N}{\pi \vec q\,{}^2} \, \hat \C_{1}^{(5)}\,, \label{eq:CNR:5}
 \\
\begin{split}\label{eq:CNR:6}
 c_{{\rm NR},6}^p&=m_N^2\bigg\{\frac{2\alpha}{\pi \vec q\,{}^2}
 \frac{\mu_p}{m_N} \, \hat \C_{1}^{(5)}
+\frac{2}{3}\frac{(\Delta u_p+\Delta d_p- 2 \Delta s)}{m_\eta^2+\vec q\,{}^2}\Big( \hat \C_{4,u}^{(6)}+ \hat \C_{4,d}^{(6)}- 2 \hat \C_{4,s}^{(6)}\Big) 
\\
&\quad + \frac{2g_A}{m_\pi^2+\vec q\,{}^2}\Big(\hat \C_{4,u}^{(6)}-\hat \C_{4,d}^{(6)}\Big) -\frac{B_0}{m_\chi} \frac{g_A}{m_\pi^2+\vec q\,{}^2}  \big( m_u\,\hat \C_{8,u}^{(7)}-m_d \,\hat \C_{8,d}^{(7)}\big)
\\
&\quad-\frac{B_0}{3 m_\chi}\frac{(\Delta u_p +\Delta d_p- 2\Delta s)}{ m_\eta^2+\vec q\,{}^2}  \big(m_u \,\hat \C_{8,u}^{(7)}+ m_d\,\hat \C_{8,d}^{(7)}-2 m_s\hat \C_{8,s}^{(7)}\big) 
\\
 &\quad -\frac{\tilde m}{m_\chi} \bigg[ \frac{\Delta u_p}{m_u} +\frac{\Delta d_p}{m_d} + \frac{\Delta s}{m_s}
  - \frac{g_A}{2} \bigg( \frac{1}{m_u} - \frac{1}{m_d} \bigg) \frac{\vec q^{\,\,2}}{m_\pi^2+\vec q^{\,\,2}}
\\
  & \qquad \qquad \quad - \frac{1}{6}\big(\Delta u_p + \Delta d_p - 2\Delta s\big) \bigg( \frac{1}{m_u} + \frac{1}{m_d} - \frac{2}{m_s} \bigg) \frac{\vec q^{\,\,2}}{m_\eta^2+\vec q^{\,\,2}}  \bigg]\hat \C_4^{(7)}\bigg\}\,, 
\end{split}
\\
   c_{{\rm NR},7}^p&=-2 \Big(\Delta u_p\, \hat \C_{3,u}^{(6)}+ \Delta d_p\, \hat \C_{3,d}^{(6)}+ \Delta s\, \hat \C_{3,s}^{(6)}\Big)\,,
   \\
    c_{{\rm NR},8}^p&=4\,\hat \C_{2,u}^{(6)}+2 \,\hat \C_{2,d}^{(6)}\,,
    \\
     \label{eq:CNR9:app}
     c_{{\rm NR},9}^p&=
      4\hat \mu_u^p\hat \C_{2,u}^{(6)} + 2\hat \mu_d^p \hat \C_{2,d}^{(6)} 
 -6 \mu_s\, \hat \C_{2,s}^{(6)} +\frac{2m_N}{m_\chi} \Big(\Delta u_p \, \hat \C_{3,u}^{(6)} +
\Delta d_p\, \hat \C_{3,d}^{(6)}+\Delta s\, \hat \C_{3,s}^{(6)}\Big)\,,
\\ 
 \begin{split}
 \label{eq:CNR10:app}
c_{{\rm NR},10}^p&=m_N\bigg\{
 \frac{B_0 \, g_A}{m_\pi^2+\vec q\,{}^2}\big( m_u\hat \C_{7,u}^{(7)}-m_d \hat \C_{7,d}^{(7)}\big)
 \\
&\qquad\quad+\frac{B_0}{3}\frac{(\Delta u_p+\Delta d_p -2 \Delta s)}{ m_\eta^2+\vec q\,{}^2} \big(m_u \hat \C_{7,u}^{(7)}+ m_d\hat \C_{7,d}^{(7)}-2 m_s\hat \C_{7,s}^{(7)}\big)
\\
&\qquad\quad -\tilde m \bigg[ \frac{\Delta u_p}{m_u} +\frac{\Delta d_p}{m_d} + \frac{\Delta s}{m_s}
  - \frac{g_A}{2} \bigg( \frac{1}{m_u} - \frac{1}{m_d} \bigg) \frac{\vec q^{\,\,2}}{m_\pi^2+\vec q^{\,\,2}} \\
  & \qquad \qquad \quad - \frac{1}{6}\big(\Delta u_p + \Delta d_p - 2\Delta s\big) \bigg( \frac{1}{m_u} + \frac{1}{m_d} - \frac{2}{m_s} \bigg) \frac{\vec q^{\,\,2}}{m_\eta^2+\vec q^{\,\,2}}  \bigg]\hat \C_3^{(7)}\bigg\}\,,
\end{split}
\\
\label{eq:CNR11:app}
  c_{{\rm NR},11}^p&=\frac{m_N}{m_\chi}
  \Big(- \sigma_u^p \hat \C_{6,u}^{(7)} - \sigma_d^p \hat \C_{6,d}^{(7)}- \sigma_s \hat \C_{6,s}^{(7)} 
+\frac{2 m_G}{27}
\hat \C_2^{(7)} + \frac{2\alpha Q_p m_\chi}{\pi \vec q\,{}^2} \hat \C_{2}^{(5)}\Big) \,,
\end{align}
while the remaining coefficients are zero. The coefficients for
neutrons are obtained by replacing $p\to n$, $u\leftrightarrow d$.
The Wilson coefficients of ${\mathcal O}_2^{N}, {\mathcal O}_3^{N}$,
${\mathcal O}_{13}^{N}$, ${\mathcal O}_{14}^{N}$ are zero in our
framework as a result of the fact that we limited our discussion to
the operators \eqref{eq:dim5:HDM:nf5:Q1Q2}-\eqref{eq:dim6EW:Q5Q6:HDM1}
that can be generated from UV physics described by dimension-five and
dimension-six operators above the electroweak scale~\cite{BBGZ:2016b}.
These Wilson coefficients are expected to be generated if either one
works to higher orders in $q$ or if higher dimension operators are
included in the UV.

The above expressions extend the results in
Ref.~\cite{DelNobile:2013sia}, where estimates for $ c_{{\rm NR},6}^N$
and $ c_{{\rm NR},10}^N$ were obtained without using the chiral
expansion and thus do not contain the pion pole contributions. 
A chiral expansion was performed in
Ref.~\cite{Hoferichter:2015ipa}. Our expressions involving the axial
quark current agree with Ref.~\cite{Hoferichter:2015ipa}, as do the
expressions for the pseudoscalar quark currents in the limits where
the results of Ref.~\cite{Hoferichter:2015ipa} are applicable, i.e.,
for either isospin triplet or flavor octet flavor structures.

\section{Further details on chiral dark matter interactions}
\label{app:furtherDMChPT}
In this appendix we give further details on the ChPT and HBChPT
Lagrangians that describe DM interactions.
\subsection{HBChPT Lagrangian at second order}
\label{app:HBChPT2}
We first give the full form of the HCBhPT Lagrangian at ${\mathcal
  O}(p^2)$, including DM interactions. The terms relevant for our
analysis were shown already in Eq.~\eqref{eq:HBChPT2-vec}. The
complete form of the Lagrangian is (see also \cite{Jenkins:1990jv,
  Brown:1993yv})
\begin{equation}
\label{eq:appC:HBChPT2}
{\cal L}_{\rm HBChPT}^{(2)}={\cal L}_{ps}^{(2)}+{\cal L}_{V}^{(2)}+{\cal L}_{A}^{(2)}+{\cal L}_{S}^{(2)},
\end{equation}
where we split the contributions proportional to different spurions as
denoted by subscripts (except for ${\cal L}_{S}^{(2)}$ that collects
terms that involve the nuclear spin operator). The ${\cal
  L}_{ps}^{(2)}$ contains the scalar and pseudoscalar spurions,
\begin{equation}
{\cal L}_{ps}^{(2)}=b_D \Tr\bar B_v\{ s_+,B_v\} +b_F\Tr \bar B_v [s_+, B_v]+b_0\Tr\big(\bar B_vB_v\big)\Tr\big(s_+^\chi\big)\,.
\end{equation}
The terms with the vector current, $V^\mu$, are 
\begin{equation}
\begin{split}\label{eq:LV2}
{\cal L}_{V}^{(2)}=&c_1 \Tr\big(\bar B_v \nabla^2 B_v\big) -c_1' \Tr\big(V_\mu\big)\Tr\big(\bar B_v\,  i \lrnabla^\mu B_v\big)
-c_2' \Tr\big(\partial\ncdot V \big)\Tr\big(\bar B_v  B_v\big)\\
&-c_3' \Tr\big(\bar B_v  B_v\big) v\ncdot \partial\Tr\big(v \ncdot V \big)
+ c_4' \big(\Tr V_\mu)^2\Tr(\bar B_v  B_v\big)\\& +c_5' \big(\!\Tr v \ncdot V \big)^2\Tr(\bar B_v  B_v)\,,
\end{split}
\end{equation}
where\footnote{We use the convention that $\lrnabla^\mu$ acts only
  inside the brackets. In $c_1'$ there is thus not derivative acting
  on $\Tr(V_\mu)$.} $(\bar B_v \lrnabla^\mu B_v) \equiv \bar B_v
\nabla^\mu B_v - \bar B_v \lnabla^\mu B_v$.  The $c_i'$ terms vanish
if there are no DM currents in $V_\mu$, as then $\Tr V_\mu=\Tr
V_\mu^\xi=0$; they were thus omitted in \cite{Jenkins:1990jv,
  Brown:1993yv}. Note also that, in general, the DM current
$\nu_{\chi, \mu}$, Eq.~\eqref{eq:vmu}, is not conserved, $\partial_\mu
\nu_\chi^\mu\ne 0$, so that $\partial \cdot V\ne 0$. Because the
vector quark currents are conserved, one does have $c_2'=c_3'=0$, see
Section~\ref{app:expanded:meson:fields}. The terms involving the
axial-vector current $A^\mu$ but not the spin operator are 
\begin{equation}
\begin{split}\label{eq:LA2}
{\cal L}_{A}^{(2)}=&
d_1 \Tr\big(\bar B_v A^2 B_v\big)+d_2 \Tr\big(\bar B_v  (v\ncdot A )^2 B_v\big)+d_3 \Tr\big(\bar B_v B_v A^2\big)\\
 &+d_4 \Tr\big(\bar B_v B_v (v\ncdot A)^2\big)+d_5 \Tr\big(\bar B_v  B_v\big) \Tr \big(A^2\big)
 + d_6 \Tr\big(\bar B_v  B_v\big) \Tr \big((v \ncdot A)^2\big)
 \\
 &+ d_7 \Tr\big(\bar B_v A_\mu\big)\Tr\big(A^\mu B_v\big)+
 d_8 \Tr\big(\bar B_v v\ncdot A\big) \Tr\big(v\ncdot A B_v\big)
 \\
 &+ d_9 \Tr\big(\bar B_v A_\mu  B_v A^\mu\big)+
 d_{10} \Tr\big(\bar B_v v\ncdot AB_v v\ncdot A  \big)
 \\
 &+ d_1' \Tr\big(\bar B_v  B_v\big) \big(\Tr  A_\mu\big)^2+ d_2' \Tr\big(\bar B_v  B_v\big) \big(\Tr v\ncdot A\big)^2
\\
&+ d_3' \Tr\big(\bar B_v  A_\mu B_v\big) \Tr \big(A^\mu\big)+d_4' \Tr\big(\bar B_v  v\ncdot A B_v\big) \Tr\big(v\ncdot A \big)
\\
&+ d_5' \Tr\big(\bar B_v   B_v A_\mu\big) \Tr \big(A^\mu\big)+d_6' \Tr\big(\bar B_v   B_v v\ncdot A\big) \Tr\big(v\ncdot A \big)\,.
\end{split}
\end{equation}
For the $d_7$ and $d_8$ terms, the contraction of Dirac indices is
understood across the two traces. The $d_i'$ terms vanish in the limit
of vanishing DM currents, and were omitted in~\cite{Brown:1993yv}.

Finally, the terms involving the spin operator $S_v^\mu$ are
\begin{equation}
\begin{split}\label{eq:LA2:S}
{\cal L}_{S}^{(2)}={\cal L}_{S_0}^{(2)}+{\cal L}_{S'}^{(2)}+{\cal L}_{S,\epsilon}^{(2)}\,,
\end{split}
\end{equation}
where 
\begin{equation}
\begin{split}\label{eq:LS0}
{\cal L}_{S_0}^{(2)}=
&  f_1 \Tr\big(\bar B_v \{v\ncdot i \lrnabla, S_N\ncdot A\} B_v\big)+
 f_2 \Tr\big(\bar B_v \{S_N \ncdot i \lrnabla, v \ncdot A\} B_v\big)\\
&+i f_3 \Tr\big(\bar B_v [S_N \ncdot i \lrnabla, v \ncdot A] B_v\big)
+ f_4 \Tr\big[(\bar B_v  v\ncdot  i \lrnabla S_N^\nu  B_v) A_\nu
\big]\\
&+ f_5 \Tr\big[(\bar B_v S_N\ncdot i \lrnabla B_v) (v \ncdot A) \big]
+ f_6 \Tr\big(\bar B_v (S_N \ncdot A)^2 B_v\big)
\\
&+ f_{7} \Tr\big(\bar B_v  S_N^\nu  B_v v\ncdot  \nabla A_\nu
\big)+f_{8} \Tr\big(\bar B_v S_N^\mu B_v\,\nabla_\mu v \ncdot A
\big)\,, 
\end{split}
\end{equation}
while 
\begin{equation}
\begin{split}
{\cal L}_{S'}^{(2)}=
& f_1' \Tr \big(\bar B_v v\ncdot i \lrnabla S_N^\mu B_v\big) \Tr \big(A_\mu\big) + f_2' \Tr\big(\bar B_v S_N\ncdot i \lrnabla B_v\big) \Tr\big(v \ncdot A\big)
\\
& +f_3' \Tr \big(\bar B_v  (S_N \ncdot  A) B_v\big) \Tr\big(v\ncdot V\big) + f_4' \Tr\big(\bar B_v S_N^\mu (v \ncdot A) B_v \big)\Tr\big(V_\mu \big) 
\\
& + f_5' \Tr \big(\bar B_v  S_N^\mu B_v\big) \Tr\big(A_\mu\big)\Tr\big(v\ncdot V\big) +f_6' \Tr\big(\bar B_v S_N^\mu B_v \big)\Tr\big(V_\mu\big) \Tr\big(v \ncdot A\big) 
\\
&+  f_{7}' \Tr \big(\bar B_v \, S_N^\mu B_v\big) v\ncdot \partial \Tr \big(A_\mu\big) +f_{8}' \Tr\big(\bar B_v S_N^\mu \, B_v\big)\partial_\mu \Tr\big(v \ncdot A\big)\,,
\end{split}
\end{equation}
and 
\begin{equation}
\begin{split}\label{eq:LSepsilon2}
{\cal L}_{S,\epsilon}^{(2)}=
 &-i \epsilon^{\alpha\beta \lambda\sigma} v_{\alpha} \Big[g_1 \Tr \big(\bar B_v S_{N\beta}A_\lambda A_\sigma B_v\big) +ig_2 \Tr \big(\bar B_v S_{N\beta}A_\lambda  B_v A_\sigma \big)
 \\
 & + g_3 \Tr \big(\bar B_v S_{N\beta}  B_v A_\lambda A_\sigma \big)+ g_4 \Tr \big(\bar B_v S_{N\beta}   \nabla_\lambda \nabla_\sigma B_v \big)
 \\
 &-i g_5  \Tr \big(\bar B_v S_{N\beta}  B_v \nabla_\lambda V_\sigma \big) +g_6\Tr \big(\bar B_v S_{N\beta}A_\lambda\big)\Tr\big( A_\sigma B_v\big) 
 \\
 &+ig_1' \Tr \big(\bar B_v S_{N\beta}A_\lambda  B_v\big) \Tr (A_\sigma)
 +ig_2' \Tr \big(\bar B_v S_{N\beta} B_v A_\lambda \big) \Tr (A_\sigma)
 \\
 &+ g_3' \Tr \big(\bar B_v S_{N\beta}\lrnabla_\lambda  B_v\big) \Tr (V_\sigma)
 +ig_4' \Tr \big(\bar B_v S_{N\beta}  B_v\big) \partial_\lambda \Tr (V_\sigma)\Big]\,.
\end{split}
\end{equation}

 In writing the above Lagrangian
we imposed invariance of QCD under parity. Equations of motion for the
baryon fields were used to trade $\Tr\big(\bar B_v (v \ncdot \nabla)^2
B_v\big)$, $\Tr\big(v \ncdot V\big)\Tr\big(\bar B_v v \ncdot \nabla
B_v\big)$, $\Tr\big(\bar B_v (S_N\ncdot A) (v\ncdot \nabla) B_v\big)$
in favor of the other terms in
\eqref{eq:LV2}-\eqref{eq:LSepsilon2}. This differs from the convention
used in \cite{Jenkins:1990jv, Brown:1993yv}.  We also used the
relations in Eq.~\eqref{eq:Svprop} to simplify the terms involving the
spin operators. Note that the $f_{1,2}$ terms in~\eqref{eq:LS0} are
multiplied by $i$, correcting a typographical error in
\cite{Jenkins:1990jv} (see also \cite{Park:1993jf}).

Note that the last term in the first line of \eqref{eq:appC:HBChPT2}
contains only the DM part of the scalar current, while the QCD part
has already been absorbed in the definition of the $B_v$ masses. This
term and the $d_{i}'$, $f_{i}', g_1'$ terms do not appear in
\cite{Jenkins:1990jv} since the traces of QCD vector and axial
currents vanish 

The terms that contain at most one insertion of DM current are
\begin{align}
\begin{split}\label{eq:appC:HBChPT2-vec}
{\cal L}_{\rm HBChPT}^{(2)}&\supset  
b_D \Tr\bar B_v\{ s_+,B_v\} +b_F\Tr \bar B_v [s_+, B_v]+b_0\Tr\big(\bar B_vB_v\big)\Tr\big(s_+^\chi\big),
\\
&+c_1 \Tr\bar B_v \nabla^2 B_v  -c_1' \Tr\big(V_\mu\big)\Tr\big(\bar B_v  i\lrnabla^\mu B_v\big)
-c_2' \Tr\big(\partial\ncdot V \big)\Tr\big(\bar B_v  B_v\big)
\\
&-c_3' \Tr\big(\bar B_v  B_v\big) v\ncdot \partial\Tr\big(v \ncdot V \big)+  f_1 \Tr\big(\bar B_v \{v\ncdot i\lrnabla, S_N\ncdot A\} B_v\big)
\\
&+f_2 \Tr\big(\bar B_v \{S_N\ncdot i\lrnabla, v \ncdot A\} B_v\big)+i f_3 \Tr\big(\bar B_v [S_N\ncdot i\lrnabla, v \ncdot A]   B_v\big)
\\
&+ f_4 \Tr\big[\big(\bar B_v v\ncdot  i\lrnabla S_N^\nu  B_v \big)A_\nu \big]+ f_5 \Tr\big[(\bar B_v S_N\ncdot i\lrnabla B_v) (v \ncdot A) \big]
\\
&+ f_{7} \Tr\big(\bar B_v  S_N^\nu  B_v v\ncdot  \nabla A_\nu
\big)+ f_{8} \Tr\big(\bar B_v S_N^\mu B_v\,\nabla_\mu v \ncdot A
\big)
\\
& + f_1' \Tr \big(\bar B_v v\ncdot i\lrnabla S_N^\mu B_v\big) \Tr \big(A_\mu\big) + f_2' \Tr\big(\bar B_v S_N\ncdot i\lrnabla B_v\big) \Tr\big(v \ncdot A\big)
\end{split}
\\ \notag 
\begin{split}
&+ f_{7}' \Tr \big(\bar B_v \, S_N^\mu B_v\big) v\ncdot \partial \Tr \big(A_\mu\big) + f_{8}' \Tr\big(\bar B_v S_N^\mu \, B_v\big)\partial_\mu \Tr\big(v \ncdot A\big)
\\
 &-i \epsilon^{\alpha\beta \lambda\sigma} v_{\alpha} \Big[ g_4 \Tr \big(\bar B_v S_{N\beta}   \nabla_\lambda \nabla_\sigma B_v \big) -i g_5  \Tr \big(\bar B_v S_{N\beta}  B_v \nabla_\lambda V_\sigma \big)
\\
 & \qquad + g_3' \Tr \big(\bar B_v S_{N\beta}\lrnabla_\lambda  B_v\big) \Tr (V_\sigma)
+i g_4' \Tr \big(\bar B_v S_{N\beta}  B_v\big) \partial_\lambda \Tr (V_\sigma)\Big]
+\cdots,
\end{split}
\end{align}
where we kept only the terms that are nonzero once expanded up to
linear order in the meson fields. Not all of these terms are needed
for our ChEFT analysis, though. The reduced set of relevant terms is
given in \eqref{eq:HBChPT2-vec}.

The coefficients $m_G, D, F, G, b_i, b_i', c_i, c_i', d_i, d_i', f_i,
f_i', g_i, g_i'$ are real low-energy constants. Some of these
coefficients are fixed by the fact that the theory needs to be
invariant under infinitesimal Lorentz
transformations~\cite{Heinonen:2012km, Luke:1992cs},
\begin{equation}
\begin{split}\label{eq:RPI}
B_v&\to e^{i\varepsilon\cdot x}B_v, \qquad v\cdot \nabla\to v\cdot \nabla +\frac{1}{m_N} \varepsilon\cdot \nabla_\perp, \qquad \nabla_\perp^\mu\to \nabla_\perp^\mu -\frac{1}{m_N} \varepsilon^\mu (v\cdot \nabla).
\end{split}
\end{equation}
To lowest order the above transformation effectively corresponds to
reparametrization invariance under the shift of the label momentum
$v^\mu\to v^\mu+ \varepsilon^\mu/m_N$, but they also shift the
external currents,\footnote{This can also be used to show the equality
  in \eqref{eq:rel:C:dim6} imposed by reparametrization invariance,
  but now $\bar f \gamma_\mu f$ and $\bar f \gamma_\mu \gamma_5 f$ are
  to be treated as external currents.}  
\begin{equation}\label{eq:RPI:external}
A^\mu\to A^\mu + v^\mu \frac{\varepsilon\cdot A}{m_N}-\varepsilon^\mu \frac{v\cdot A}{m_N}, \qquad 
V^\mu\to V^\mu + v^\mu \frac{\varepsilon\cdot V}{m_N}-\varepsilon^\mu \frac{v\cdot V}{m_N}.
\end{equation}
Reparametrization invariance then leads to the
relations~\cite{Bos:1996ja} (see also \cite{Hill:2014yxa,
  Heinonen:2012km}) 
\begin{equation}\label{eq:RPI:c's}
\begin{split}
c_1&=c_1'=-\frac{1}{2m_N}\,, \qquad
f_2=-\frac{1}{2m_N}(D+F)\,, \qquad f_5=-\frac{1}{m_N}(D-F)\,,
\\ 
f_2'&=-\frac{1}{m_N}G\,, \qquad\qquad g_3'=0\,.
\end{split}
\end{equation}
In addition, the conservation of the quark vector current and the
Lorentz structure of the matrix element for quark axial vector current
give
\begin{equation}\label{eq:app:c2':etc}
c_2'=c_3'=0, \qquad f_3=f_8=f_8'=0,
\end{equation}
respectively, see Section~\ref{app:expanded:meson:fields}.

We discuss the numerical values of the remaining parameters that are
relevant for DM phenomenology in Sec.~\ref{app:low:eng:const}.

\subsection{ChPT and HBChPT Lagrangians expanded in meson fields}
\label{subsec:DMinteraction}
Here we give the DM interaction Lagrangian in ChPT,
Eqs. \eqref{eq:chiCHPT1}-\eqref{eq:chiCHPT3} and HBChPT,
Eqs. \eqref{eq:HBChPT0}-\eqref{eq:HBChPT2}, expanded up to linear
order in the meson fields. Unlike in the main text, the expressions in
this subsection are valid beyond tree level matching onto HDMET.  The ${\mathcal O}(p)$ ChPT Lagrangian for
DM interactions with mesons is 
\begin{equation}\label{eq:L1chiChPT:expand}
\begin{split}
{\cal L}_{\chi, {\rm ChPT}}^{(1)}&=2 f (\bar\chi_v iq \ncdot S_\chi
\chi_v) \Big[\big(-\hat \C_{4,u}^{(6,0)}+\hat \C_{4,d}^{(6,0)}\big)
  \pi^0-\big(\hat \C_{4,u}^{(6,0)}+\hat \C_{4,d}^{(6,0)}-2\hat
  \C_{4,s}^{(6,0)}\big)\frac{\eta}{\sqrt 3}\Big]+\cdots,  
\end{split}
\end{equation}
with $q^\mu=p_1^\mu-p_2^\mu$ the difference of incoming and outgoing
DM momenta, while the ellipses denote terms with two or more
mesons. It comes from the product $J_\chi^A\ncdot J_q^A$
in~\eqref{eq:chiCHPT1}, while the contributions from $J_\chi^V\ncdot
J_q^V$ and $J_\chi^A\ncdot J_q^V$ start only at ${\mathcal O}(\pi^2)$.
Note that the formally leading term in~\eqref{eq:L1chiChPT:expand}
from $J_\chi^V\ncdot J_q^A$ in~\eqref{eq:chiCHPT1}, i.e., from the
first line in \eqref{eq:chiCHPT1:full}, cancels exactly due to vector
current conservation against the corresponding $1/m_\chi$ suppressed
contribution to~\eqref{eq:L2chiChPT:expand} from the third line in
\eqref{eq:chiCHPT2:full}. We thus do not display these two
contributions.

The ${\mathcal O}(p^2)$ and ${\mathcal O}(p^3)$ DM ChPT Lagrangians
are 
\begin{align}
\begin{split}\label{eq:L2chiChPT:expand}
{\cal L}_{\chi, {\rm ChPT}}^{(2)}&=(\bar \chi_v \chi_v)\Big\{B_0 f  \Big[\big(m_u\hat \C_{7,u}^{(7,0)}-m_d \hat \C_{7,d}^{(7,0)}\big)  \pi^0
\\
&\qquad\qquad\qquad\qquad\qquad+\big(m_u \hat \C_{7,u}^{(7,0)}+m_d
\hat \C_{7,d}^{(7,0)}-2m_s \hat \C_{7,s}^{(7,0)}\big)\frac{\eta}{\sqrt
  3}\Big]\\
&\qquad\qquad  -\frac{f \tilde m}{2} q^2 \Big[  
  \Big( \frac{1}{m_u} - \frac{1}{m_d} \Big)\pi^0 + \Big( \frac{1}{m_u} + \frac{1}{m_d} - \frac{2}{m_s} \Big) \frac{
    \eta}{\sqrt{3}} \Big] \hat \C_{3}^{(7,0)} \Big\}+ \cdots\,,
\end{split}
\\
\begin{split}\label{eq:L3chiChPT:expand}
{\cal L}_{\chi, {\rm ChPT}}^{(3)}&=(\bar \chi_v i q \ncdot S_\chi \chi_v) \Big\{B_0 f  \Big[\big(m_u\hat \C_{8,u}^{(8,1)}-m_d \hat \C_{8,d}^{(8,1)}\big)  \pi^0
\\
&\qquad\qquad\qquad\qquad\qquad+\big(m_u \hat \C_{8,u}^{(8,1)}+m_d \hat \C_{8,d}^{(8,1)}-2m_s \hat \C_{8,s}^{(8,1)}\big)\frac{\eta}{\sqrt 3}\Big]+
\\
&\qquad\qquad + \frac{f \tilde m}{2} q^2 \Big[
  \Big( \frac{1}{m_u} - \frac{1}{m_d} \Big)\pi^0 + \Big( \frac{1}{m_u} + \frac{1}{m_d} - \frac{2}{m_s} \Big) \frac{
    \eta}{\sqrt{3}} \Big] \hat \C_{4}^{(8,1)} \Big\}+ \cdots\,.
\end{split}
\end{align}
The terms shown above come from $J_\chi^S\cdot J_q^P$ and
$J_\chi^{S}\cdot J^\theta$ in \eqref{eq:chiCHPT2}, and from
$J_\chi^P\cdot J_q^P$ and $J_\chi^{P}\cdot J^\theta$ in
\eqref{eq:chiCHPT3}, respectively.  The contributions from
$J_\chi^{S,P}\cdot J_q^S$ and $J_\chi^{S,P}\cdot J^G$, by contrast,
start only at ${\mathcal O}(\pi^2)$. The unexpanded versions of
\eqref{eq:L1chiChPT:expand}-\eqref{eq:L3chiChPT:expand} are given in
\eqref{eq:chiCHPT1:full}-\eqref{eq:chiCHPT3:full}.

Expanding the ${\mathcal O}(p^0)$ HBChPT interaction Lagrangian with
DM, Eq.~\eqref{eq:HBChPT0}, to lowest order in meson fields gives
\begin{equation}
\begin{split}\label{eq:HBChPT0:expand}
{\cal L}_{\chi, {\rm HBChPT}}^{(0)}&\supset (\bar \chi_v \chi_v)(\bar p_v p_v)\Big( 2 \hat \C_{1,u}^{(6,0)}+\hat \C_{1,d}^{(6,0)} -\frac{2m_G}{27}
\hat \C_{1}^{(7,0)}\Big)
\\
&+ 4 (\bar \chi_v S_{\chi,\mu} \chi_v)(\bar p_v S_N^\mu p_v) \Big((D+F+G)\hat \C_{4,u}^{(6,0)} +
G \hat \C_{4,d}^{(6,0)}+(D-F+G)\hat \C_{4,s}^{(6,0)}\Big)
\\
&+ (p_v \leftrightarrow n_v\,, u \leftrightarrow d)\,.
\end{split}
\end{equation}
Here the $p_v$ and $n_v$ are the HBChPT fields for protons and
neutrons. The ${\mathcal O}(p)$ Lagrangian, Eq.~\eqref{eq:HBChPT1}, is
given by 
\begin{equation}
\begin{split}\label{eq:HBChPT1:expand}
{\cal L}_{\chi, {\rm HBChPT}}^{(1)}&\supset -(\bar \chi_v v_\perp \ncdot S_\chi  \chi_v)(\bar p_v p_v)\times 2 \Big[ 2 \hat \C_{2,u}^{(6,0)}+ \hat \C_{2,d}^{(6,0)} \Big]
\\
& +(\bar \chi_v  \chi_v)(\bar p_v v_\perp\ncdot S_N  p_v)\times 2 m_\chi \Big[(D+F+G) \hat \C_{3,u}^{(7,1)}+G\hat \C_{3,d}^{(7,1)}+(D-F+G) \C_{3,s}^{(7,1)}\Big]
\\
&+(\bar \chi_v i q\ncdot S_\chi  \chi_v)(\bar p_v p_v)\times \frac{2 }{27}m_G \hat \C_2^{(8,1)}
\\
&-(\bar \chi_v \chi_v)(\bar p_v i q\ncdot S_N p_v) \times
\bigg[D \Big( \frac{\tilde m}{m_u} + \frac{\tilde m}{m_s} \Big) + F \Big( \frac{\tilde m}{m_u} - \frac{\tilde m}{m_s} \Big) + G\bigg] \hat \C_3^{(7,0)} 
\\
&+ 2 i \epsilon^{\alpha\beta\mu\nu} v_{\alpha} q_\beta (\bar \chi_v S_{\chi,\mu} \chi_v)(\bar p_v S_{N,\nu} p_v)\times\Big[ \big(g_4 -g_4'\big)\hat \C_{2,u}^{(6,0)} -g_4'\hat \C_{2,d}^{(6,0)}
\\
&-\big(g_4-g_5+g_4'\big)\hat \C_{2,s}^{(6,0)} +  \Big((D+F+G)\hat \C_{6,u}^{(7,1)} +
G \hat \C_{6,d}^{(7,1)}+(D-F+G)\hat \C_{6,s}^{(7,1)}\Big) \Big]
\\
&+ (p_v \leftrightarrow n_v\,, u \leftrightarrow d)\,,
\end{split}
\end{equation}
where we used that some terms vanish due to
Eq.~\eqref{eq:app:c2':etc}.  The four-component perpendicular relative
velocity $v_\perp^\mu$ is defined in \eqref{eq:vperp}. In the
derivation of the above HBChPT Lagrangian we also used the relations
\begin{align}
p_{1,2}^\mu&=\frac{1}{2}\big[\pm q^\mu +\big(p_1^\mu+p_2^\mu \big)\big]=\frac{1}{2}\big[\pm q^\mu +2 m_\chi v_\perp^\mu +\frac{m_\chi}{m_N}\big(k_1^\mu+k_2^\mu\big) \big],
\label{eq:p12rel}\\
k_{1,2}^\mu&=\frac{1}{2}\big[\mp q^\mu +\big(k_1^\mu+k_2^\mu \big)\big]=\frac{1}{2}\big[\mp q^\mu -2 m_N v_\perp^\mu +\frac{m_N}{m_\chi}\big(p_1^\mu+p_2^\mu\big) \big],\label{eq:k12rel}
\end{align}
the relation~\eqref{eq:sigma-to-epsilon-S}, as well as the relation
\eqref{eq:rel:C:dim6} imposed by reparametrization invariance.

The ${\mathcal O}(p^2)$ DM--nucleon interaction Lagrangian,
Eq.~\eqref{eq:HBChPT2}, expanded for each of the Wilson coefficients
to the first nontrivial order in meson fields, is given by 
\begin{equation}
\begin{split}\label{eq:HBChPT2:expand}
{\cal L}_{\chi, {\rm HBChPT}}^{(2)}&\supset (\bar \chi_v  \chi_v)(\bar p_v p_v)\times 2 \Big[- \big(b_0+b_D+b_F\big)m_u\hat \C_{5,u}^{(7,0)}-b_0 m_d \hat \C_{5,d}^{(7,0)}
\\
&\qquad\qquad\qquad\qquad\qquad\qquad\qquad\qquad-\big(b_0+b_D-b_F\big) m_s \hat \C_{5,s}^{(7,0)} \Big]
\\
&-(-1)^{Q_p}(\bar \chi_v  \chi_v)(\bar p_v p_v)\frac{\pi^0}{f}\times 2\Big[ \big(b_0+b_D+b_F\big)m_u\hat \C_{7,u}^{(7,0)}
-b_0 m_d \hat \C_{7,d}^{(7,0)} \Big]
\\
&+(\bar \chi_v  \chi_v)(\bar p_v p_v)\frac{\eta}{\sqrt{3}f}\times 2\Big[\big(b_0+b_D+b_F\big)m_u\hat \C_{7,u}^{(7,0)}+b_0 m_d \hat \C_{7,d}^{(7,0)}
\\
&\qquad\qquad\qquad\qquad\qquad\qquad\qquad\qquad-2\big(b_0+b_D-b_F\big) m_s  \hat \C_{7,s}^{(7,0)} \Big]
\\
&+(\bar \chi_v i q\ncdot S_\chi  \chi_v)(\bar p_v i q \ncdot S_N p_v)\times 
\bigg[D \Big( \frac{\tilde m}{m_u} + \frac{\tilde m}{m_s} \Big) + F \Big( \frac{\tilde m}{m_u} - \frac{\tilde m}{m_s} \Big) +  G\bigg] \hat \C_4^{(8,1)}
\\
&+ (p_v \leftrightarrow n_v\,, u \leftrightarrow d)\,,
\end{split}
\end{equation}
while the ${\mathcal O}(p^3)$ DM--nucleon interaction Lagrangian,
Eq.~\eqref{eq:HBChPT3}, is given by 
\begin{equation}
\begin{split}\label{eq:HBChPT3:expand}
{\cal L}_{\chi, {\rm HBChPT}}^{(3)}&\supset  (\bar \chi_v i q\ncdot S_\chi \chi_v)(\bar p_v p_v)\times 2 \Big[ \big(b_0+b_D+b_F\big)m_u\hat \C_{6,u}^{(8,1)}+b_0 m_d \hat \C_{6,d}^{(8,1)}
\\
&\qquad\qquad\qquad\qquad\qquad\qquad\qquad\qquad+\big(b_0+b_D-b_F\big) m_s \hat \C_{6,s}^{(8,1)} \Big]
\\
&-(-1)^{Q_p} (\bar \chi_v i q\ncdot S_\chi \chi_v)(\bar p_v p_v)\frac{\pi^0}{f}\times 2\Big[ \big(b_0+b_D+b_F\big)m_u\hat \C_{8,u}^{(8,1)}
-b_0 m_d \hat \C_{8,d}^{(8,1)} \Big]
\\
&+(\bar \chi_v i q\ncdot S_\chi \chi_v)(\bar p_v p_v)\frac{\eta}{\sqrt{3}f}\times 2\Big[\big(b_0+b_D+b_F\big)m_u\hat \C_{8,u}^{(8,1)}+b_0 m_d \hat \C_{8,d}^{(8,1)}
\\
&\qquad\qquad\qquad\qquad\qquad\qquad\qquad\qquad-2\big(b_0+b_D-b_F\big) m_s  \hat \C_{8,s}^{(8,1)} \Big]
\\
&+ (p_v \leftrightarrow n_v\,, u \leftrightarrow d)\,.
\end{split}
\end{equation}

The final ingredient that we need is the leading HBChPT chiral
Lagrangian without DM fields~\eqref{eq:LChPT:expand} 
\begin{equation}
{\cal L}_{\rm HBChPT}^{(1),{\rm QCD}}\supset \frac{(D+F)}{f}\big(\bar p_v \, iq \cdot S_N p_v-\bar n_v \, iq \cdot S_N n_v\big) \pi^0 - \frac{D-3F}{\sqrt{3} f}\big(\bar n \, iq \cdot S_N n+\bar p \, iq \cdot S_N p\big) \eta+\cdots\,,
\end{equation}
where we expanded to linear order in meson fields, and only display
the couplings to the neutral mesons. As in the rest of the paper,
$q^\mu=k_2^\mu-k_1^\mu$ is the difference of final and initial nucleon
momenta. The single photon interactions with neutrons and protons are
given by
\begin{equation}\label{eq:HBChPTQED}
{\cal L}_{\rm HBChPT}^{{\rm QED}}\supset -e\,\bar p_v \Big(v^\mu + \frac{\tilde k_{12}^\mu}{2m_N} \Big) A_\mu^e \, p_v + \frac{e}{2 m_N}\Big( \mu_p \bar p_v (\sigma_{\perp}^{\mu\nu} iq_\mu) p_v+ \mu_n \bar n_v (\sigma_{\perp}^{\mu\nu} iq_\mu) n_v\Big) A_{\nu}^e\,,
\end{equation}
where $\mu_p=2.79$ and $\mu_n=-1.91$ are the proton and neutron
magnetic moments in units of nuclear magnetons, respectively, and
$\tilde k_{12}^\mu$ is defined below in Eq.~\eqref{eq:ktilde:def}.

\subsection{Explicit form of hadronic currents}
It is straightforward to give the explicit expressions for the
different currents appearing
in~\eqref{eq:chiCHPT1}-\eqref{eq:chiCHPT3},
\begin{align}
J_q^S&=-\frac{B_0f^2}{2}\Tr\big[(U+U^\dagger)m_q \mathbb{1}_q\big], \qquad\qquad\,\,\, J_q^P=-\frac{B_0f^2}{2}\Tr\big[i (U-U^\dagger)m_q \mathbb{1}_q\big],\label{eq:JSP}
\\
J_{q,\mu}^V&=-\frac{if^2}{2}\Tr\big[(U\partial_\mu U^\dagger+U^\dagger\partial_\mu U) \mathbb{1}_q\big], 
\qquad \,\, J_{q,\mu}^A=-\frac{if^2}{2}\Tr\big[(U\partial_\mu U^\dagger-U^\dagger\partial_\mu U) \mathbb{1}_q\big],\label{eq:JVA}
\\
J^G&=\frac{f^2}{27} \Big[  \Tr\big(\partial_\mu U^\dagger \partial^\mu
U\big) + 3 B_0 \Tr\big[{\cal M}_q(U+U^\dagger)\big] \Big]
\,,
\\ 
\label{eq:JGtheta}
J^\theta&=\frac{if^2}{4\Tr({\cal M}_q^{-1})}
\Tr\Big[\partial^\mu\big(U\partial_\mu U^\dagger-U^\dagger
  \partial_\mu U\big) {\cal M}_q^{-1} \Big]\,.
\end{align}
Expanding the currents to first nonzero order in meson fields and
dropping the constant terms in $J_q^S$ gives 
\begin{align}
J_{q,\mu}^V&=i \Tr\big(\big[\partial_\mu \Pi, \Pi\big] \mathbb{1}_q\big)+\cdots\,, \qquad &J_{q,\mu}^A&=-\sqrt2 f \Tr\big(\partial_\mu \Pi \,\1_q\big)+\cdots\,,\\
J_q^S&=
B_0 \Tr \big(\Pi^2 m_q\1_q)+\cdots, \qquad &J_q^P&=\sqrt2 B_0 f
m_q\Tr(\Pi\, \1_q) +\cdots\,, \\
J^G&=\tfrac{2}{27} \Tr\big(\partial_\mu \Pi \partial^\mu\Pi\big)-
\tfrac{6}{27} B_0 \Tr \big({\cal M}_q\Pi^2\big)+\cdots\,, &&\\ 
J^\theta&=\frac{f}{\sqrt{2}\Tr({\cal M}_q^{-1})}
\Tr\Big[\partial^2 \Pi {\cal M}_q^{-1} \Big] +\cdots\,.
&&
\end{align}
Here we defined $\1_u=\diag(1,0,0)$, $\1_d=\diag(0,1,0)$,
$\1_s=\diag(0,0,1)$.  The explicit forms of the pseudoscalar and
axial-vector currents in terms of the $\pi^0$ and $\eta$ fields are
given in \eqref{eq:JAmuexpand} and \eqref{eq:JPexpand}.

In \eqref{eq:hiHBChPTexpan} we have expanded the DM-nucleon
interactions in terms of their chiral scaling. The LO expressions for
the currents in~\eqref{eq:HBChPT0},~\eqref{eq:HBChPT1},
\eqref{eq:HBChPT2}, are
\begin{align}
\begin{split}\label{eq:tildeJV}
\tilde J_{q}^{V\mu,\text{LO}}=& \frac{1}{2} \Tr\bar B_v \big[v^\mu (\xi^\dagger\mathbb{1}_q\xi+\xi\mathbb{1}_q \xi^\dagger),B_v\big]
 +D  \Tr \bar B_v S_N^\mu \big\{\xi^\dagger\mathbb{1}_q\xi-\xi\mathbb{1}_q \xi^\dagger, B_v\big\}\\
& +F  \Tr \bar B_v S_N^\mu \big[\xi^\dagger\mathbb{1}_q\xi-\xi\mathbb{1}_q \xi^\dagger, B_v\big]+\Tr\bar B_v v^\mu B_v\,,
\end{split}
\\
\begin{split}\label{eq:tildeJA}
\tilde J_{q}^{A\mu,\text{LO}}=& \frac{1}{2} \Tr\bar B_v \big[v^\mu (\xi^\dagger\mathbb{1}_q\xi-\xi\mathbb{1}_q \xi^\dagger),B_v\big]
 +D  \Tr \bar B_v S_N^\mu \big\{\xi^\dagger\mathbb{1}_q\xi+\xi\mathbb{1}_q \xi^\dagger, B_v\big\}\\
& +F  \Tr \bar B_v S_N^\mu \big[\xi^\dagger\mathbb{1}_q\xi+\xi\mathbb{1}_q \xi^\dagger, B_v\big]+ 2 G \Tr \bar B_v S_N^\mu B_v\,,
\end{split}
\\
\begin{split}\label{eq:tildeJG}
\tilde J^{G,\text{LO}}=&-\frac{2 m_G}{27}
\Tr \bar B_v  B_v\,,
\end{split}
\\
\begin{split}\label{eq:tildeJtheta}
\tilde J^{\theta,\text{LO}}=&-\frac{1}{2\Tr ({\cal M}_q^{-1}) }
\biggr\{\frac{1}{2} v\ncdot \partial \Tr\bar B_v \big[ (\xi^\dagger  {\cal M}_q^{-1} \xi-\xi  {\cal M}_q^{-1}  \xi^\dagger),B_v\big]
\\
&\quad+ \partial_\mu \Big(D  \Tr \bar B_v S_N^\mu
\big\{\xi^\dagger {\cal M}_q^{-1} \xi+\xi {\cal M}_q^{-1} \xi^\dagger, B_v\big\}\\
& \quad  +F  \Tr \bar B_v S_N^\mu \big[\xi^\dagger
  {\cal M}_q^{-1} \xi+\xi {\cal M}_q^{-1} \xi^\dagger, B_v\big]+ 2 G  \Tr ({\cal M}_q^{-1})
\Tr \bar B_v S_N^\mu B_v \Big)\bigg\}  \,,
\end{split}
\\
\begin{split}\label{eq:tildeJS}
\tilde J_q^{S,\text{LO}}=&- b_0 \Tr(\bar B_v B_v)\Tr\big( (U^\dagger +U)m_q \mathbb{1}_q\big) -b_D \Tr\bar B_v \big\{\xi^\dagger m_q \mathbb{1}_q\xi^\dagger+\xi m_q \mathbb{1}_q\xi, B_v\big\}\\
 &-b_F \Tr \bar B_v\big[\xi^\dagger m_q \mathbb{1}_q\xi^\dagger+\xi m_q \mathbb{1}_q\xi, B_v\big]\,,
\end{split}
\\
\begin{split}\label{eq:tildeJP}
\tilde J_q^{P,\text{LO}}=& b_0 \Tr(\bar B_v B_v)\Tr\big( (U^\dagger -U)i m_q \mathbb{1}_q\big) +b_D \Tr \bar B_v \big\{\xi^\dagger i m_q \mathbb{1}_q\xi^\dagger-\xi i m_q \mathbb{1}_q\xi, B_v\big\}\\
& +b_F \Tr \bar B_v \big[\xi^\dagger i m_q \mathbb{1}_q\xi^\dagger-\xi im_q \mathbb{1}_q\xi, B_v\big]\,.
\end{split}
\end{align}

When contracting with the nonrelativistic DM currents we also need the
expression for the QCD vector current $\tilde J_{q}^{V\mu}$ and
axial-vector current $\tilde J_{q}^{A\mu}$ to NLO in the chiral
expansion, i.e., to ${\mathcal O}(p)$.  The NLO contributions to
$\tilde J_{q}^{V\mu}$ are 
\begin{equation}
\begin{split}\label{eq:JVNLO}
\tilde J_q^{V\mu, {\rm NLO}} \supset& -i c_1 \big(\Tr \bar B_v \big[\mathbb{1}_q^\xi , \nabla^{\xi,\mu} B_v\big]-  \Tr \bar B_v \overset{\leftarrow}{\nabla}{}^{\xi,\mu} \big[  \mathbb{1}_q^\xi , B_v\big]\big)
   - c_1'\Tr\big(\bar B_v {i\lrnabla}{}^{\xi,\mu} B_v\big)
   \\
  & + c_2' \partial^\mu \Tr\big(\bar B_v  B_v\big)+ c_3'  v^\mu v\ncdot \partial \Tr\big(\bar B_v  B_v\big)
   \\
   &+\epsilon^{\alpha \beta \lambda\mu} v_{\alpha} 
   \Big(g_4 \Tr\bar B_v S_{N\beta} \overset{\leftarrow}{\nabla}{}_\lambda^\xi[\mathbb{1}_q^\xi,B_v]
  +g_4\Tr \bar B_v S_{N\beta} [ \mathbb{1}_q^\xi,\nabla_\lambda^\xi B_v]
 \\
   & \qquad\qquad\qquad + g_5\, \partial_\lambda \Tr \big( \bar B_v S_{N\beta} B_v \mathbb{1}_q^\xi \big)-ig_3'\,\Tr \bar B_v S_{N\beta} \,{\lrnabla}{}_\lambda^\xi B_v 
 \\
   &\qquad\qquad\qquad - g_4'\partial_\lambda \Tr \bar B_v S_{N\beta}  B_v
   \Big)+\cdots,
\end{split}
\end{equation}
where $\bar B_v \overset{\leftarrow}{\nabla}{}_\mu^\xi=\partial_\mu
\bar B_v - [\bar B_v, V_\mu^\xi]=\partial_\mu \bar B_v +
     [V_\mu^\xi,\bar B_v]$, and we used the abbreviation
     $\mathbb{1}_q^\xi \equiv \tfrac{1}{2}(\xi^\dagger \mathbb{1}_q
     \xi+\xi \mathbb{1}_q \xi^\dagger)$. The ellipses denote terms
     that, when expanded in terms of meson fields, start at linear
     order or higher.  Keeping only the terms that do not involve the
     meson fields gives 
\begin{equation}\label{eq:JVNLOexpand}
\begin{split}
\tilde J_q^{V\mu, {\rm NLO}} \supset& - i c_1\Tr \bar B_v \overset{\leftrightarrow}{\partial^\mu}\big[\mathbb{1}_q, B_v\big]
-i c_1'   \Tr \bar B_v  \lrpartial^\mu  B_v +c_2'  \partial^\mu \Tr \bar B_v    B_v+c_3'  v^\mu v\ncdot \partial \Tr \bar B_v    B_v
\\
&+ \epsilon^{\alpha\beta\lambda\mu}v_{\alpha}\Big[ g_4 \partial_\lambda \Tr \bar B_v S_{N\beta}[\mathbb{1}_q,B_v] 
+ g_5\, \partial_\lambda \Tr \big( \bar B_v S_{N\beta} B_v \mathbb{1}_q \big)
\\
& \qquad\qquad - i
g_3' \Tr \bar B_v S_{N\beta} \lrpartial_\lambda B_v- g_4'  \partial_\lambda \Tr \bar B_v S_{N\beta}  B_v\Big]\,,
\end{split}
\end{equation}
with $\overset{\leftrightarrow}{\partial_\mu}$ defined through
$\phi_1\overset{\leftrightarrow}{\partial_\mu} \phi_2=\phi_1
\partial_\mu \phi_2- (\partial_\mu \phi_1) \phi_2$, as before. The
axial current at NLO is 
\begin{equation}
\begin{split}\label{eq:JANLO}
\tilde J_q^{A\mu, {\rm NLO}} \supset
&\,\,  2i f_1 \Big[\Tr \big(\bar B_v \mathbb{1}_q^\xi v \ncdot \nabla^\xi S_N^\mu   B_v\big)- \Tr \big(\bar B_v S_N^\mu v \ncdot \overset{\leftarrow}{\nabla}{}^\xi   \mathbb{1}_q^\xi  B_v\big)\Big]
\\
&+2i f_2 v^\mu \Big[ \Tr\big( \bar B_v  \mathbb{1}_q^\xi  S_N \ncdot {\nabla}{}^\xi B_v\big)-  \Tr\big( \bar B_v S_N \ncdot \overset{\leftarrow}{\nabla}{}^\xi  \mathbb{1}_q^\xi B_v\big)\Big]
\\
&+  2 f_3 v^\mu \Big[\Tr \big(\bar B_v  \mathbb{1}_q^\xi  S_N \ncdot \nabla^\xi B_v\big) +  \Tr \big(\bar B_v   S_N \cdot \lnabla^\xi \mathbb{1}_q^\xi B_v\big)\Big]
\\
& +  i f_4  \Tr \big[\big(\bar B_v  v \cdot  \lrnabla^\xi S_N^\mu B_v\big)  \mathbb{1}_q^\xi \big]
+ i f_5 v^\mu  \Tr \big[\big(\bar B_v  S_N \cdot\, \lrnabla^\xi B_v \big) \mathbb{1}_q^\xi \big]
\\
& - f_{7}   \Tr\big(\mathbb{1}_q^\xi v\ncdot  \nabla^\xi \bar B_v  S_N^\mu  B_v    \big)
- f_{8} v^\mu \Tr\big( \mathbb{1}_q^\xi \nabla_\nu^\xi \bar B_v S_N^\nu B_v \big)
\\
&+i f_1' \Tr \big(\bar B_v v \cdot \lrnabla^\xi S_N^\mu B_v \big)
+ i f_2' v^\mu \Tr \big(\bar B_v  S_N \cdot \lrnabla^\xi B_v\big)
\\
 &- f_{7}' v \ncdot \partial \Tr \big(\bar B_v \, S_N^\mu B_v\big)
- f_{8}' v^\mu\, \partial_\nu \Tr\big(\bar B_v S_N^\nu \, B_v\big) + \cdots\,.
\end{split}
\end{equation}
Expanding in meson fields gives
\begin{equation}
\begin{split}\label{eq:JANLO:expanded}
 \tilde J_q^{A\mu, {\rm NLO}} \supset
&  2i f_1 \Tr \big(\bar B_v v \cdot \lrpartial\, S_N^\mu\,  \mathbb{1}_q B_v\big)+  i f_4  \Tr \big(\bar B_v  v \cdot \lrpartial S_N^\mu B_v  \mathbb{1}_q \big)- f_{7}   \Tr\big(  v\ncdot  \partial \bar B_v  S_N^\mu  B_v \mathbb{1}_q \big)
\\
&+
i f_1' \Tr \big(\bar B_v v \cdot {\lrpartial} S_N^\mu B_v \big) - f_{7}' v \ncdot \partial \Tr \big(\bar B_v \, S_N^\mu B_v\big)
\\
&+  v^\mu \Big[2 i f_2  \Tr\big( \bar B_v S_N \cdot \lrpartial\, \mathbb{1}_q B_v\big)+2 f_3  \partial_\nu\Tr \big(\bar B_v  \mathbb{1}_q  S_N^\nu B_v\big) 
+ i f_5  \Tr \big(\bar B_v  S_N \cdot \lrpartial B_v  \mathbb{1}_q \big) 
\\
&- f_{8} \partial_\nu \Tr\big(  \bar B_v S_N^\nu B_v \mathbb{1}_q \big)+  i f_2' \Tr \big(\bar B_v  S_N \cdot \lrpartial B_v\big)
- f_{8}'  \partial_\nu \Tr\big(\bar B_v S_N^\nu \, B_v\big)
\Big]+\cdots\,.
\end{split}
\end{equation}

\subsection{Quark currents expanded in meson fields}
\label{app:expanded:meson:fields}

In this subsection we collect the expanded results for the nucleon and
meson currents in term of meson fields, keeping only the lowest
orders. The expanded expressions have been used in
Section~\ref{eq:pionlessEFT} to match onto the chiral effective theory
of nuclear forces. For this calculation we need single meson exchanges
for the hadronized versions of the $\bar q \gamma_\mu \gamma_5 q$,
$\bar q i\gamma_5 q$, and $G_{\mu\nu}^a G^{a\mu\nu}$ currents. The
corresponding DM-meson interactions in ${\cal L}_{\chi, {\rm
    ChPT}}^{(1,2,3)}$, Eqs.~\eqref{eq:chiCHPT1}-\eqref{eq:chiCHPT3},
contain the mesonic currents given
in~\eqref{eq:JSP}-\eqref{eq:JGtheta}. Expanding in meson fields to the
first nonzero order one has for the axial currents 
\begin{equation}\label{eq:JAmuexpand}
(J_{u,d}^{A})_\mu=f\Big(\mp\partial_\mu \pi^0-\frac{\partial_\mu \eta}{\sqrt 3}\Big)+\cdots\,,\quad (J_{s}^{A})_\mu=\frac{2 f}{\sqrt 3} \partial_\mu \eta+\cdots\,.
\end{equation}
while the pseudoscalar currents are 
\begin{equation}\label{eq:JPexpand}
J_{u,d}^P=B_0 f m_{u,d} \Big(\pm \pi^0+\frac{1}{\sqrt 3} \eta\Big)+\cdots\,,
\qquad J_{s}^P=-\frac{2}{\sqrt3} B_0 f m_{s} \eta+\cdots\,.
\end{equation}
The contribution of the $J^\theta$ current is
\begin{equation}\label{eq:Jthetaexpand}
J^\theta = \frac{f}{2} \Big[ \Big( \frac{\tilde m}{m_u} - \frac{\tilde
    m}{m_d} \Big) \partial^2 \pi^0 + \Big( \frac{\tilde m}{m_u} +
  \frac{\tilde m}{m_d} - \frac{2\tilde m}{m_s} \Big) \frac{\partial^2
    \eta}{\sqrt{3}} \Big]\,.
\end{equation}

Expanding the nucleon currents \eqref{eq:tildeJV}-\eqref{eq:tildeJP}
to the first nonzero order in meson fields gives for the $q=u,d$ quark
currents 
\begin{align}
\begin{split}
\tilde J_{q}^{V,\mu} &=\Big(v^\mu+\frac{\tilde k_{12}^\mu}{2m_N}\Big) \big(\bar N_q N_q+\bar N N\big) +i c_2' q^\mu \bar N N
\\
 &\quad+i \epsilon^{\alpha \beta \lambda \mu} v_{\alpha} q_\lambda \big(g_4\bar N_q S_{N\beta}N_q -g_4'\bar N S_{N\beta}N \big) + \cdots\,, \label{eq:tildeJqV}
\end{split}
\\
\begin{split}
\tilde J_{q}^{A,\mu} &=2 (D+F) \bar N_q \Big(S_N^\mu-\frac{v^\mu}{2 m_N}\tilde k_{12}\ncdot S_N\Big) N_q+2 G \bar N  \Big(S_N^\mu-\frac{v^\mu}{2 m_N}\tilde k_{12}\ncdot S_N\Big) N
\\
 & \quad + i v^\mu \big(2 f_3   \bar N_q q\ncdot S_N N_q
- f_8' \, \bar N q\ncdot S_N N \big) +\cdots\,,\label{eq:tildeJqA}
\end{split}
\\
\tilde J_q^S&=-2 b_0 m_q \bar N N - 2 (b_D+b_F) m_q \bar N_q N_q +\cdots\,,\label{eq:tildeJqS}\\
\tilde J_{q}^P&=\frac{2m_q}{f} \Big[b_0 \bar N N\Big(\pm \pi^0+\frac{\eta}{\sqrt 3}\Big)+ (b_D+b_F) \bar N_q N_q \Big(\pm \pi^0+\frac{\eta}{\sqrt 3}\Big)\Big]+\cdots\,.\label{eq:tildeJqP}
\end{align}
In $\tilde J_{q, \mu}^V$ and $\tilde J_{q, \mu}^A$ we keep the
${\mathcal O}(p)$ terms from \eqref{eq:JVNLO}, \eqref{eq:JVNLOexpand},
and do not display the $v\cdot q$-suppressed terms, while for $\tilde
J_{q}^P$ we display only the couplings to neutral mesons.  The
plus(minus) sign in \eqref{eq:tildeJqP} is for $q=u(d)$, and there is
no summation over repeated $q$ indices.  Here $N=(p_v,n_v)$ is the
nucleon isospin doublet, so that the up and down components are
\begin{equation}
N_u=p_v, \quad N_d=n_v, 
\quad{\rm while }\quad
\bar NN=\bar p_vp_v+\bar n_vn_v.
\end{equation}
To shorten the notation we introduced 
\begin{equation}\label{eq:ktilde:def}
\tilde k_{12}^\mu=\tilde k_1^\mu+\tilde k_2^\mu, \qquad q^\mu=-\tilde k_1^\mu+\tilde k_2^\mu,
\end{equation}
with $\tilde k_{1,2}^\mu =k_{1,2}^\mu - m_N v_N^\mu$ the soft nucleon
momenta. The expression for the momentum transfer $q^\mu$ coincides
with the definition in \eqref{eq:momentum:exchange}.

The corresponding strange-quark currents are given by 
\begin{align}
\begin{split}
\tilde J_{s}^{V,\mu}&=  i c_2' q^\mu \bar N N
-i (g_4-g_5+g_4') \epsilon^{\alpha \beta \lambda \mu} v_{\alpha} q_\lambda \bar N S_{N\beta}N + \cdots\,,
\label{eq:tildeJV:exp:s}
\end{split}
\\
\begin{split}
\tilde J_{s}^{A,\mu}&=2 (D-F+G) \bar N \Big(S_N^\mu-\frac{v^\mu}{2 m_N}\tilde k_{12}\ncdot S_N\Big) N -i (f_8+f_8') v^\mu \bar N q\ncdot S_N N
+ \cdots\,, \label{eq:tildeJA:exp:s}
\end{split}
\\
\tilde J_s^S&=-2\big(b_0 +b_D-b_F\big) m_s \bar N N +\cdots\,,  \label{eq:tildeJS:exp:s}\\
\tilde J_s^P&=-\frac{4}{f}\big(b_0+b_D-b_F\big) m_s \bar N N \frac{ \eta}{\sqrt3}+\cdots\,, \label{eq:tildeJP:exp:s}
\end{align}
where for $\tilde J_s^P$ we again display only the couplings to the
neutral mesons, and do not show the $v\ncdot q$-suppressed terms in
$\tilde J_{s}^{V,\mu}$ and $\tilde J_{s}^{A,\mu}$. Note that in order
to obtain the above expressions we have used the
reparametrisation-invariance relations~\eqref{eq:RPI:c's}.

The conservation of the vector current, $q\ncdot \tilde
J_{q,s}^{V}=0$, requires $c_2'=0$. Comparing the most general
parametrization of the matrix element for the axial-vector current,
$\langle N(k_2)| \bar q\gamma^\mu \gamma_5 q|N(k_1)\rangle=\bar u_N
\big(F_A(q^2) \gamma^\mu \gamma_5+F_{P'}(q^2) q^\mu/(2m_N)
\gamma_5\big)u_N$, with its nonrelativistic
decomposition~\eqref{eq:HDMETlimit:pscalar},
\eqref{eq:axialDM:expand}, requires $f_3=f_8=f_8'=0$. 

The gluonic $GG$ and $G\tilde G$ currents hadronize to 
\begin{align}
\tilde J^G=&-\frac{2 m_G}{27}
\bar N N,
\\
\begin{split}
\tilde J^\theta=&
-\bigg[D \Big( \frac{\tilde m}{m_u} + \frac{\tilde m}{m_s} \Big) +F \Big( \frac{\tilde m}{m_u} - \frac{\tilde m}{m_s} \Big)
 + G\bigg] \bar p_v i q\cdot S_N p_v\\
 &+p\to n, u\to d+\cdots.
\label{eq:Jtildetheta}
\end{split}
\end{align}
The values of the low-energy constants $D, F, G, m_G, b_0, b_D, b_F,
g_4, g_4', g_5$ are discussed in the following section and are
collected in Table~\ref{tab:LEC}.

\section{Values of low energy constants}
\label{app:low:eng:const}
In this appendix we derive numerical values for the low-energy
coefficients $m_G, D, F, G, b_D, b_F, b_0, g_4, g_4', g_5$.  The
nonperturbative coefficient $m_G$ is the gluonic contribution to the
nucleon mass,
\begin{equation}
m_G \bar u_B u_B=- \frac{9\alpha_s}{8\pi}\langle B_v |G_{\mu\nu}G^{\mu\nu}|B_v \rangle.
\end{equation}
This can be estimated from the trace of the stress-energy tensor $
\theta_\mu^\mu=-9\alpha_s/(8\pi) G_{\mu \nu}
G^{\mu\nu}+\sum_{u,d,s}m_q\bar q q$, giving
\begin{equation}
m_G=m_B-\sum_q \sigma_q^B, 
\end{equation}
where $ \sigma_q^B \bar u_B u_B=\langle B_v |m_q\bar q q|B_v \rangle$,
with $u_B$ the heavy baryon spinor.\footnote{We use the conventional
  HQET normalization for the fields and states, $\langle B(\vec
  k')_{v'}| B(\vec k)_v\rangle= 2 v^0 \delta_{vv'} (2\pi)^3
  \delta^3(\vec k - \vec k')$, so that $\langle B_{v}|\bar B_v
  \gamma^\mu B_v| B_v\rangle=\bar u_B(v) \gamma^\mu u_B(v)= 2 v^\mu$,
  where the heavy fermion spinors $u_B$ are related to relativistic
  spinors through $u(p)=\sqrt{m_N} u_B(v)$, see also
  \cite{Manohar:2000dt}. Similarly, one has $\langle B_{v}|\bar B_v
  \gamma^\mu \gamma_5 B_v| B_v\rangle=\bar u_B(v) \gamma^\mu \gamma_5
  u_B(v)= 2 s^\mu$, where $s^\mu$ is the heavy baryon spin.} A common
notation is also $\sigma_q^N=m_Nf_{Tq}^{(N)}$, where $N=p,n$. Taking
the naive average of the most recent lattice QCD
determinations~\cite{Junnarkar:2013ac, Yang:2015uis, Durr:2015dna}, we
find $\sigma_s=(41.3\pm 7.7)$~MeV The matrix elements of $u$ and $d$
quarks are related to the $\sigma_{\pi N}$ term, defined as
$\sigma_{\pi N} = \langle N | \bar m (\bar u u + \bar d d) | N
\rangle$, where $\bar m = (m_u + m_d)/2$. A HBChPT analysis of the
$\pi N$ scattering data gives $\sigma_{\pi N} =
59(7)$\,MeV~\cite{Alarcon:2011zs}, in agreement with $\sigma_{\pi
  N}=52(3)(8)$ MeV obtained from a fit to world lattice $N_f=2+1$ QCD
data \cite{Alvarez-Ruso:2014sma}. Including, however, both $\Delta
(1232)$ and finite spacing in the fit shifts the central value to
$\sigma_{\pi N}=44$ MeV.  We thus use a conservative estimate
$\sigma_{\pi N}=(50\pm15)$ MeV. Using the expressions
in~\cite{Crivellin:2013ipa} gives $\sigma_u^p=(17\pm 5)$ MeV,
$\sigma_d^p=(32\pm 10)$ MeV, $\sigma_u^n=(15\pm 5)$ MeV,
$\sigma_d^n=(36\pm 10)$ MeV.  From there we get
\begin{equation}\label{eq:mG:app}
m_G=(848\pm14) {\rm ~MeV},
\end{equation}
in the isospin limit. While the isospin violation in the $\sigma_q^N$
values, factoring out the masses, is of ${\mathcal O}(10\%)$, this
translates to a very small isospin violation in $m_G$, of less than 1
MeV. The obtained value of $m_G$ thus applies to both $p$ and $n$,
while for other members of $B_v$ octet it is correct up to flavor
$SU(3)$ breaking terms.

The $\sigma_q^N$ are related to the low-energy constants $b_{0}$,
$b_{D}$, and $b_{F}$ through
\begin{align}
\sigma_u^p&=-2 m_u (b_0 +b_D+b_F)\,, &\sigma_d^n&=-2 m_d (b_0 +b_D+b_F)\,,\\
\sigma_u^n&=-2 m_u b_0 \,,  &\sigma_d^p&=-2 m_d b_0\,,
\end{align}
while
\begin{equation}
\sigma_s = -2 m_s (b_0 +b_D-b_F)\,.
\end{equation}
The combinations that are well determined are 
\begin{align}
\bar m (b_D+b_F)&=(-1.41\pm0.24){\rm MeV},\label{eq:mbDbF}\\
2 \bar m (2 b_0+b_D+b_F)&=-\sigma_{\pi N}=(-50\pm15){\rm MeV},\label{eq:2b0bDbF}\\
2m_s(b_0+b_D-b_F) &=-\sigma_s=(-41.3\pm7.7){\rm MeV}.\label{msb0bDbF}
\end{align}
where we used $\bar m=3.5^{+0.7}_{-0.2}$MeV~\cite{Agashe:2014kda}.
For the first line we used the results from~\cite{Crivellin:2013ipa}
\begin{equation}\label{eq:mbbBc5}
\bar m (b_D+b_F)= B c_5 (m_d -m_u)\frac{m_d+m_u}{m_d-m_u},
\end{equation}
with $B c_5 (m_d -m_u)=(-0.51\pm0.08)\,$MeV \cite{Crivellin:2013ipa,
  Gasser:1982ap}, and $m_u/m_d=0.47\pm0.04$ \cite{Agashe:2014kda}. The
above results can be transcribed to 
\begin{equation}\label{eq:mbarbbb}
\begin{split}
\bar m (b_D+b_F)&=(-1.41\pm0.24){\rm MeV}, \\
\bar m b_0&=(-11.8\pm 3.8){\rm MeV}, \\
\bar m (b_0+b_D-b_F)&=(-0.75\pm 0.14){\rm MeV},
\end{split}
\end{equation}
where we used $m_s/\bar m = 27.5\pm1.0$~\cite{Agashe:2014kda}. From
here we get 
\begin{equation}
\bar m b_0=(-12.5\pm 3.8){\rm MeV}\,,\qquad \bar m b_D=(4.8\pm1.9){\rm MeV}\,, \qquad \bar m b_F=(-6.2\pm 1.9){\rm MeV}\,.
\end{equation}
Using symmetrized errors on $\bar
m=3.5^{+0.7}_{-0.2}$\,MeV~\cite{Agashe:2014kda} this gives at the
renormalization scale $\mu=2$\,GeV 
\begin{equation}\label{eq:b0bDbF}
b_0=-3.7\pm 1.4\,,\qquad b_D=1.4\pm0.8\,, \qquad b_F=-1.8\pm 0.8\,.
\end{equation}
Note that the errors in the last set of relations are large because of
the relatively poorly known $\bar m$.

The low-energy constants $D$, $F$, $G$ multiplying the axial vector
currents can be expressed in terms of the matrix elements
\begin{equation}\label{eq:def:Deltaq}
2 s^\mu \Delta q_p = \langle p_v |\bar q \gamma^\mu \gamma_5 q|p_v \rangle_Q\,,
\end{equation}
where $p$ is a proton state at rest, $s^\mu$ is the proton spin (or
polarization) vector such that $s^2=-1, s\cdot p=0$, see,
e.g. \cite{Barone:2001sp}, and the matrix element is evaluated at
scale $Q$. We work in the isospin limit so that \eqref{eq:def:Deltaq}
gives also the matrix elements for neutrons with $d \leftrightarrow u$
exchanged,
\begin{equation}
\Delta u\equiv \Delta u_p=\Delta d_n\,, \qquad
\Delta d\equiv \Delta d_p=\Delta u_n\,.
\end{equation}
The matrix elements $\Delta q$ are scale dependent. The non-isosinglet
combinations $\Delta u-\Delta d$ and $\Delta u +\Delta d- 2 \Delta s$
are scale independent, since they are protected by non-anomalous Ward
identities. The isovector combination
\begin{equation}\label{eq:App:u-d}
\Delta u-\Delta d=g_A=1.2723(23),
\end{equation}
is determined precisely from nuclear $\beta$ decays
\cite{Agashe:2014kda}. For the remaining two combinations we use
lattice QCD determinations \cite{QCDSF:2011aa, Engelhardt:2012gd,
  Abdel-Rehim:2013wlz, Bhattacharya:2015gma, Abdel-Rehim:2015owa,
  Abdel-Rehim:2015lha}. Following \cite{diCortona:2015ldu}, the
averages of lattice QCD results give $\Delta u+\Delta d=0.521(53)$ and
$\Delta s=-0.026(4)$ in $\overline{\rm MS}$ at $Q=2\,$GeV. Combining
with Eq.~\eqref{eq:App:u-d} this gives~\cite{diCortona:2015ldu}
\begin{equation}
\Delta u=0.897(27), \qquad \Delta d=-0.376(27), \qquad \Delta s=-0.026(4),
\end{equation}
all at the scale $Q=2\,$GeV. At LO in the chiral expansion we have
then 
\begin{equation}
2D=\Delta u-2 \Delta d +\Delta s\,, \qquad 2F=\Delta u-\Delta s\,, \qquad G=\Delta d\,,
\end{equation}
so that at $Q=2$ GeV
\begin{equation}\label{eq:DFG}
D=0.812(30)\,, \qquad F=0.462(14)\,, \qquad G=-0.376(28)\,.
\end{equation}
Note that the scale invariant combination 
\begin{equation}
D+F=g_A=1.2723(23),
\end{equation}
is determined more precisely than $D$ and $F$ separately.

Proton and neutron magnetic moments fix the values of the coefficients
$g_{4}$, $g_5$, $g_4'$ in Eq.~\eqref{eq:HBChPT2-vec}. Using the NLO
quark vector currents \eqref{eq:JVNLO} (cf. also~\eqref{eq:tildeJqV},
\eqref{eq:tildeJV:exp:s}, \eqref{eq:HBChPTQED}) one obtains 
\begin{align}
 \frac{2}{3}(g_4-g_4')_u& + \frac{1}{3}(g_4')_d -
\frac{1}{3}(g_5-g_4-g_4')_s =
\frac{\mu_p}{m_N}\,,
\\
-\frac{1}{3}(g_4-g_4')_d& - \frac{2}{3}(g_4')_u -
\frac{1}{3}(g_5-g_4-g_4')_s = 
\frac{\mu_n}{m_N}\,,
\end{align}
where $\mu_p=2.79$ and $\mu_n=-1.91$ are the values for proton and
neutron magnetic moments in units of nuclear magnetons $\hat\mu_N =
e/(2m_N)$~\cite{Agashe:2014kda}. Above we denoted with subscripts
which quark current $\tilde J_q^{V,\mu}$ the contributions originate
from. The $s$ quark contributions to the proton and neutron magnetic
moments are the same in the isospin limit,
giving~\cite{Sufian:2016pex} (see also~\cite{Green:2015wqa}) 
\begin{equation}\label{eq:mus}
 -\frac{1}{3}(g_5-g_4-g_4')= \frac{\mu_s}{m_N}= \frac{-0.073(19)}{m_N}\,.
\end{equation}
We then have 
\begin{equation}
g_4 = \frac{\mu_p-\mu_n}{m_N}=\frac{4.70}{m_N}\,, \qquad g_4'=-\frac{\mu_p+2\mu_n}{m_N}=\frac{1.03}{m_N}\,,
\end{equation}
neglecting the small corrections due to $\mu_s$. For notational
convenience we also define
\begin{equation}
\mu_p=\frac{4}{3}\hat \mu_u^p-\frac{1}{3}\hat \mu_d^p\,, \qquad 
\mu_n=-\frac{2}{3}\hat \mu_d^n+\frac{2}{3}\hat \mu_u^n\,,
\end{equation}
where $\hat \mu_{u,d}^p$ and $\hat \mu_{u,d}^n$ are the contributions
to the proton and neutron magnetic moments from the $u$- and $d$-quark
currents (the hats indicate that the quark charges have been factored
out from the definitions). Isospin relates contributions to neutron
and proton, giving 
\begin{equation}\label{eq:nuc:mm:quarks}
\hat \mu_u^p=\hat \mu_d^n=1.84\,, \qquad \hat \mu_d^p=\hat \mu_u^n=-1.03\,.
\end{equation}

Note that in the numerics it is advantageous not to use $B_0$
directly, but rather the numerical values for the products $B_0
m_q$. We can use the relation $2 B_0 \bar m= m_{\pi}^2$ to write
\begin{equation}
\begin{split}\label{eq:B0mq:num}
B_0 m_u & = \frac{m_\pi^2}{1+m_d/m_u}=  (6.2\pm0.4) \times 10^{-3} \,\text{GeV}^2\,,\\
B_0 m_d & = \frac{m_\pi^2}{1+m_u/m_d} =  (13.3\pm0.4) \times 10^{-3} \,\text{GeV}^2\,,\\
B_0 m_s & =\frac{m_\pi^2}{2}\frac{m_s}{\bar m} = (0.27 \pm0.01) \,\text{GeV}^2\,,
\end{split}
\end{equation}
using the ratios $m_u/m_d=0.47\pm0.04$, $m_s/\bar m =
27.5\pm1.0$~\cite{Agashe:2014kda}, and the charged-pion mass for
$m_\pi$.

\section{HDMET for Majorana fermions}
\label{app:Majorana}

In this appendix we list the changes that need to be made in our
results if DM is a Majorana fermion. The changes to the final results,
given for Dirac fermions in \eqref{eq:CNR1:app}-\eqref{eq:CNR11:app},
are straightforward. The Wilson coefficient $\hat \C_{1,2}^{(5)}$ and
$\hat \C_{1,q;3,q}^{(6)}$ are zero, because the corresponding
operators vanish for Majorana DM. All the other Wilson coefficients in
\eqref{eq:CNR1:app}-\eqref{eq:CNR11:app} need to be multiplied by an
extra factor of 2 that arises from matching onto HDMET, as we explain
below.\footnote{We adopt the same notation, $\chi$, for Majorana DM as
  we did for Dirac DM. No confusion should arise as this abuse of
  notation is restricted to this appendix.}

We construct the HDMET for Majorana fermions,
following~\cite{Kopp:2011gg}, by splitting the small and large
components according to\footnote{Alternatively, one could instead
  impose $\chi= e^{-i m_\chi v \cdot x} \sqrt{2} \big[ \chi_v (x) +
    X_v (x) \big]=e^{i m_\chi v \cdot x} \sqrt{2} \big[ \chi_v^c (x) +
    X_v^c (x) \big]$, with $ \chi_v=\frac{1}{2}(1 + {\slashed v})
  \chi_v $, $ X_v=\frac{1}{2}(1 - {\slashed v}) X_v $
  \cite{Hill:2011be,Biondini:2013xua}. The tree-level matching leads
  to the same results.}
\begin{equation}\label{eq:chi-field-def:Maj}
  \chi (x) = e^{-i m_\chi v \cdot x} \big[ \chi_v (x) + X_v (x) \big] +e^{i m_\chi v \cdot x} \big[ \chi_v^c (x) + X_v^c (x) \big]\,,
\end{equation}
where $ \chi_v=\frac{1}{2}(1 + {\slashed v}) \chi_v $, $
X_v=\frac{1}{2}(1 - {\slashed v}) X_v $, while the charge conjugated
fields are given by $\chi_v^c=C \bar \chi_v^T $, $X_v^c=C \bar X_v^T
$. Here $C$ is the charge conjugation matrix satisfying
$C^\dagger=C^T=C^{-1}=-C$ and $C \gamma_\mu^T C^{-1}=-\gamma_\mu$,  for instance,
one can choose $C=-i \gamma_2 \gamma_0$.  The ``small-component''
field $X_v$ carries momenta of order ${\mathcal O}(2 m_\chi)$ and is
integrated out. At tree-level one has the relation 
\begin{equation}
\chi=e^{-i m_\chi v \cdot x} \Big(1 +\frac{i \slashed
  \partial_\perp}{i v\cdot \partial+2 m_\chi-i \epsilon}\Big)
\chi_v+
e^{i m_\chi v \cdot x} 
\Big(1 -\frac{i \slashed
  \partial_\perp}{i v\cdot \partial-2 m_\chi+i \epsilon}\Big) \chi_v^c 
\,.\label{eq:chi:rel:Maj}
\end{equation}
Note that $\chi$ is self-conjugate, $\chi^c=\chi$, while the HDMET
field $\chi_v$ is not self-conjugate.  The HDMET field describing
Dirac fermion is also not self-conjugate. Still, there is a difference
between HDMET describing Majorana and Dirac fermions, since for
Majorana fermion the HDMET Lagrangian is symmetric under $v^\mu\to
-v^\mu$, $\chi_v\to \chi_v^c$~\cite{Hill:2011be,Kopp:2011gg}.

The relativistic Lagrangian for Majorana DM, ${\cal L}\supset
\frac{1}{2}\bar \chi i \slashed \partial \chi- \frac{1}{2} m_\chi \bar
\chi \chi$, then already leads to the canonically normalized HDMET
Lagrangian
\begin{equation} 
\label{eq:HDMET:Maj}
{\cal L}_{\rm HDMET}= \bar \chi_v (i v \cdot \partial) \chi_v
+\frac{1}{2m_\chi}\bar \chi_v (i \partial_\perp)^2 \chi_v+\cdots+{\cal
  L}_{\chi_v}.
\end{equation}
The higher dimension interaction Lagrangian, ${\cal L}_{\chi_v}$, is
given still by \eqref{eq:ewDM:Lnf5}. Due to the Majorana nature of
$\chi$, however, not all operators enter: the operators that are odd
under $v^\mu\to -v^\mu$, $\chi_v\to \chi_v^c$ vanish. Neglecting
radiative corrections to the matching conditions we have 
\begin{align}
\label{eq:HDMETlimit:scalar:Maj}
\bar \chi \chi&\to 2 \bar \chi_v \chi_v+\cdots, 
\\
\label{eq:HDMETlimit:pscalar:Maj}
 \bar \chi i \gamma_5
\chi &\to \frac{2}{ m_\chi}\partial_\mu \big(\bar \chi_v S_\chi^\mu \chi_v \big) + \ldots\,,
\\
\label{eq:axialDM:expand:Maj}
\bar \chi \gamma^\mu \gamma_5 \chi & 
\to  4 \bar \chi_v S_\chi^\mu \chi_v -2 \frac{i}{m_\chi} v^\mu \bar \chi_v S_\chi\cdot \lrpartial \chi_v
+\cdots,
\end{align}
which differs by an extra factor of 2 compared to the Dirac fermion
case, while the remaining currents vanish, $\bar \chi \gamma^\mu \chi
\to 0$, $\bar \chi \sigma^{\mu\nu} \chi \to 0$, $\chi \sigma^{\mu\nu}
i\gamma_5 \chi\to 0$.

\bibliography{paper}

\end{document}